\documentclass[useAMS,usenatbib]{mn2e}
\usepackage[dvips]{graphicx}
\usepackage{amssymb}
\usepackage{color}

\title[Break down of optical-IR relations]{Resolved optical-infrared
  SEDs of galaxies: universal relations and their break-down on local
  scales}

\author[Zibetti \& Groves]{Stefano Zibetti$^{1,2}$\thanks{E-mail:
    zibetti@dark-cosmology.dk} \& Brent Groves$^{2}$ \\
  $^{1}$Dark Cosmology Centre, Niels Bohr Institute - University of
  Copenhagen Juliane Maries Vej 30, DK-2100 Copenhagen, Denmark\\
  $^{2}$Max-Planck-Institut f\"ur Astronomie, K\"onigstuhl 17, D-69117 Heidelberg, Germany\\
}
\begin{document}
\citestyle{mn2e}
\bibliographystyle{mn2e}

\date{Accepted . Received ; in original form }

\pagerange{\pageref{firstpage}--\pageref{lastpage}} \pubyear{2011}

\maketitle

  \label{firstpage}

\begin{abstract}
  A large body of evidence has demonstrated that the global rest-frame
  optical and infrared colours of galaxies correlate well with each
  other (i.e.~$(u-g)$, with $(r-8\mu\mathrm{m})$), as well as with
  other galactic properties such as surface brightness and morphology.
  However the processes that lead to the observed correlations are
  contrary; the stellar light that contributes to the optical is
  readily absorbed by dust which emits in the IR. Thus on small scales
  we expect these correlations to break down. We examine here seven
  nearby galaxies ranging from early- to late-types, which have all
  been smoothed to the same physical scale and signal-to-noise ratio,
  using data from the optical to the mid-IR ($u$-band to $8\mu$m). By
  examining these galaxies on a pixel-by-pixel basis we demonstrate
  that there is disconnect between the optical and IR when normalized
  to the near-IR ($H$-band). For five of the seven galaxies we can
  decompose this disconnect into two distinct components through a
  Principal Component Analysis of the $H$-band normalized spectral
  energy distribution of the pixels: one mainly correlated with
  variations in the IR, the other correlated with variations in the
  optical. The two exceptions are the elliptical galaxy NGC\,4552,
  whose SED can be well reproduced using a single principal component,
  and the highly inclined spiral NGC\,3521, due to its complex dust
  geometry. By mapping these two components in the five ``regular''
  galaxies, it is clear they arise from distinct spatial regions. By
  comparing these components with the surface brightnesses in
  H$\alpha$ and $H$-band we demonstrate that the IR dominated
  component is strongly associated with the specific star-formation
  rate, while the optical-dominated component is broadly associated
  with the stellar mass density. However, when the pixels of all
  galaxies are compared, the well known optical--IR colour
  correlations return, demonstrating that the variance observed within
  galaxies is around a mean which follows the well-known trend. As a
  final step, we extend this work by examining the extremely tight
  correlations observed between the IRAC--near-IR colours, and
  demonstrate that these correlations are tight enough to use a single
  IRAC--near-IR colour (i.e.~$8\mu{\rm m}-H$) to determine the fluxes
  in the other IRAC bands. These correlations arise from the differing
  contribution of stellar light and dust to the IRAC bands, enabling
  us to determine pure ``stellar'' colours for these bands, but still
  demonstrating the need for dust (or stellar) corrections in these
  bands when being used as stellar (dust) tracers.
\end{abstract}


\section{Introduction}\label{intro_sec}
One of the remarkable aspects of astronomy is the strong correlations
that exist between the broad observable properties of galaxies. The
morphological sequence of galaxies first noted by \citet{hubble26}
correlates well with surface brightnesses, sizes, and optical colours.
These in turn correlate with and arise from the intrinsic physical
properties of these galaxies, such as the total stellar masses
($M_{*}$), star formation rates (SFR), and mean stellar ages and
metallicities (for reviews on these correlations see
\citealt{roberts_haynes94} and \citealt{kennicutt_98}).

The rest-frame optical colours of galaxies broadly correlate with each
other (i.e.~$(u-g)$, with $(g-r)$), as well as with surface
brightness. While forming continuous sequence in colours, galaxies can
be broadly classified into `blue' and `red' galaxies, especially when
compared with luminosity or absolute magnitude, with a strong
correlation with late- and early-type respectively
\citep{blanton03_sersic}. The variation in optical colours arise from
differences in the star formation histories of galaxies, with UV to
blue light dominated by short-lived massive stars, while the red to
near infrared light is more sensitive to the total amount of
stars. This is complicated however by the presence of interstellar
dust, which attenuates and reddens the stellar light \citep[see
e.g.][]{kennicutt_98,bc03}.  Thus the colours of a galaxy broadly
measure the ratio of young to old stars, which is generally
parameterized by the ratio of current to past star formation
($b=$SFR/$<$SFR$>$), or the ratio of the current star formation rate
to the total stellar mass, the specific star formation rate
(sSFR=SFR/$M_{*}$). In the local universe, these intrinsic physical
parameters, and thus a galaxy's colours, are related to the total
stellar mass of a galaxy \citep[e.g.][]{brinchmann+04}, with more
massive galaxies having lower sSFRs and thus redder colours. The
stellar mass, and thus the sSFR and colours, are related to the size,
concentration, surface brightness, and mean metallicity of a galaxy
\citep[e.g.][]{kauffmann+03b,tremonti+04,gallazzi+05}.

In a similar manner, the IR colours of galaxies are also strongly
associated with the intrinsic properties of the galaxies, with the IR
fluxes increasing relative to the near-IR fluxes (i.e.~becoming
redder) from early- to late-type galaxies \citep[see e.g.][especially
fig.\ 4]{kennicutt_98}, and with increasing sSFR
\citep{dacunha+08}. As with the optical colours, the infrared is
sensitive to the younger stellar populations, as the IR emitting dust
is heated predominantly by UV--blue light. In fact, the fluxes in both
of these wavelength regions can be used as measures of a galaxy's SFR
\citep{kennicutt_98} and a clear correlation between the
\textit{global} optical and IR colours is observed, in the sense that
optically blue galaxies are also found to have enhanced IR emission
\citep[e.g.][]{hogg+05}.

However on a physical basis, the processes that lead to the fluxes in
these two wavelength regimes are contrary: the radiation from young
stars that dominate the UV-blue optical colours is also that
preferentially absorbed by dust which then re-emits in the IR
\citep[see e.g.][]{draine_03}. This contrary nature is exacerbated by
the association of the youngest stars with the clouds of gas and dust
from which they form, leading to relatively higher attenuation
observed in star forming regions \citep[see
e.g.][]{calzetti97,charlot_fall}. This contradiction of the observed
galaxy scale correlation of both blue colours and IR excess and the
physical contrary nature of dust absorption and UV-blue light emission
indicate that on some spatial scale \emph{within} galaxies this
correlation must break down.

In some respects this difference is expected as the blue light and
infrared emission arise from physically distinct components
(i.e.~stars and dust), and this correlation will break down once these
different components are resolved out, as seen within our own
Galaxy. However, this distinction between IR and blue light exists on
larger scales within galaxies, as we show here by examining seven
nearby galaxies that span a range of galaxy types from early to
late. By examining all galaxies on the same spatial scales we
demonstrate that the variation of the IR light and the blue light
occurs in spatially distinct regions, and show the scales over which these
regions occur. Importantly, these distinct components appear to be
associated with different physical quantities that are correlated on
galaxy scales: the specific star formation rate (sSFR) and the stellar
mass density.

In Section \ref{sample_sec}, we introduce the sample of galaxies, the
multiwavelength dataset and the relative reduction. In section
\ref{sec_SEDcorr}, we analyse the resolved SEDs of the individual
galaxies, showing the correlations that exist to different degrees
between the IR and optical colours, analyzing them by means of
principal component analysis, and showing how they depend upon other
local observables, namely the NIR and H$\mathrm{\alpha}$ surface
brightness. In Section \ref{globalcorr_sec} we show how these
\textit{local} colour correlations (and lack thereof) result in
stronger \textit{global} correlations. Section \ref{sec_IRcolors} is
devoted to an in-depth investigation of the SEDs between 1.65 and
8$\mu$m, which we describe as a universal 1-parameter family: fitting
formulae to obtain the luminosity at intermediate wavelengths given
$H$-band and 8$\mu$m and estimates of purely stellar SEDs are
provided. Section \ref{sec_conclusions} concludes the paper with a
summary of the principal results.

\section{Sample and imaging data}\label{sample_sec}
The present sample of 7 galaxies is drawn from the original sample of
nine galaxies analyzed in \citet[][ZCR09 hereafter]{ZCR09}. We require
galaxies to be part of the Spitzer Infrared Nearby Galaxies Survey
\citep[SINGS,][]{SINGS_kennicutt+03}, a comprehensive imaging and
spectroscopic study of 75 nearby galaxies (D $< 30$ Mpc) conducted in
the IR with the Spitzer Space Telescope, and have SDSS imaging
\citep{SDSS} as well as deep NIR imaging in H band (1.65$\mu$m) either
from GOLDMine \citep{goldmine} or UKIDSS \citep{lawrence+07}.  In
addition we require the galaxies to be at most $\approx 17$ Mpc away
(corresponding to the assumed distance to the Virgo cluster,
\citealt{gavazzi+99}), in order to resolve scales of approximately 200
pc at all wavelengths, from SDSS $u$-band to IRAC 8$\mu$m. This leaves
us with a small, yet representative, sample of regular galaxies in the
local Universe. The 7 galaxies are listed in Table \ref{sample_tab}
along with their coordinates (according to NED/RC3), assumed
distances, corresponding angular scale, source and depth of the NIR
imaging (see below).
\begin{table*}
\begin{minipage}{\textwidth}
\caption{The sample.}\label{sample_tab}
\begin{tabular}{lrrrrrccr}
  \hline
  Name      & RA & Dec  & Distance & Angular scale & NIR source & Limiting SB$(H)$\\
  & (J2000.0)  & (J2000.0) & Mpc & pc arcsec$^{-1}$ && MJy Sr$^{-1}$       \\
  (1) & (2) & (3)  & (4) & (5) & (6) & (7) \\
  \hline

  NGC\,3521 & 11h05m48.6s & $-$00d02m09s  & 9.2 &  45 & UKIDSS   & 0.30\\
  NGC\,4254 & 12h18m49.6s & +14d24m59s    &17.1 &  82 & GOLDMine & 0.21\\
  NGC\,4321 & 12h22m54.9s & +15d49m21s    &17.1 &  82 & GOLDMine & 0.32\\
  NGC\,4450 & 12h28m29.6s & +17d05m06s    &17.1 &  82 & GOLDMine & 0.35\\
  NGC\,4536 & 12h34m27.0s & +02d11m17s    &17.1 &  82 & GOLDMine & 0.39\\
  NGC\,4552 & 12h35m39.8s & +12d33m23s    &17.1 &  82 & UKIDSS   & 0.47\\
  NGC\,4579 & 12h37m43.5s & +11d49m05s    &17.1 &  82 & GOLDMine & 0.53\\
\end{tabular}
\end{minipage}
\end{table*}
With these seven galaxies we cover the full morphological range of
regular ``giant'' ($3.9~10^9<M^*/M_\odot<9.3~10^{10}$) galaxies, from
elliptical to Sc spiral, also corresponding to a range of specific
star formation rates from $<10^{-11}$ to $\approx 10^{-9}$
yr$^{-1}$. In Table \ref{sample_phys_tab} we present these and other
essential morphological and physical properties of the galaxies in the
sample. The inferred metallicities also span the full range for
galaxies of such stellar mass. Only two spirals have an apparent
inclination in excess of 45 degrees, thus allowing us to study not
only a ``clean'' sample of low inclination discs, for which the
radiative transfer in the dusty ISM is expected to be relatively
simple, but also to explore possible complications due to a larger
inclination. We note that nuclear activity from AGN does not influence
our analysis significantly, as we include only weak or obscured AGNs
(no type 1) which affect only a very limited number of central
pixels. The statistical effect of these AGN affected pixels over the
several thousands of pixels in each galaxy is negligible.

\begin{table*}
\begin{minipage}{\textwidth}
\caption{Physical properties of the sample galaxies.}\label{sample_phys_tab}
\begin{tabular}{lcrrrrrrrl}
 \hline
 Denomination & Morph. type & Inclination & $M^*$  & SFR & sSFR & Z(gas) & Nuclear classification\\
 &             & degrees     & log M$_\odot$& log M$_\odot$ yr$^{-1}$&log yr$^{-1}$& 12 + log (O/H) &\\
 (1) & (2) & (3)  & (4) & (5) & (6) & (7) & (8)\\
 \hline

 NGC\,3521 & SAB(rs)bc  & 64 & $10.52$& $0.33$&$-10.19$& $9.01$ & LINER \\
 NGC\,4254 & SA(s)c     & 30 & $9.71$ & $0.82$ & $-8.89$& $9.17$ & LINER \\
 NGC\,4321 & SAB(s)bc   & 32 &$10.46$ &$0.55$ & $-9.91$& $9.11$ & LINER \\
 NGC\,4450 & SA(s)ab    & 43 &$10.80$ &$0.03$ &$-10.77$ & $9.13$ & LINER\\
 NGC\,4536 & SAB(rs)bc  & 67 &$9.59$ & $0.49$&$-9.10$ & $9.00$& None\footnote{\cite{moustakas+10}
find line ratios consistent with very low-level AGN activity.}\\
 NGC\,4552 & E0-1       & $\ldots$ &$10.97$ &$\ldots$ &$\ldots$ & $9.12$ & LINER\\
 NGC\,4579 & SAB(rs)b   & 38 & $10.06$ & $-0.06$ & $-10.12$& $9.22$ & LINER\\
\hline
\end{tabular}\\
{\bf Notes.} Columns: (1) NGC name. (2) Morphological type according
to NED/RC3 \citep{RC3}, following the classification scheme of
\cite{devaucouleurs_59}: S indicates spiral, E elliptical followed by
a number indicating the ellipticity (0 means perfectly round); A is
normal unbarred spiral, B barred, AB weakly barred; (s) is s-shaped, (r)
indicates the presence of a ring, (rs) both ringed and s-shaped; a, b,
c, d are the same as in the classical Hubble spiral classification.
(3) Inclination (for spirals only), according to
\cite{RC3,jarrett+03}. (4) Stellar mass from Skibba et al. (ApJ in
press) assuming a \cite{kroupa_01} IMF, except NGC\,4450 and
NGC\,4552. Stellar masses for these two galaxies have been computed
multiplying the total $H$-band luminosity given by 2MASS-LGA
\citep{jarrett+03} times the $M^*/L_{H}$ computed according to ZCR09
using the $g-i$, $i-H$ colours measured from our images and corrected
to a \cite{kroupa_01} IMF. (5) Star formation rate from Skibba et al.
(ApJ in press), based on H$\alpha+24\mu$m luminosity, except for  NGC\,4450 which is from
\cite{calzetti+10}. (6) Specific
star formation rate, sSFR$=$SFR$/M^*$. (7) Gas phase metallicity, from
the characteristic oxygen abundance from \cite{moustakas+10} based on the
\cite{KK04} calibration, except for NGC 4450, NGC 4552, and NGC 4579,
which are use the luminosity-metallicity relation. (8) Nuclear activity, from NED.
\end{minipage}
\end{table*}

The SDSS and $H$-band
images are processed, mosaiced (when required) and calibrated as
explained in ZCR09. Spitzer IRAC images are publicly available from
SINGS, already calibrated in MJy/Sr and sky-subtracted. For part of
our analysis we also use narrow-band H$\alpha$ imaging, provided as an
ancillary dataset of SINGS: details about the flux calibration of the
``net'' (i.e. continuum-subtracted) images is given in the technical
documentation that accompanies the data release. No H$\alpha$ image is
provided for the elliptical galaxy NGC\,4552, which allegedly does not
have any detectable HII regions.

For all images the properties of the background, which is the dominant
contributor to the noise budget at all wavelengths, are carefully
computed. The mean background is subtracted by fitting a linear
polynomial to the regions surrounding the galaxies, which are selected
not to display any strong contamination from bright sources nor to be
affected by obvious artifacts. Noise properties in terms of
pixel-based r.m.s. and fluctuations on scales of the order of a
fraction of the galaxy's effective radius are also determined in these
regions and are then used to determine the adaptive smoothing strategy
and the surface brightness cuts to be applied in the following
analysis.

We manually remove all obvious contaminating sources (foreground
stars, background galaxies) by replacing those pixels with the
surrounding average background plus noise, using the task
\texttt{imedit} in IRAF. For more extended or unamendable artifacts
(e.g. the extended blooming due to the bright nucleus of NGC\,4536 in
IRAC bands) we simply mask out those pixels (and the corresponding
ones in all bands) and discard them from the following analysis.

Pixel-by-pixel SED analysis requires two more steps: the placement of
all images on a common plate scale, and to ensure that in each pixel
at all wavelengths a sufficiently high signal-to-noise ratio (S/N) is
provided. The imaging data involved in this analysis have point spread
functions (PSF, for optical or $H$-band) or point response functions
(for Spitzer IRAC bands) FWHM ranging from approximately 1 arcsec
(some SDSS images in good seeing conditions) to 2 arcsec (IRAC
8$\mu$m), while pixel scales range from 0.4 (SDSS, UKIDSS) to 1.6
(some GOLDMine $H$-band images) arcsec per pixel. By re-sampling the
images with a common pixel scale of $>2.5$~arcsec pix$^{-1}$ we ensure
that most of the flux of a point-like source is contained in one
pixel. We can therefore avoid convolving all images to a common PSF,
which would be hardly feasible for a number of H-band images where the
PSF is severely under-sampled and not enough stars are available to
compute it properly. We choose therefore to resample the images of the
6 galaxies belonging to the Virgo cluster to a pixel scale of 2.5
arcsec, corresponding to 207 pc pix$^{-1}$ for the assumed distance to
Virgo of 17.1 Mpc \citep{gavazzi+99}. For the remaining galaxy,
NGC\,3521, we resample to a pixel scale of 4.65 arcsec pix$^{-1}$,
corresponding to the same physical resolution of 207 pc pix$^{-1}$ as
for the other galaxies.

In order to extend our analysis with a sufficiently high
signal-to-noise ratio (S/N) out to a large fraction of the optical
radius, as in ZCR09, we perform an adaptive median smoothing of the
images using the code ADAPTSMOOTH \citep{adaptsmooth}, which also
allows the matching of the variable smoothing kernel among all bands. A
minimum S/N=20 per pixel is required in all bands, except $u$-SDSS,
for which a minimum S/N of 10 is deemed as sufficient. In fact, this
band is imaged with a significantly lower sensitivity in the SDSS,
such that a more massive smoothing would be required and a smaller
radial extent could be covered if the same S/N=20 had to be
required. On the other hand the large leverage of this band on the SED
makes it possible to study pixel-to-pixel variations even with an
accuracy of 10\% only, as we show in the following analysis.  We note
that the H$\alpha$ images do not contribute to determine the size of
the smoothing kernel required at each pixel (as this would imply
missing a significant fraction of pixels where no substantial
H$\alpha$ emission is observed), but they are nevertheless smoothed
according to the common kernel as for all other
bands. Background-dominated noise is assumed for all images, with a
uniform r.m.s. equal to the one measured in the background regions. A
maximum smoothing radius of 4 pixels (2 for NGC\,3521) corresponding
to $\approx 800$ pc is allowed. All pixels for which the surface
brightness (SB) cannot be computed with the minimum required S/N even
after smoothing over the maximum radius are flagged and discarded from
the rest of the analysis. Furthermore, we also discard pixels whose SB
is less than 10 times the typical large scale background fluctuations,
in order to avoid spurious effects at the low SB end of the pixel
distribution. The limiting SB$(H)$ that is reached as a result of
these cuts is given in column 8 of Table \ref{sample_tab}.

Images of the galaxies in the ten broad bands and in H$\alpha$ after
editing, adaptive smoothing and SB cuts are shown in
Fig. \ref{fig_A_N03521} to \ref{fig_A_N04579} in Appendix
\ref{append_images_sed}, coded in grey intensity according to SB in
units of MJy Sr$^{-1}$.  In the rest of the paper we discuss the
analysis and the results for the full sample, but we show specific
figures and plots only for two galaxies which are representative of
late star-forming spirals (NGC\,4254) and early type galaxies
(NGC\,4450), with the remaining five being presented in online
appendices.

\section{Spatially resolved SEDs: internal trends within
  galaxies}\label{sec_SEDcorr}
\subsection{Pixel-by-pixel SEDs and colour-colour
  relations}\label{subsec_resolved_colors}
The processing described in the previous Section provides us with an
optical to mid-IR SED for each valid pixel, described by an array of
values for the SB in MJy Sr$^{-1}$ in ten broad bands ($u$, $g$, $r$,
$i$, $z$, $H$, 3.6, 4.5, 5.7 and 8$\mu$m) plus the surface brightness
in erg cm$^{-2}$ sec$^{-1}$ arcsec$^{-2}$ for the H$\alpha$ line
emission.  At each pixel we normalize $\log \nu f_\nu$ in each band to
$H$ band in order to be able to compare {\it on the same scale} the
SED of pixels spanning about two orders of magnitude in SB and hence
to study the spatial variations of the SED {\it shape} across the
galaxies.

The choice of $H$ band is dictated by three main reasons which make it
the natural point of normalization: \textit{i)} 1.65$\mu$m divides the
SED into one part of pure stellar emission (possibly dust extincted),
shortward of it, and one part affected by dust emission, longward of
it. \textit{ii)} the $H$ band is the one that minimizes the effects of
dust both in emission \citep[see e.g. Fig. 5 of][]{dacunha+08} and in
absorption, as the extinction is always smaller at longer
wavelengths. The $z$ band is equally unaffected by dust emission, but
is more sensitive to extinction; on the other hand, 3.6$\mu$m,
although even less affected by extinction than the $H$ band, is
severely affected by dust emission, as we show below. \textit{iii)}
$H$-band luminosity is probably the best proxy to stellar mass, as
shown e.g. in ZCR09, showing the minimum dependence of $M/L$ on age,
metallicity and dust; therefore normalizing to $H$-band is close to
normalizing to the stellar mass density at each pixel, a much more
fundamental parameter than any surface brightness.

The distribution of pixels in SED for each galaxy is given for
reference in the bottom right panels of Figs. \ref{fig_A_N03521} to
\ref{fig_A_N04579}, respectively, in the online Appendix
\ref{append_images_sed}. The number of pixels per bin of normalized
$\log \nu f_\nu$ in each band is represented by the colour
coding. Galaxy SED distributions display different amount of scatter,
which vary upon morphology and wavelength considered. As we show in
more detail in the following analysis, most of the scatter is
contributed by the mid-IR and the $u$ bands. It is interesting to note
that the integrated SED of a galaxy (flux-weighted average of pixels,
`+' signs in the figures) does not perfectly correspond to the
(number-weighted) mean pixel SED (`$\times$') or the most frequent SED
(peaks of the distributions). A similar effect was already noted by
ZCR09, where it turned out to cause a significant bias in stellar mass
estimates, and can be simply explained by the fact that regions of
different brightness are often characterized by systematically
different SEDs\footnote{A typical example occurs in spiral galaxies
  with a bright central bulge: the pixels dominated by the red bulge
  light are fewer but brighter than the more numerous but dimmer
  pixels in the disc where the SED is bluer. As a result, the
  flux-weighted SED will be redder than the number-weighted one.}.

\begin{figure*}
\includegraphics[width=\textwidth]{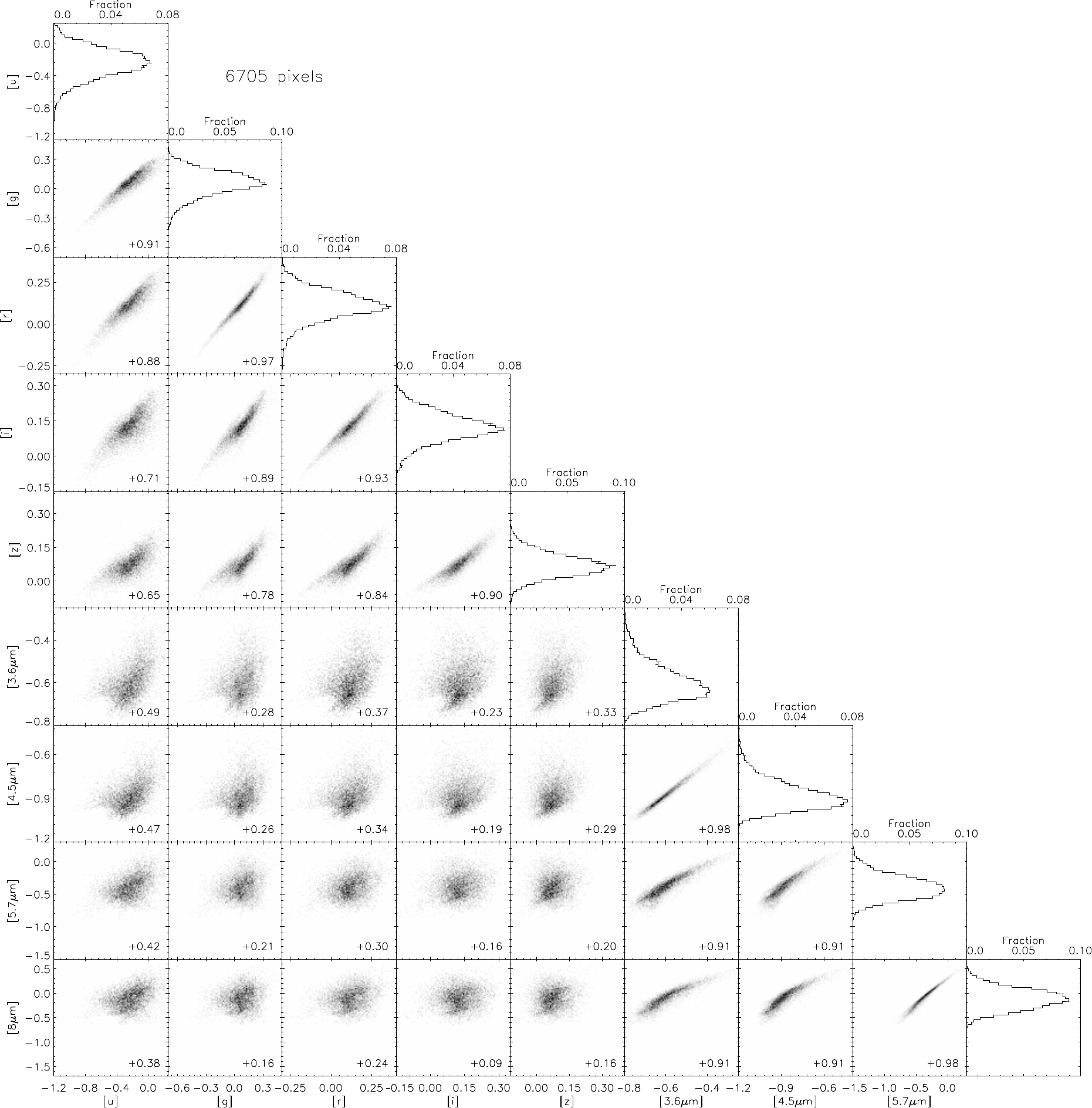}
\caption{Distribution of the pixels of NGC\,4254 in the ``colour-colour''
  spaces defined by the $\log \nu f_\nu$ fluxes normalized to
  $H$-band, for the $u$, $g$, $r$, $i$, $z$, 3.6, 4.5, 5.7 and 8$\mu$m
  bands. The grey-scale intensity linearly traces the density of pixels
  at each point in the colour-colour spaces. The Spearman rank
  correlation coefficient for the distribution of pixels is written at
  the bottom right corner of each corresponding panel. 
  The histograms on the diagonal show the
  distribution of pixels in normalized $\log \nu f_\nu$ colour space as
  labeled on the y-axis. The total
  number of pixels analyzed in NGC\,4254 is given at the top of the
  figure.}\label{fig_SEDcorr_N04254}
\end{figure*}

\begin{figure*}
\includegraphics[width=\textwidth]{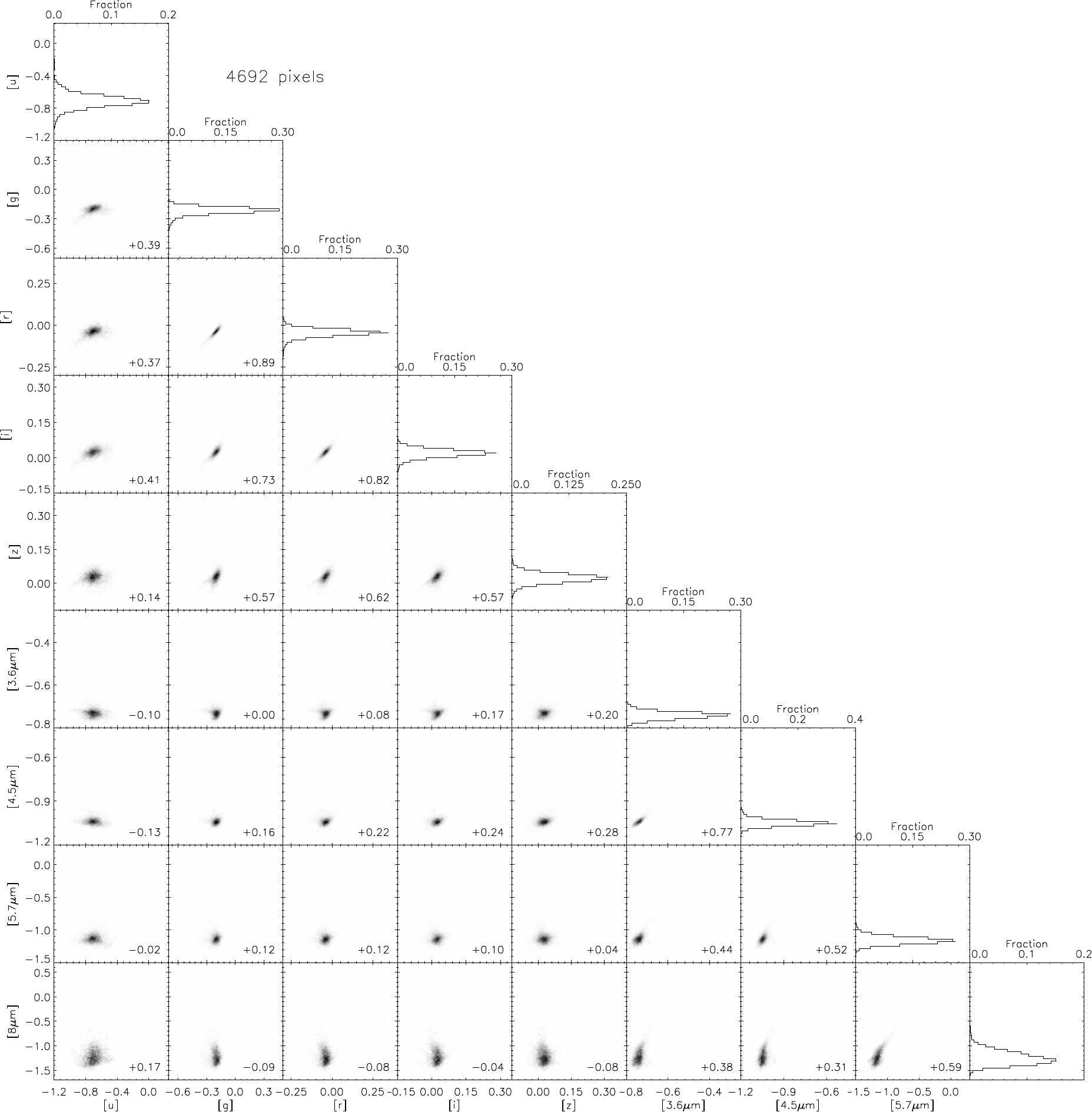}
\caption{Same as Fig. \ref{fig_SEDcorr_N04254}, but for
  NGC\,4450.}\label{fig_SEDcorr_N04450}
\end{figure*}

As a first step, we study the correlations between the normalized
fluxes in all possible pairs of bands for all pixels within each
galaxy. This is shown in Fig. \ref{fig_SEDcorr_N04254} and
\ref{fig_SEDcorr_N04450} for NGC\,4254 and NGC\,4450, respectively,
with the remaining galaxies shown in the online figures
\ref{fig_SEDcorr_N03521} to \ref{fig_SEDcorr_N04552}. In each panel of
the figures we plot the distribution of pixels in the $(\log \nu_Y
f_{\nu_Y} - \log \nu_H f_{\nu_H})$ vs. $(\log \nu_X f_{\nu_X} - \log
\nu_H f_{\nu_H})$ space (where $X$ and $Y$ are any two broad bands
among $u$, $g$, $r$, $i$, $z$, 3.6, 4.5, 5.7 and 8$\mu$m), which is
effectively a colour-colour space $(Y-H)$ vs. $(X-H)$, modulo a linear
transformation. The panels on the diagonal display the distribution of
the fraction of pixels as a function of $(\log \nu_Y f_{\nu_Y} - \log
\nu_H f_{\nu_H})$, which allows to directly derive a more quantitative
information about the relative contributions of pixels at different
colour. In all following discussion we refer to $(\log \nu_X f_{\nu_X}
- \log \nu_H f_{\nu_H})$ as ``$[X]$ colour'' for brevity.

By comparing Fig. \ref{fig_SEDcorr_N04254} and
\ref{fig_SEDcorr_N04450} it is immediately clear that the late-type,
star-forming galaxy NGC\,4254 spans a much broader range of colours
than the early type NGC\,4450, which is overall very uniform. Apart
from this, both galaxies display common features in the colour-colour
distributions, such that one can divide up these figures into three
distinct regions, corresponding to different spectral correlations. In
the upper left panels we observe very significant correlations between
optical colours, from $[u]$ to $[z]$. Similarly, the bottom right
panels display significant correlations between IRAC colours, from 3.6
to 8$\mu$m. The third region includes the bottom left corner and shows
IRAC colours vs. optical colours: correlations in this region are very
weak if present at all. In star forming galaxies an exception to this
is the $[u]$ colour, which appears to (weakly) correlate with IRAC
colours as well. This visual impression is quantitatively confirmed by
the Spearman rank correlation coefficients, which we computed for each
colour-colour distribution and are listed in the lower-right corner of
each corresponding panel.

From this analysis we can conclude that, on the local scales probed by
our actual pixellation (from 200 pc up to $\lesssim 1$kpc when the
effects of the smoothing are maximal), the shapes of the SED in a
given galaxy in the optical range can be described, to first order, as
a 1-parameter family and the same holds true for the IR range. The two
parts of the SED shortwards and bluewards of 1.65$\mu$m appear to be
largely uncorrelated. The optical SED is mainly driven by
light-weighted mean stellar age (and secondarily by dust extinction
and metallicity, see e.g. Fig. A1 in ZCR09 and \citealt{macarthur+04})
while the mid-IR SED probed by IRAC is determined by PAH features and
hot dust, which trace on-going star formation
\citep[e.g.][]{calzetti+07}. This tells us that on local scales
light-weighted mean stellar ages and actual on-going star formation
are largely uncorrelated, most likely because stellar aging and star
formation are subject to very different time scales, namely $\approx
10^{8-9}$ and $\approx 10^7$ years, respectively
\citep[e.g.][]{bruzual_charlot_93,kennicutt_98}.

It is also very notable that in star-forming galaxies the colours of
the two shortest-wavelength IRAC bands at 3.6 and 4.5$\mu$m correlate
extremely well with those of the longer-wavelength bands, which are dominated
by PAH features, but are very weakly correlated with the $z$ and $i$
bands which are very good tracers of the stellar mass
(e.g. ZCR09). This is in agreement with the finding of
\cite{mentuch+10}, who show, using pixel-by-pixel analysis, that the
\textit{colour excess} at 3.6 and 4.5$\mu$m with respect to pure
stellar emission extrapolated from $J$-band (1.25$\mu$m) is related to
star formation. Therefore caution should be used when adopting
images at 3.6$\mu$m alone to trace the stellar mass distribution in
galaxies with no correction for the `dust' contribution.

\subsection{Principal Component Analysis of resolved SEDs}\label{subsec_PCA}
In this section we use principal component analysis (PCA) of the SEDs
of all pixels, galaxy by galaxy, to show {\it i)} that the variation
of the optical part and of the IR part of the SED are uncorrelated and
{\it ii)} that these independent variations are spatially related to
the structure of a galaxy.

We refer the reader to \cite{jolliffe_02} for a thorough description
of principal component analysis. For the present purpose it is
sufficient to say that the goal of PCA is to ``reduce the
dimensionality of a data set consisting of a large number of
interrelated variables, while retaining as much as possible of the
variation present in the data set. This is achieved by transforming to
a new set of variables, the principal components (PCs), which are
uncorrelated, and which are ordered so that the first few retain most
of the variation present in all of the original variables''
\citep{jolliffe_02}. In our specific application, for each pixel $i$
of the $N_{\mathrm{pix}}$ pixels in a galaxy we consider the
9-dimensional SED vector given by the $H$-band normalized $\log \nu
f_\nu$, i.e. $\mathrm{SED}_i=\{\log \nu_X f_{\nu_X,i} - \log \nu_H
f_{\nu_H,i}\}_{X=u,g,r,i,z,3.6,4.5,5.7,8\mu\mathrm{m}}$. The set of
these $N_{\mathrm{pix}}$ SED vectors is then decomposed into principal
components using the \texttt{pcomp} function in IDL. Each SED$_i$ can
be then represented as
$\mathrm{SED}_i=\mathrm{SED_{mean}}+\sum_{j=1,9} a_{i,j}
\mathrm{PC}j$, where $\mathrm{SED_{mean}}$ is the arithmetic mean of
$\mathrm{SED}_i$ taken over all pixels, $\mathrm{PC}j$ are the
principal component vectors, which are obtained as eigenvectors of the
correlation matrix of the dataset, and $a_{i,j}$ are the corresponding
coefficients for $i$th pixel and $j$th component.
\begin{figure*}
\includegraphics[width=0.95\textwidth]{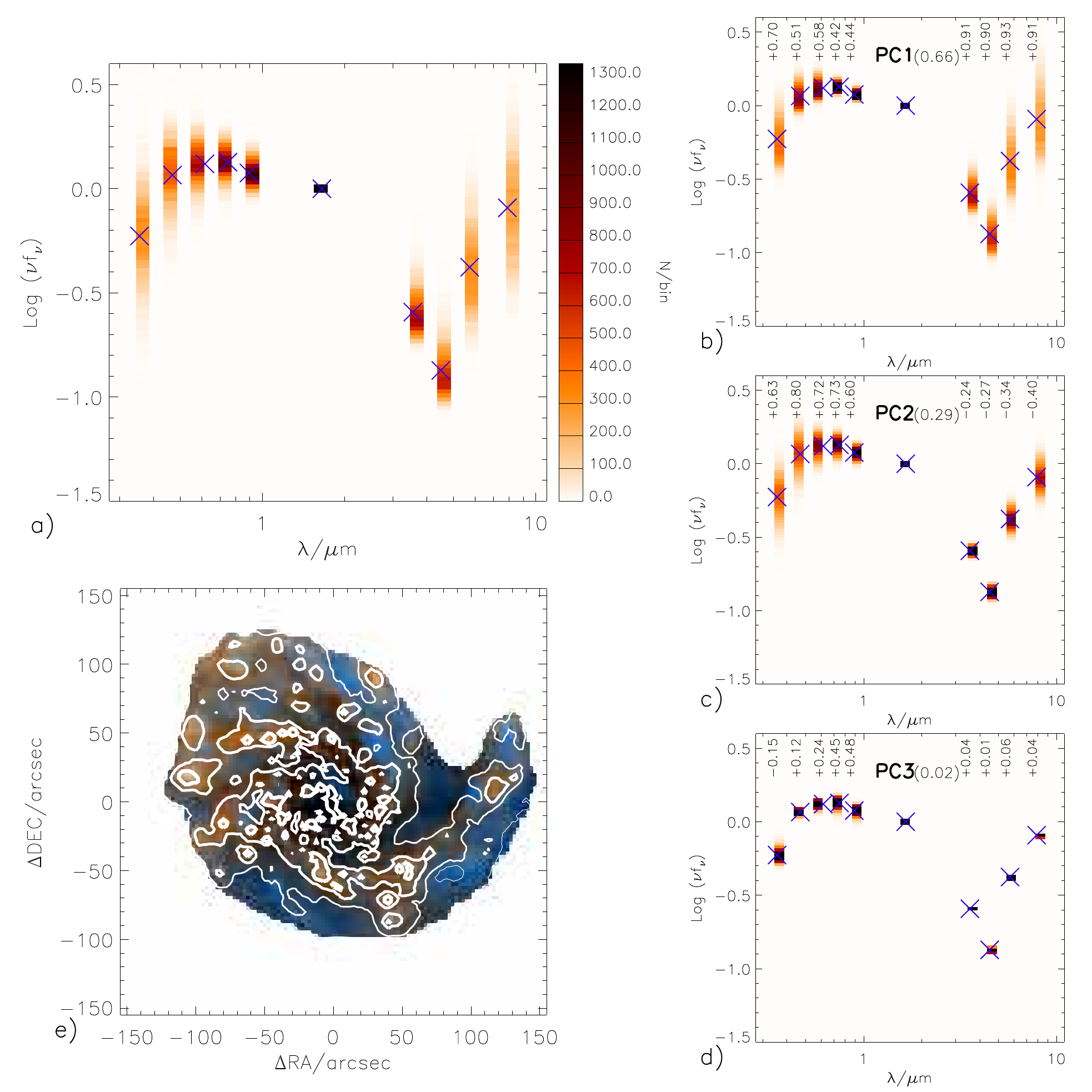}
\caption{SED pixel distribution and PCA for NGC\,4254. \textit{a)}
  displays the distribution of all pixels in terms of $H$-band
  normalized SED. At each wavelength, corresponding to the ten
  optical-IR broad bands, the different colour intensities represent
  the number of pixels in bins of 0.02 dex of $\log \nu f_\nu$, as
  shown by the key on the right side. The mean SED (i.e. the
  arithmetic mean of $\log \nu f_\nu$ for all pixels at each
  wavelength) is marked with blue crosses. \textit{b), c), d)} are
  analogous to figure \textit{a)}, except that the SED of each pixel
  is not the observed one, but the one reconstructed by using only the
  mean plus one single PC (1, 2 and 3 respectively) times the
  corresponding coefficient. The colour key is the same as in figure
  \textit{a)}. For each wavelength the Spearman rank correlation
  coefficient between the corresponding flux and the strength of the
  PC is written at the top of the graph. This number indicates both
  the relative strength and sign of the PC contribution to that
  wavelength.  The number in parenthesis beside the PC label is the
  fractional amount of variance contributed by the PC. Blue crosses
  are the mean SED, as in figure \textit{a)}. In \textit{e)} we
  simultaneously map the intensity of PC1 and PC2 on the galaxy
  image. In the RGB rendition, the intensity of PC1 is mapped to the
  red channel (spanning the range corresponding to the range between
  the 5th and 95th percentile of the distribution in PC1 coefficient)
  and PC2 is similarly mapped to the blue channel. The green channel
  intensity is just the mean of the blue and the red ones. Note that
  the range of PC coefficients is centered on $0$ and extends both
  positive (bright on the map) and negative (dark on the map). The
  overlaid contours correspond to different isophotal levels of
  H$\alpha$ emission: the five levels, ordered by increasing line
  thickness, correspond to $\log \mathrm{SB(H\alpha)}$ of $-17.00$,
  $-16.25$, $-15.50$, $-14.75$, $-14.00$ in units of $\mathrm{erg
    ~s^{-1}~cm^{-2}~arcsec^{-2}}$.}\label{fig_PCA_N04254}
\end{figure*}
\begin{figure*}
\includegraphics[width=0.95\textwidth]{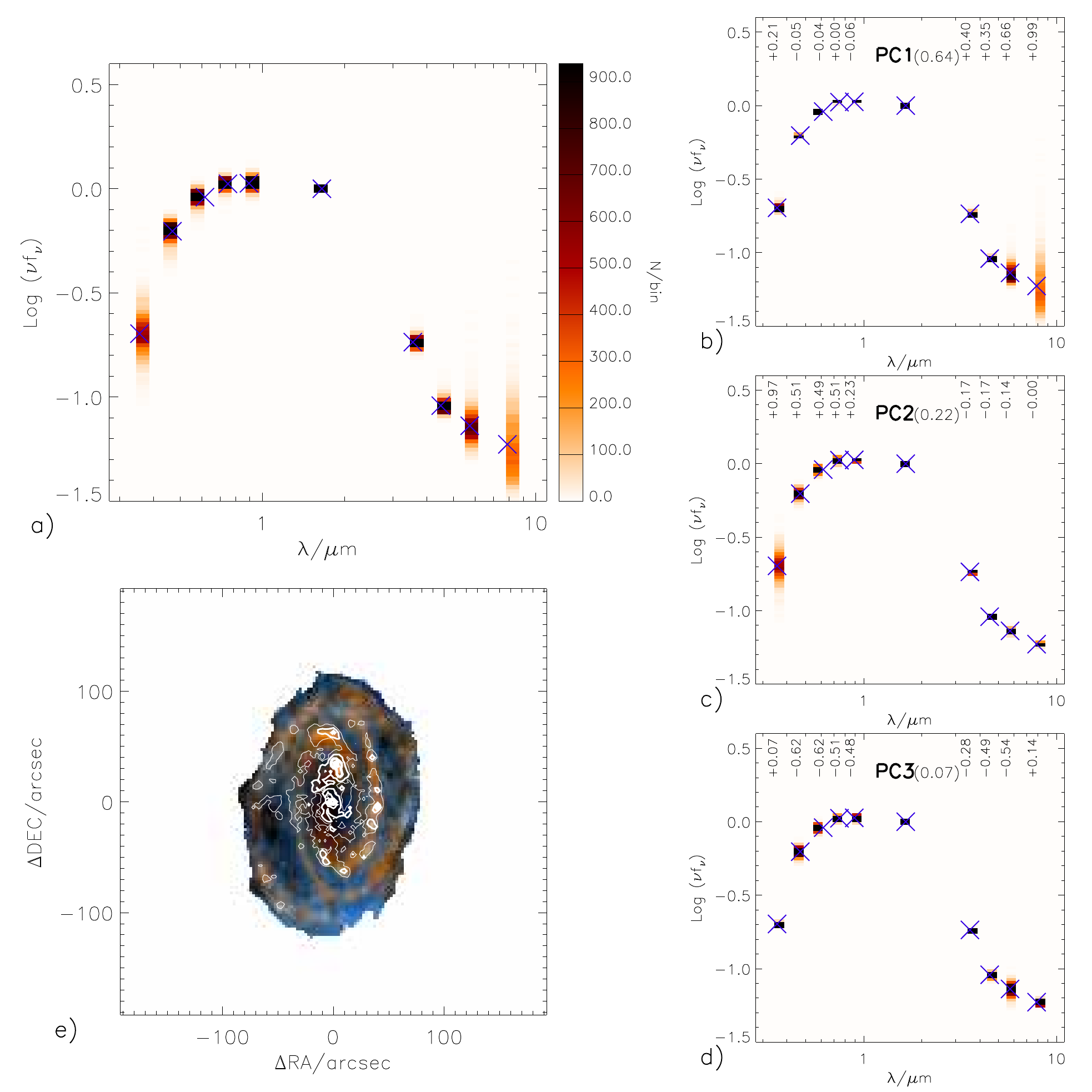}
\caption{Same as Fig. \ref{fig_PCA_N04254}, but for
  NGC\,4450.}\label{fig_PCA_N04450}
\end{figure*}
In Fig. \ref{fig_PCA_N04254} and \ref{fig_PCA_N04450} we illustrate
the results of the PCA on our reference galaxies (corresponding
figures for the complete sample are available in the online only
appendix, figures \ref{fig_PCA_N03521} to \ref{fig_PCA_N04579}).
Figure \textit{a)} (top left) shows the pixel distribution of SEDs,
with the arithmetic mean marked with blue crosses for
reference. Figures \textit{b)}, \textit{c)} and \textit{d)} (to the
right) display the pixel distribution in the SED space when only PC1,
PC2 and PC3, respectively, are used in addition to the mean, such that
$\mathrm{SED}_{i,j}=\mathrm{SED_{mean}}+ a_{i,j} \mathrm{PC}j$, for
$j=1,2,3$ respectively. As apparent from these plots, the majority of
the scatter is accounted for by the first two PCs. This holds true for
the two examples shown here and in general for the full sample. The
variance in each PC is given by the number in parenthesis next to the
PC label in each plot. We find that typically $\sim$90\% of the
variance is contained in the first 2 PCs (with PC1 typically
contributing at least roughly twice as much as PC2) and only 2 to 7\%
in PC3. The remaining variance is contributed in decreasing order by
the other PCs, and is basically at the level of noise.  While in
general PCs are not expected to have any obvious physical
interpretation (this is certainly the case for PC$i$, $i>3$), these
plots show that PC1 and PC2 mainly affect the IR and the optical part
of the SED very differently, but in a very consistent way for
different morphological types and levels of star formation
activity. In particular, PC1 accounts for most of the variations in
the IR part of the SED and, in some cases, for part of the variations
in the optical, whereas PC2 mainly reflects variations in the optical
regime. This is seen more quantitatively with the Spearman rank
correlation coefficients between colors and PC coefficients reported
for each band at the top of figures \textit{b)}, \textit{c)} and
\textit{d)}, with these numbers indicating the relative contribution
and direction to each band. This relation of PC1 and PC2 holds true
for five of the seven galaxies in our sample\footnote{For NGC\,4536
  the role of the first two PCs is switched: PC1 correlates best with
  optical colours, whereas PC2 \textit{anti}-correlates with IR
  colours. However, modulo index exchange and sign of the correlation,
  the same observations and conclusions as for the other four
  ``regular'' galaxies apply to NGC\,4536 as well.}, with the only
exceptions being the inclined spiral NGC\,3521 and the elliptical
galaxy NGC\,4552, which we discuss below in
Sec. \ref{subsec_PCA_anolamous}. The third principal component, though
apparently dominated by variance in the optical bands for most
galaxies, is only a low level contribution and is different enough
between galaxies that it is of very uncertain physical interpretation
and thus will not be discussed further in this work.

The PCA demonstrates that the optical and IR parts of the SED are to
large extent independent and their linear combination can reproduce
pixels SED to high accuracy. As already noted in the previous section,
this shows that on local scales (0.2 to $\lesssim1$ kpc) the PAH
emission that drives IR colours is not related to the combination of
mean light-weighted age of stars and extinction, which is traced by
the optical colours. Apparently any combination of PAH emission and
optical colours is allowed \textit{within} a galaxy.

The decomposition into PCs provides a very compact representation of
the SEDs, which is essentially given by the two coefficients
describing the intensity of the two main PCs. In
Fig. \ref{fig_PCA_N04254}\textit{e)} and
\ref{fig_PCA_N04450}\textit{e)} we show how the intensities of PC1 and
PC2 affect different regions of the galaxies. The intensity of the red
tint represents the intensity of PC1, while the blue one is for
PC2. Regions dominated by either PCs in fact extend over a broad range
of scales and trace a structure in which spiral arms, bright
star-forming knots therein, interarm regions, bulge can be
identified.

PC1 (mainly IR-related) is typically strong (positive, enhanced IR
emission) where bright star forming regions are found and is faint
(negative, diminished IR emission) in the bulge and in the galaxy
outskirts without ongoing star formation. The very close link between
star formation and PC1 is analyzed in more detail in the following
Sec. \ref{subsec_SB_SFR} and is expected from the observed relations
between PAH, mainly responsible for enhanced emission in the IRAC
bands, and SFR \citep[see, e.g.,][]{calzetti+07}. By overlaying
H$\alpha$ isophotal contours on the maps of
Fig. \ref{fig_PCA_N04254}\textit{e)} and
\ref{fig_PCA_N04450}\textit{e)} we show that PC1-bright regions are
actually coincident with H$\alpha$-bright regions and therefore PC1
can be broadly interpreted as an indicator of SFR density. On the
other hand, PC2 (mainly optical-related) is strong (i.e. positive,
enhanced optical luminosity with respect to NIR) where young,
unattenuated stellar associations are found along the spiral arms, but
also in the disk outskirts, as an effect of gradients in age (and
possibly also in metallicity), but is weak (i.e. negative, suppressed
optical emission with respect to NIR) in the bulge, in the interarm
space and in heavily attenuated regions in the arms\footnote{Note that
  the interarm regions have a slightly \textit{negative} PC2 and a
  largely negative PC1, hence the resulting dark blue tint.}. This
complex interplay between regions of different physical conditions
results in the observed lack of correlation between the optical and
the IR part of the SED. As can be inferred from figures
\ref{fig_PCA_N04254}\textit{e)} and \ref{fig_PCA_N04450}\textit{e)},
this result appears to be robust against the choice of resolution
scale, as long as a sufficient number of elements (pixels) can be
analyzed and the limit at which the internal structure of the star
forming regions start to be resolved is not reached (in fact, these
scales are never reached in the galaxies explored here as they are
almost a factor of 10 below our resolution).

It is notable that in the galaxies of later type (hence having higher
specific SFR) PC1 and PC2 appear to contribute very similar amount of
variance in the $[u]$ colour, which has, in fact, a very similar degree
of correlation with the two main PCs. This can be easily accommodated
in the interpretation given in the previous paragraph by considering
that the flux in $u$ band is contributed both by star forming regions,
hence correlated with PC1, and by the regular diffuse stellar
population in relation to its age and relative attenuation, hence correlated with PC2.

\subsubsection{PCA of the ``anomalous'' galaxies NGC\,4552 and
  NGC\,3521}\label{subsec_PCA_anolamous}

Contrary to all other galaxies, the PCA of the elliptical NGC\,4552 results in PC1
being responsible for most (89\%) of the total SED variance. All
pixels in NGC\,4552 have SEDs very close to a pure stellar continuum,
with no evidence for dust/PAH emission and the IR part being
consistent with a Rayleigh-Jeans tail. PC1 affects the SED in the
sense of changing the ``curvature'', i.e. of enhancing or decreasing
at the same time all bands relative to $H$. PC1 has a clear gradient,
such that the central regions have redder optical-$H$ colours and
bluer $H$-IR colours, while the outskirts display the opposite. This
kind of colour gradients are well known in the optical and NIR
\citep[see, e.g.,][]{michard_00,michard_05} and are usually well
explained by metallicity gradients \citep{kobayashi_arimoto_99,
  henry_worthey_99}, plus possibly minor contributions of age
gradients as well. Our analysis shows that there is only one parameter
driving the SED variations in NGC\,4552, and this is likely to be
linked to the metallicity of the stellar population.

The case of NGC\,3521 is opposite and much more complicated. First of
all, the scatter in SED from pixel to pixel is evidently larger than
in any other galaxy. The two main PCs of this galaxy contribute very
similar amount of variance (54 and 40\% respectively), at odds with
the fact that PC1 is clearly the dominant contributor to the variance
in all other galaxies. PC1 appears to be more closely correlated to
optical colours than PC2, which, as opposed, has stronger correlations
with the IR bands; however PC1 and PC2 do not separate optical and IR
in a clear way. The spatial distribution of PC1 and PC2 intensity
(Fig. \ref{fig_PCA_N03521}) shows a surprising result. In the inner
parts PC1 is approximately axissymmetric and reflects in part the
spiral structure of the galaxy; beyond $\approx$ 50 arcsec, the
symmetry is broken, with the western part of the galaxy displaying
more intense, positive PC1, and the eastern part having mainly
negative PC1. By looking at the colour-composite $giH$ image of this
galaxy (see ZCR09, Fig. 5), we see that this reflects the differential
attenuation and the scattering of light in the inclined disc. Thick
dust lanes on the near side (to the west) strongly attenuate all
optical wavelengths, and even attenuate the $H$ band as well, meaning
that our normalization will be biased for all these pixels. The
opposite scenario is observed on the far side of the galaxy (to the
east) where the dust back-scatters the light coming from the central
regions, thus resulting in much bluer colours than expected from the
intrinsic stellar population.\\
PC2 is clearly strong in the central concentrated bulge and in the
very faint outskirts: this can be explained by the fact that PC2
anti-correlates with the colours which are most sensitive to current
star-formation and young stars.

In NGC\,3521 the interplay between star-formation, stellar age and
dust appears much more complex than in other galaxies and this would
explain why the PCA in this case produces very different results than
in other galaxies. As we show in the following sections, this is also
reflected in NGC\,3521 being an outlier in a number of other relations.

\subsection{Dependence of SEDs on surface brightness and star formation
  rate density}\label{subsec_SB_SFR}
The PCA performed in the previous section allows us to condense the
properties of pixel SEDs into just two parameters (the intensity of
PC1 and PC2) instead of 9, thus making it much easier to study the
dependence of such properties on other local parameters. In particular
we analyze here the dependence of SED properties on stellar density
and star formation rate density.

\begin{figure*}
\includegraphics[width=0.9\textwidth]{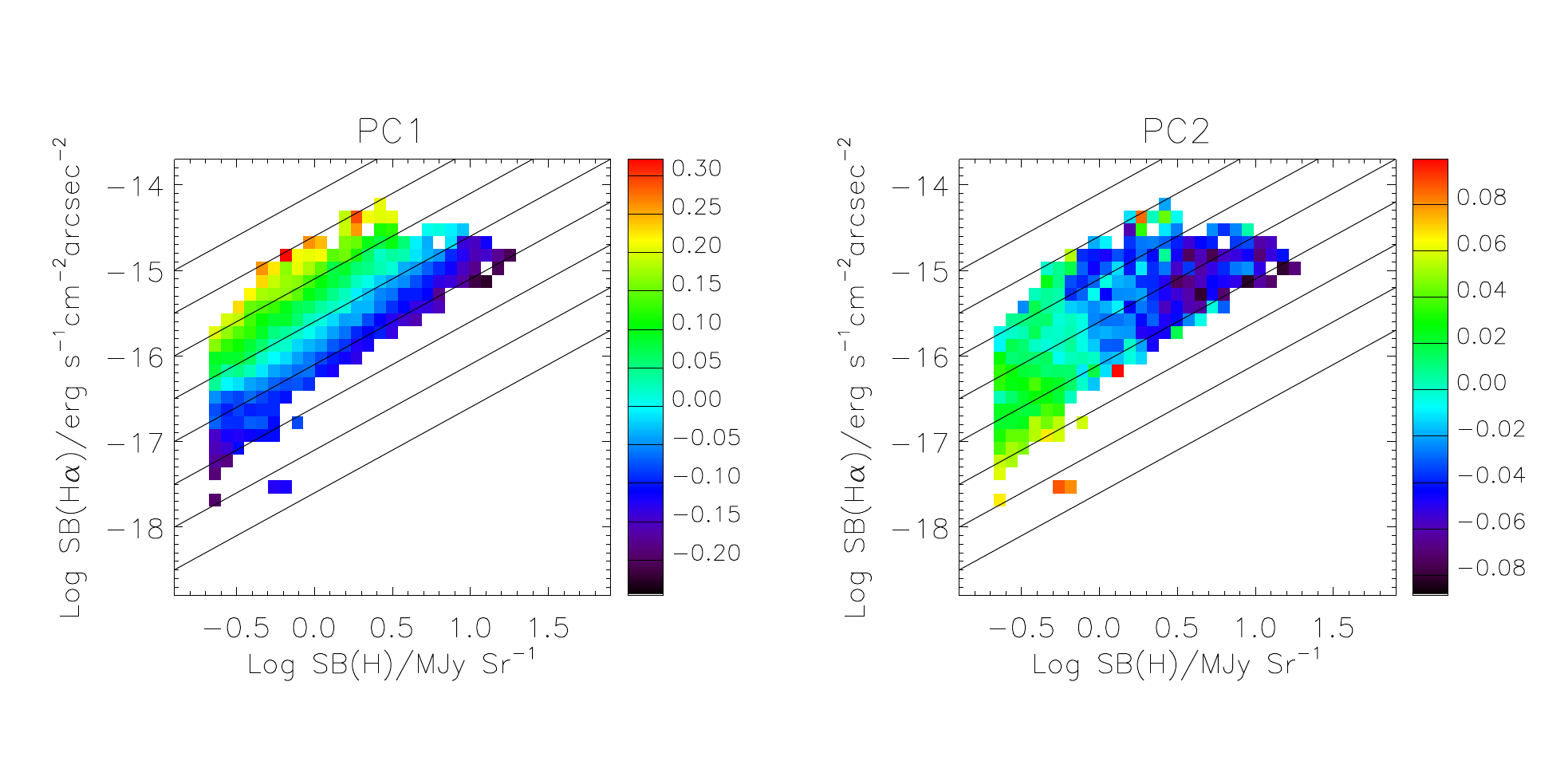}
\caption{Dependence of PC1 and PC2 intensity on $H$-band and H$\alpha$
  surface brightness in NGC\,4254. \textit{Left}: median intensity of
  PC1 for the pixels in bins of $\log \mathrm{SB}(H)$-$\log
  \mathrm{SB(H\alpha)}$ (width is 0.075 and 0.15 dex respectively),
  colour coded according to the key on the right side. The diagonal
  lines are lines of constant $\mathrm{SB(H\alpha)}/\mathrm{SB}(H)
  \approx \mathrm{sSFR}$. Bins with less than three contributing
  pixels are not displayed. \textit{Right}: same as the left, but for
  PC2.}\label{fig_SBPC_N04254}
\end{figure*}

\begin{figure*}
\includegraphics[width=0.9\textwidth]{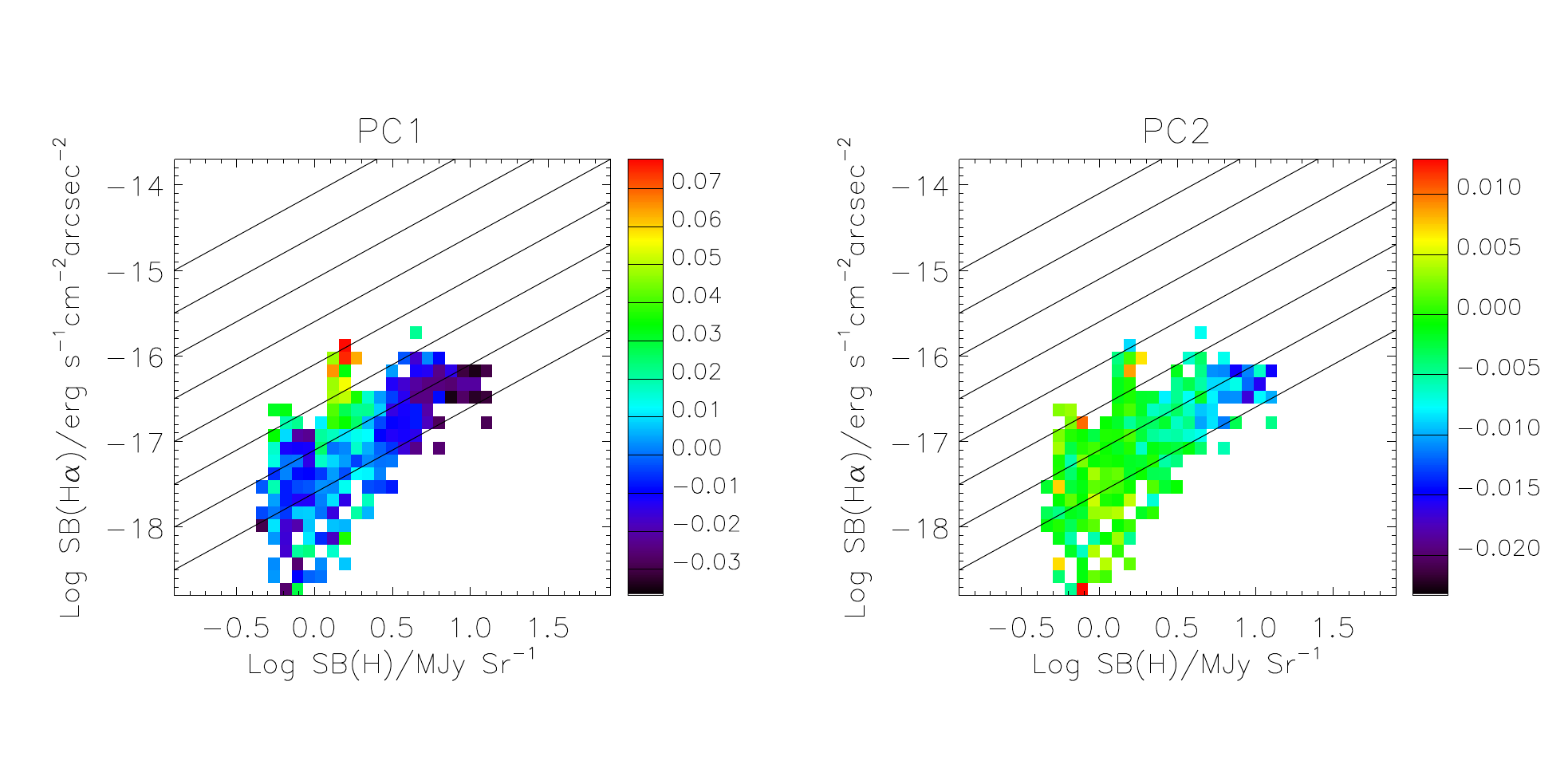}
\caption{Same as Fig. \ref{fig_SBPC_N04254}, but for
  NGC\,4450.}\label{fig_SBPC_N04450}
\end{figure*}

To this goal, for each galaxy we bin all pixels in the 2-dimensional
space of $\log \mathrm{SB}(H)$-$\log \mathrm{SB}(\mathrm{H}\alpha)$
and compute the median value of the coefficients of PC1 and PC2 in
each bin. This is shown in the left and right diagrams, respectively,
of Fig. \ref{fig_SBPC_N04254} for NGC\,4254 and
Fig. \ref{fig_SBPC_N04450} for NGC\,4450 (figures for the other
galaxies are provided in the online appendix \ref{append_SBPCA},
except for NGC\,4552, which has no H$\alpha$ imaging). Importantly,
these quantities have direct physical interpretations, with H$\alpha$
SB often used as a proxy for star formation rate density
\citep{kennicutt_98}, while the $H$-band SB can be considered a rough
proxy for stellar mass density (e.g. ZCR09).

Regular face-on, late type, star forming galaxies, such as NGC\,4254,
display clear trends for PC1 and PC2 intensity in this SB space. At
any given SB($H$) PC1 increases with increasing SB(H$\alpha$), while a
reversed trend is seen with SB($H$) at given SB(H$\alpha$). The
diagonal lines drawn in each figure mark the loci of constant
SB(H$\alpha$)/SB($H$). Constant PC1 intensity levels are almost
parallel to these lines, thus indicating that PC1 is effectively a
measure of SB(H$\alpha$)/SB($H$) or, in more physical terms, of the
local specific star formation rate $\mathrm{sSFR}=\mathrm{SFR}/M_\star
\approx \mathrm{SB}(\mathrm{H}\alpha)/\mathrm{SB}(\mathrm{H})$. On the
other hand, this is not surprising, given the very tight correlation
that we show to exist between PC1 and the 8$\mu$m color (which is
proportional to SB(8$\mu$m)/SB($H$)) and the fact that the 8$\mu$m
emission is a very good tracer of SFR, similar to H$\alpha$.  The PC2
intensity trends are somewhat more noisy, with PC2 decreasing at
increasing SB($H$), while no significant trend with SB(H$\alpha$) is
observed. Following the physical interpretation given in the previous
section of PC2 as a tracer of optical colors and hence of mean stellar
age (and dust extinction) and considering that SB($H$) is with good
approximation a monotonically decreasing function of radius, this PC2
trend with SB($H$) can be broadly interpreted as reflecting stellar
age gradients in late-type galaxies \citep[e.g.][]{macarthur+04}.

Although not as clear as for NGC\,4254, PC1 trends which can be
interpreted as tracing sSFR are seen or hinted at also for earlier
type galaxies, such as NGC\,4450. Concerning PC2 it can be seen that
three of the five ``regular'' galaxies (NGC\,4254, 4321, 4450) display
a trend of PC2 decreasing at increasing SB($H$), thus consistent with
the hypothesis that PC2 reflects mean age gradients. However for the
other two such a trend is not detected, possibly due to the presence
of thick dust lanes (NGC\,4536) or very weak stellar population
gradients (NGC\,4579).

These trends of PCs with SB are not seen in NGC\,3521 and NGC\,4552,
whose first two PCs, for different reasons, do not have a
straightforward physical interpretation, as already discussed in
Sec. \ref{subsec_PCA_anolamous}.

\section{From internal to global SED
  correlations}\label{globalcorr_sec}
\begin{figure*}
\includegraphics[width=\textwidth]{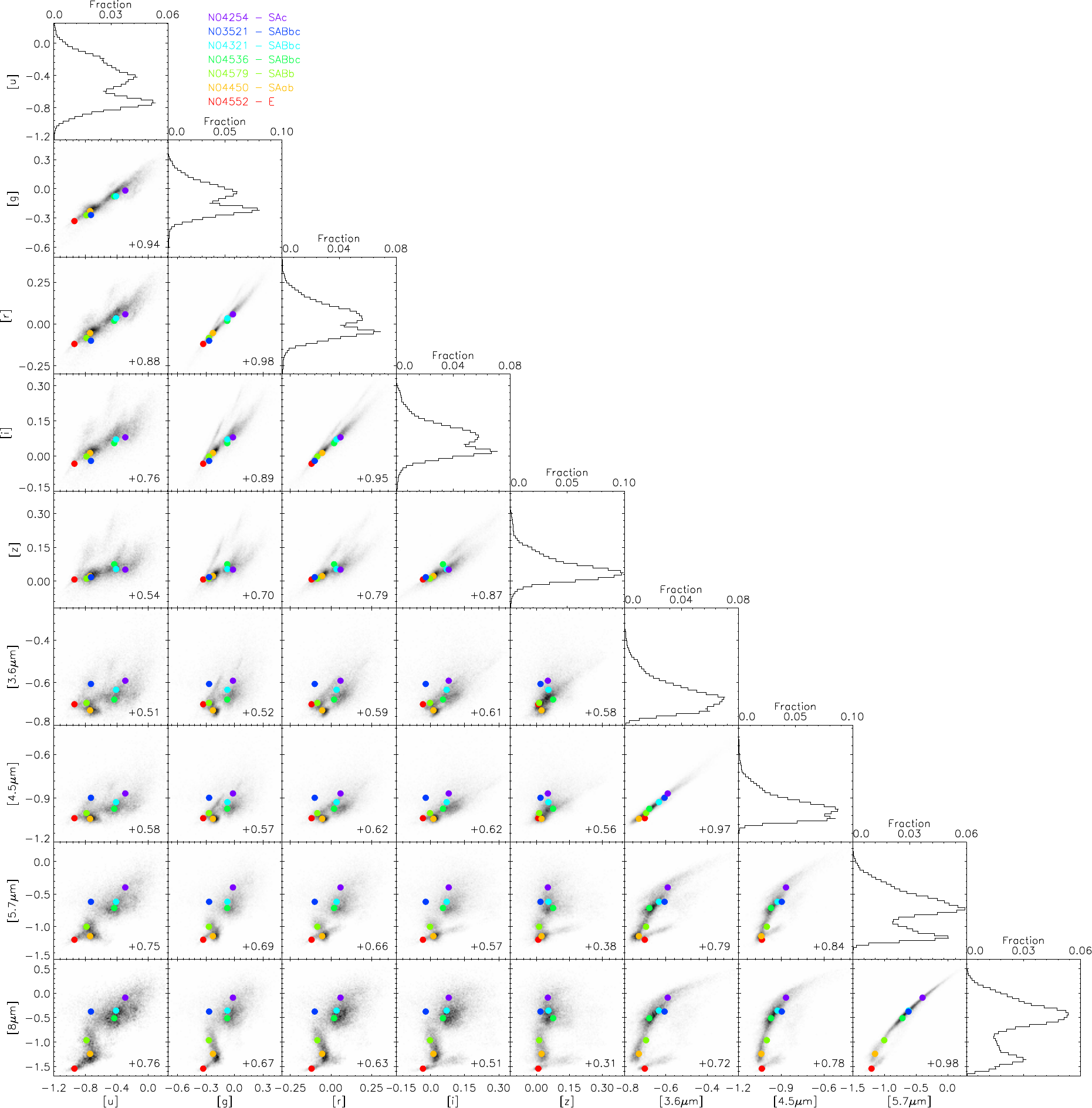}
\caption{The distribution of pixels of \textit{all galaxies} in the
  ``colour-colour'' spaces defined by the fluxes $\log \nu f_\nu$
  normalized to $H$-band, in $u$, $g$, $r$, $i$, $z$, 3.6, 4.5, 5.7
  and 8$\mu$m bands. The intensity of grey linearly traces the density
  of pixels at each point in the colour-colour spaces. The Spearman
  rank correlation coefficient among the pixels in each panel is
  written at its bottom right corner. The histograms on the diagonal
  show the distribution of pixels in normalized $\log \nu f_\nu$ for
  each band as labeled on the y-axis. Coloured filled dots indicate
  the \textit{integral} colours of individual galaxies, according to
  the legend at the top. Colours are chosen such that the sequence
  from red to purple is sorted by the morphological type of the
  galaxy, i.e.~from early- to late-type.}\label{fig_stack_SEDcorr}
\end{figure*}

In this section we show how to reconcile the lack of correlation
between variations in optical and IR colours observed on local scales
with the existence of such correlations between integrated colours of
galaxies. In Fig. \ref{fig_stack_SEDcorr} we combine the pixels from
all seven galaxies in the sample and show the combined colour-colour
plots, like those analyzed in Sec. \ref{sec_SEDcorr}.
The coloured symbols identify the colours from the integral photometry
of each galaxy, as indicated in the legend. Symbol colours from red to
purple are used to represent ``increasing'' morphological types, from
early to late. As for individual galaxies, the optical colours of
pixels are typically correlated to very high degree, as are the IR
colours. However, contrary to what we observe in individual galaxies,
when the pixels from all galaxies are combined, the optical colours
correlate with IR colours as well, albeit with a larger scatter than
that observed between optical colours or between IR colours alone. The
correlation is quantified in each panel by the Spearman rank
correlation coefficient, which is typically $\gtrsim +0.8$ for
optical-optical and IR-IR colours, and $\gtrsim +0.5$ for optical-IR
colours. The global correlations in the optical vs. IR colours result
from the fact that, although each galaxy covers a limited region in
the colour-colour diagrams with highly scattered clouds of points, the
barycentres of the distributions follow well defined relations. This
can be seen in the colour-colour relations of the integrated points,
where all galaxies, with the only notable exception of NGC\,3521 (the
blue point), follow very well defined sequences.  In other words, we
see that the shapes of the integrated SEDs are very correlated between
galaxies: IR colours are very tightly linked to each other; by
enhancing the IR flux relative to $H$, bluer optical-$H$ colours
result, which, in turn, are very well correlated with each other. This
is a natural consequence of galaxies' SFHs being typically smooth,
such that a higher current rate of star formation (which causes
enhanced IR flux) is generally accompanied by a higher abundance of
young blue stars, formed in the recent past and already liberated from
their dusty birth cloud \citep{charlot_fall}.  The only outlier in our
sample to most of these correlations is NGC\,3521, whose aspect and
geometrical distribution of dust result in a complex and peculiar
interplay between the attenuation and scattering of photons and hence
in the observed anomalous colours (see also
Sec. \ref{subsec_PCA_anolamous} and \ref{subsec_IR_anolamous}).

We note that not only is the relation between IR colours extremely
tight for the integrated SEDs and for the average pixel distribution,
but also almost all pixels, no matter which galaxy they belong to,
appear to follow extremely tight \textit{universal} relations, whose
observed scatter is consistent with or just slightly larger than pure
measurement errors. We analyze this in more detail in
Sec. \ref{sec_IRcolors}.

The degree of correlation between the $[8\mu {\rm m}]$ colour and
other colours reaches a minimum with the SDSS $z$ and $i$ bands
(Spearman rank correlation coefficient of $+0.31$ and $+0.51$
respectively), with the correlation with the $z$ band being the lowest
for all colour combinations. This can be understood by considering
that the $[z]$ colour ($[z]-[H]$) and the $[i]$ colour ($[i]-[H]$) are
almost insensitive to stellar age, marginally sensitive to dust
extinction and are mainly driven by metallicity (e.g. ZCR09), while
the $[8\mu {\rm m}]$ colour is essentially a sSFR indicator (as we
showed above), whose relation with metallicity is not obvious
\citep[see, e.g.,][]{calzetti+07,gordon+08}. Colours at shorter
wavelengths are increasingly related to stellar age and therefore it
is not surprising that a better correlation with the $[8\mu {\rm m}]$
colour is observed. This reaches a maximum with the correlation of the
$[8\mu {\rm m}]$ colour with the $[u]$ colour (for which a significant
correlation was already observed \textit{within} some individual
galaxies), as this colour (effectively $u-H$) can be also used as a
rough proxy of sSFR.

One last point to make about Fig. \ref{fig_stack_SEDcorr} is that in
these colour-colour diagrams galaxies appear to be ordered in their
integrated SEDs according to their morphological type (coded by the
colours of the symbols). While this is not always exactly the case
when one looks at optical colours, the order is exact in the IR
colours. \cite{li+07} also observed similar trends of $8\mu$m
(corrected for stellar contamination) vs. 3.6$\mu$m with
morphology. This might be surprising since morphologies are determined
at optical wavelengths, not IR. In physical terms, this implies that
the morphology of galaxies reflects primarily their sSFR, rather than
the properties of their optical SED.

\section{Mid-IR colour relations}\label{sec_IRcolors}
\subsection{A universal sequence of mid-IR colours and its predictive
  power}\label{subsec_universalIR}
\begin{figure*}
\includegraphics[width=0.95\textwidth]{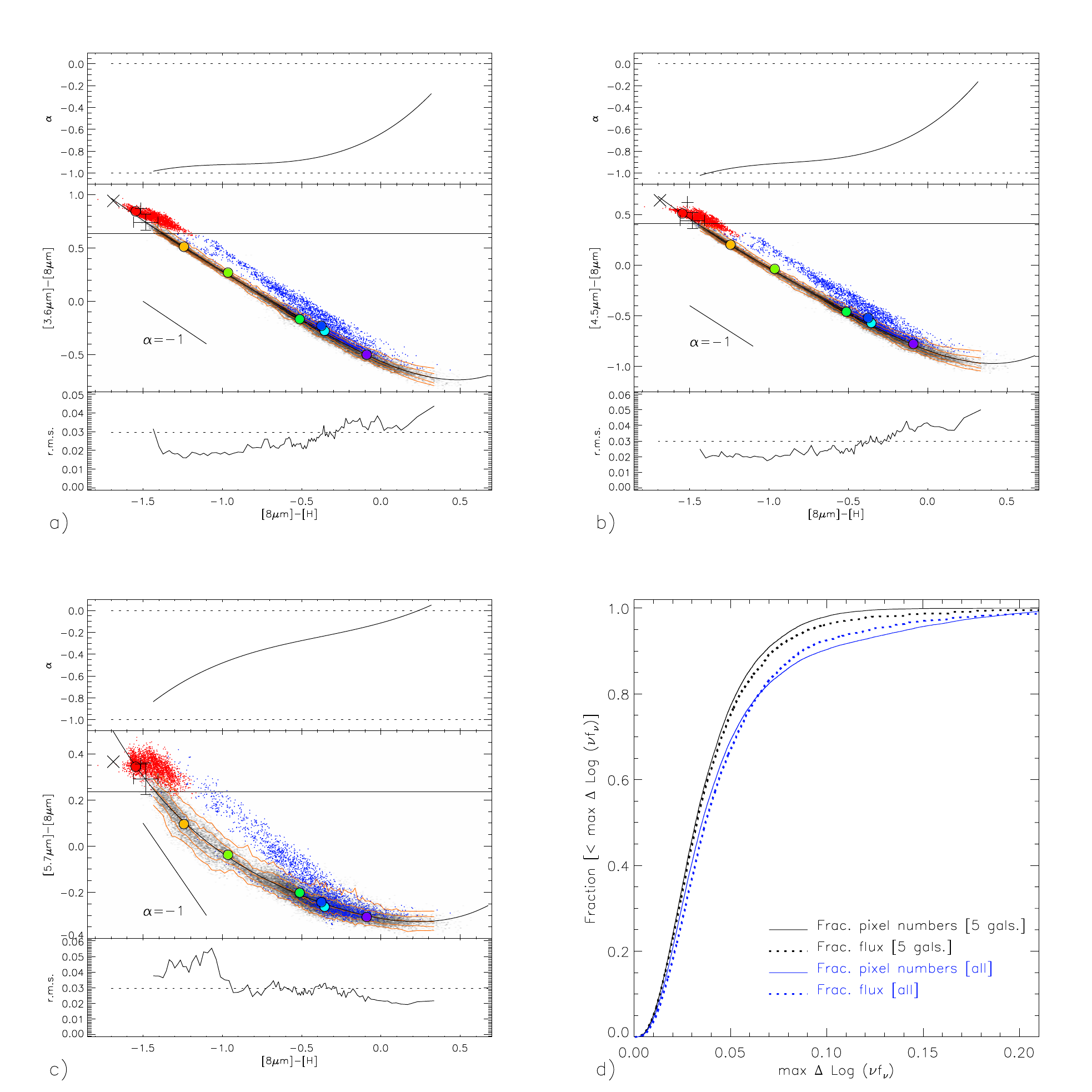}
\caption{IRAC IR colour correlations. \textit{a)}, \textit{b)} and
  \textit{c)}: \textit{mid panels} show the distributions of pixels in
  the plane given by the flux at 3.6, 4.5 and 5.7 $\mu$m (\textit{a},
  \textit{b} and \textit{c} respectively) relative to the flux at 8
  $\mu$m vs. the flux at 8 $\mu$m relative to the flux in $H$-band
  (all fluxes are expressed in units of $\log \nu f_\nu$). NGC\,3521
  and NGC\,4552 are represented by the blue and the red points
  respectively, while the density distribution of pixels belonging to
  the other five galaxies are represented by the grey scale shading
  (log scaling). For reference the \textit{global} colours of the
  seven galaxies are plotted as filled circles, coloured according to
  the key in Fig. \ref{fig_stack_SEDcorr}. The five orange lines show
  the running median and $2.5$, $16$, $84$ and $97.5$ percentiles of
  the pixel distribution for the five galaxies represented by the grey
  scale shading. The solid black curve is a 4th-degree polynomial
  fitting to these pixels, whose coefficients are given in Table
  \ref{tab_fitIRAC}. The solid black straight line to the lower-left
  shows a slope of $-1$ to guide the eye. It is immediately clear that
  NGC\,3521 and NGC\,4552 are both outliers with respect to the rest
  of the sample. The points with error bars represent the pure stellar
  colours derived in Sec. \ref{subsec_stellarSED}. Theoretical
  predictions from CB07 for stellar population of ages $>2$Gyr and
  $\approx 1$Gyr are marked with a $\times$ and a $+$, respectively,
  while the \texttt{Starburst99} predictions adopted by Helou et
  al. (2004) are marked with solid horizontal lines. In the
  \textit{bottom panels} we plot the r.m.s. (the half-widths of the
  16--84 percentile range) of the pixel distributions along the y-axis
  for the five galaxies, as a function of $[8\mu$m$]-[H]$. The
  horizontal dashed line indicates the maximum r.m.s.\ possible due to
  measurement and photometric errors. In the \textit{top panels} the
  local slope of the fitting polynomial is plotted. For reference, the
  slopes of $0$ (corresponding to constant flux ratio to $8\mu$m) and
  $-1$ (corresponding to constant flux ratio to $H$) are marked as
  dotted lines. \textit{d)}: for each pixel we define the quantity
  ``max $\Delta \log \nu f_\nu$'' as the maximum of the absolute
  deviation between $\log \nu f_\nu$ observed and predicted based on
  the respective fitted polynomial functions, at $3.6\mu$m, $4.5\mu$m
  and $5.7\mu$m (see text for details). The fraction of pixels for
  which such a maximum deviation, $\Delta$, is less than some
  difference, $x$, is plotted as solid lines, with blue showing the
  result for all galaxies, while black shows the result when the
  outliers NGC\,3521 and NGC\,4552 are excluded. The dotted lines show
  the fraction of $H$-band flux contained in pixels for which the
  maximum $\log \nu f_\nu$ deviation is less than $x$, with blue for
  all and black excluding the outliers as for the other
  curves.}\label{fig_IRAC_CHXall}
\end{figure*}

In this section we analyze in more detail the relation between IR
colours and demonstrate that the SEDs in this spectral window are a
one-parameter family to very high accuracy. In the mid panels of
Figures \ref{fig_IRAC_CHXall}\textit{a)}, \textit{b)} and \textit{c)}
we plot the flux at 3.6, 4.5 and 5.7 $\mu$m, respectively, relative to
the flux at 8 $\mu$m as a function of the $[8\mu\rm{m}]-[H]$
colour. Greyscale levels represent the log of pixel densities in the
colour-colour spaces for the five ``regular'' galaxies: NGC\,4552 and
NGC\,3521 are analyzed separately, because of their anomalous natures
(see also Sec. \ref{subsec_IR_anolamous}). The orange lines show the
running percentiles for this distribution: 2.5, 16, 50, 84 and 97.2\%,
from the bottom to the top, respectively.  The pixels of the
elliptical NGC\,4552 and the inclined dusty spiral NGC\,3521 are
represented separately by the coloured points, with NGC\,4552 the
compact cloud of red points top left in each figure (very low 8 $\mu$m
flux relative to $H$ and to all other IRAC bands), while NGC\,3521 is
the extended sequence of blue points partly offset from the sequence
defined by the main sample. These plots suggest that the ratio between
8 $\mu$m and $H$-band luminosities can be used to predict the
luminosities in the other three IRAC bands with high accuracy.

The bottom panels of Figures \ref{fig_IRAC_CHXall}\textit{a)} to
\textit{c)} show the half-width of the 16--84\% percentile range, a
robust proxy for the r.m.s., of the pixels belonging only to the five
``regular'' galaxies. The horizontal dotted line corresponds to the
maximum r.m.s. that is expected for null intrinsic scatter and the
minimum S/N of 20 in each band. At low $[8 \mu$m$]-[H]<-0.6$, the
scatter in the $[3.6 \mu$m$]-[8 \mu$m$]$ and $[4.5 \mu$m$]-[8 \mu$m$]$
colours is lower than the maximum possible contribution from
measurement error, namely 0.02 dex or 5\%, and the relation is close
to linear. At higher $[8 \mu$m$]-[H]$ the relations flatten and the
scatter increases up to 0.035--0.040 dex (8--10\%), thus requiring
significant intrinsic scatter to be explained. As far as scatter is
concerned, for the $[5.7 \mu$m$]-[8 \mu$m$]$ colour the situation is
reversed: $[8 \mu$m$]-[H]$ can be used to predict this colour to
typical accuracy better than 7\% at $[8 \mu$m$]-[H]>-1$, but its
performance is significantly worse (10--12\% accuracy) at lower
relative flux intensity. The slope of the relation also changes
significantly from almost $-1$ at low $[8 \mu$m$]-[H]$ to 0 at high
values.

\begin{figure*}
\includegraphics[width=\textwidth]{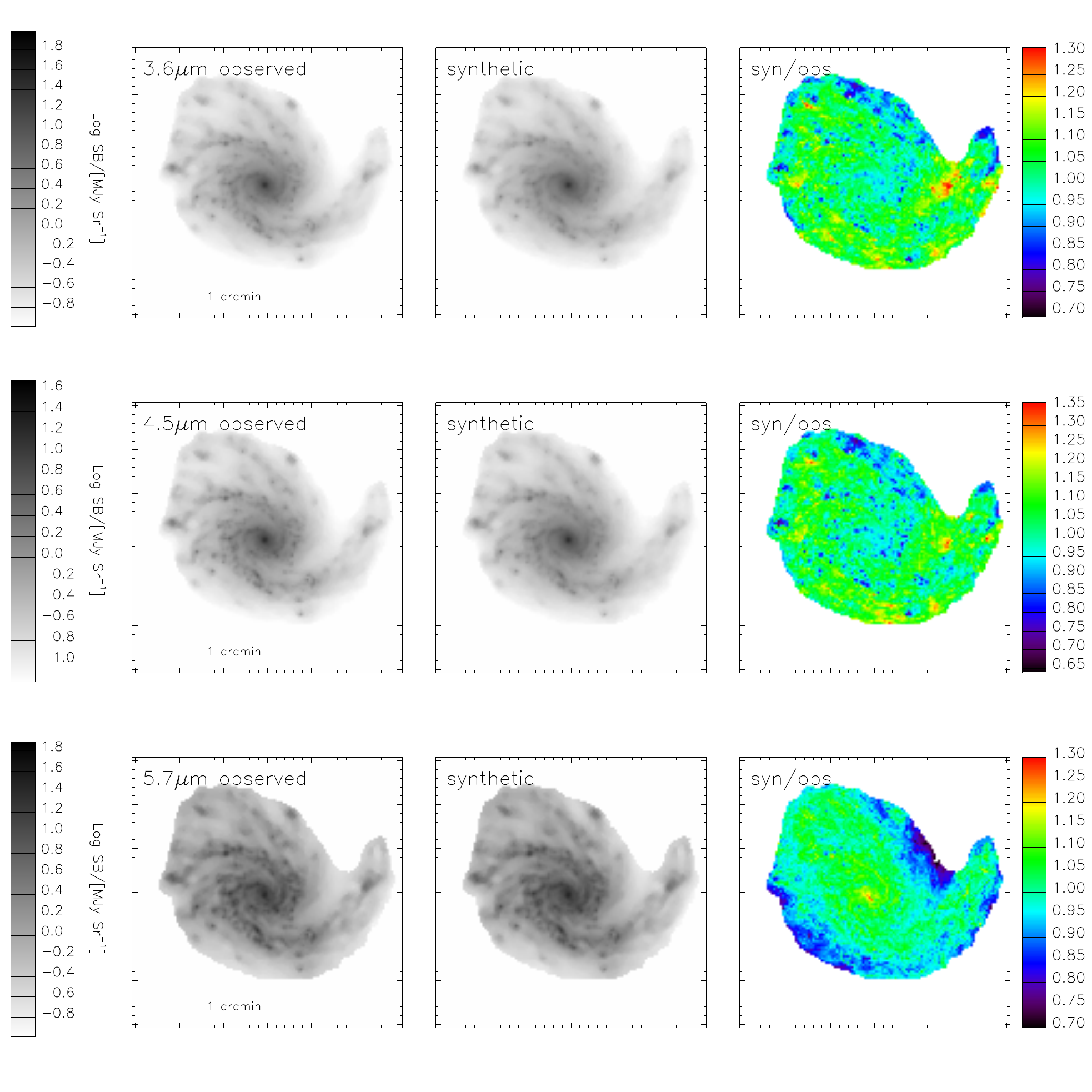}
\caption{The images of galaxies (NGC\,4254 in this case) at 3.6, 4.5
  and 5.7 $\mu$m can be reconstructed with great accuracy by combining
  the $H$-band and the 8$\mu$m images only by means of the fitting
  functions provided in Tab. \ref{tab_fitIRAC}. On each of the three
  rows, for 3.6, 4.5 and 5.7 $\mu$m from top to bottom, we reproduce,
  from left to right, the original image of the galaxy, the synthetic
  image reconstructed from $H$-band and the 8$\mu$m images, and the
  ratio between the synthetic and the observed images. While most of
  the pixels display typical deviations of a few per cent, consistent
  with or just slightly larger than photometric uncertainties, larger
  deviations up to a few 10\% display a regular structure that hints
  at systematic departures from the universal fitting functions
  related to different local physical
  conditions.}\label{fig_IRACsynth_N04254}
\end{figure*}

We fit fourth degree polynomials to the relations observed for the
five ``regular'' galaxies. The coefficients are given in Table
\ref{tab_fitIRAC}.  The fits are shown as solid black lines in the mid
panels of the Figures \ref{fig_IRAC_CHXall}\textit{a)} to \textit{c)},
where they almost perfectly overlap to the running median lines.
\begin{table}
  \caption{Polynomial fit coefficients to the IR colour-colour relations
    presented in Fig. \ref{fig_IRAC_CHXall}. The $y$ color named in column
    1 can be obtained from $[8 \mu \mathrm m] - [H]=x$ as 
    $y=c_0+c_1 x+c_2 x^2+c_3 x^3+c_4 x^4$. Note that colours relative to $[H]$,
    i.e. $[3.6 \mu \mathrm m] - [H]$, $[4.5 \mu \mathrm m] - [H]$,
    $[5.7 \mu \mathrm m] - [H]$, can be obtained from analogous polynomial 
    approximations as a function of $[8 \mu \mathrm m] - [H]$ by simply 
    replacing $c_1$ with $c_1+1$.}\label{tab_fitIRAC}
\begin{tabular}{lrrrrr}
  \hline
  Colour & $c_0$ &$c_1$ &$c_2$ &$c_3$ &$c_4$ \\
  (1) & (2) & (3) & (4) & (5) & (6) \\
  \hline
  $[3.6 \mu \mathrm m]-[8 \mu \mathrm m]$ & -0.564 & -0.639 & 0.420 & 0.291 & 0.079\\
  $[4.5 \mu \mathrm m]-[8 \mu \mathrm m]$ & -0.840 & -0.572 & 0.471 & 0.321 & 0.091\\
  $[5.7 \mu \mathrm m]-[8 \mu \mathrm m]$ & -0.312 & -0.114 & 0.201 & 0.092 & 0.060\\
  \hline
\end{tabular}
\end{table}
These relations enable the generation of maps of the galaxies at 3.6,
4.5 and 5.7 $\mu$m just from the $H$-band and 8 $\mu$m images. An
example of this is shown in Fig. \ref{fig_IRACsynth_N04254} for
NGC\,4254; the images of the remaining six galaxies are provided for
comparison in the online appendix (figures \ref{fig_IRACsynth_N03521}
to \ref{fig_IRACsynth_N04579}). The three rows of images are for the
IRAC 3.6, 4.5, and 5.7 $\mu$m bands respectively. The first image in
each row is the original image as observed, the image in the middle is
the one ``synthesized'' from the $H$-band and 8 $\mu$m images using
the fit relations given in Table \ref{tab_fitIRAC}, and the last image
is the ratio between the synthetic image and the observed one.
This figure, together with those relative to the other four
``regular'' galaxies in the online appendix
(Fig. \ref{fig_IRACsynth_N04321}, \ref{fig_IRACsynth_N04450},
\ref{fig_IRACsynth_N04536}, \ref{fig_IRACsynth_N04579}), illustrates
three points: \textit{i)} the overall accuracy of the synthetic images
is on average very good, typically within a few per cent; \textit{ii)}
the same relations can be applied to all galaxies as no significant
bias is seen from galaxy to galaxy; \textit{iii)} significant
differences between the observed and the synthetic images are observed
in some pixels, up to the level of 20--30\%. The spatial distribution
of these pixels with the largest deviations is not random, hence not
simply a consequence of noise, but rather reflects to some extent the
structure visible in the intensity images. For instance, some low-SB
regions appear to be biased low in the synthetic 5.7 $\mu$m image; arm
and interarm regions appear to be differently biased with respect to
each other in all three reconstructed bands.

To better quantify point \textit{i)} in the previous paragraph, we
measure both the fractional area and $H$-band luminosity occupied by
the pixels for which the predicted IRAC fluxes deviate from the
observed ones by more than a given amount. For each pixel we determine
the maximum absolute difference ``max $\Delta \log \nu f_\nu$''
between the predicted and observed pixel fluxes $\log \nu f_\nu$ in
the three bands, i.e.~$3.6$, $4.5$, and $5.7 \mu \mathrm m$. We plot
the fraction of pixels for which ``max $\Delta \log \nu f_\nu$'' is
less than a given amount $x$ and the fraction of luminosity in those
pixels as a function of $x$ in Fig. \ref{fig_IRAC_CHXall} \textit{d)}
as solid and dotted lines, respectively. Black lines are for the five
``regular'' galaxies only, while for the blue lines all seven galaxies
are used. If we restrict the analysis to the ``regular'' galaxies only,
in 80\% of the pixels the maximum deviation between predicted and
measured fluxes is $<0.052$ dex, i.e. $<12.7$\%; 90\% of pixels are in
agreement within 0.069 dex, i.e. 17.2\%. In terms of luminosity
fraction, the measured accuracy is slightly worse, thus indicating
that the main reason of disagreement between the observed colours and
those predicted from our simple fitting formulae does not originate
from noise only (or mainly): in fact this plot shows that, on average,
the deviating pixels have $H$-band surface brightness higher than
average. This confirms the point \textit{iii)} of the previous
paragraph, extracted from the visual analysis of
Fig. \ref{fig_IRACsynth_N04254} (and analogues in the online
Appendix).

\subsection{Relative dust and star contributions and constraints on
  the purely stellar SED}\label{subsec_stellarSED}
The IR colour-colour relations analyzed in the previous subsection
indicate that the knowledge of the flux ratio between the band mostly
affected by dust and PAHs (8 $\mu$m) and the one including stellar
emission only ($H$) is sufficient to determine the full SED in
between. To first order, we can represent the IR SED as a simple
linear combination of a stellar and a dust component, basically
templates with constant contributions to each band: at high values
of $[8 \mu \mathrm m]-[H]$ the IR SED is strongly dominated by the
dust component, while the stellar continuum dominates at low
values. The flux at the three intermediate wavelengths is expected to
asymptotically approach a constant ratio to $[8 \mu \mathrm m]$ going
toward high $[8 \mu \mathrm m]-[H]$, whereas a constant ratio to $[H]$
is expected to be approached at low values. These two regimes
translate into slopes of 0 and $-1$ respectively in the colour-colour
relations of Fig. ~\ref{fig_IRAC_CHXall}. 
In particular, the slope of $-1$ follows from the fact
that for each $[X]$ (3.6, 4.5 and 5.7 $\mu$m) $[X]-[8 \mu \mathrm
m]=([X]-[H])-([8 \mu \mathrm m]-[H])$; as we approach the purely
stellar regime, $[X]-[H]$ should become asymptotically constant and
therefore $[X]-[8 \mu \mathrm m]$ vs. $[8 \mu \mathrm m]-[H]$ should be
asymptotically approximated by a line of slope $-1$.

It is therefore interesting to show the actual slopes as a function of
$[8 \mu \mathrm m]-[H]$, which we plot in the upper panels of
Fig. \ref{fig_IRAC_CHXall} \textit{a)} to \textit{c)}. As far as 3.6
and 4.5 $\mu$m are concerned, while slopes close to $-1$ exist for a
wide range, with slopes $\alpha<-0.8$ measured up to $[8 \mu \mathrm
m]-[H]=-0.5$, a slope of $\alpha=-1$ is reached only at the very
lowest $[8 \mu \mathrm m]-[H]$ ratios. Past $[8 \mu \mathrm
m]-[H]\approx0.0$, $\alpha$ quickly approaches 0 for both bands,
though it never reaches 0 in the actual data range for either. What
these slopes imply is that, while the dust contribution is weak for
most values of $[8 \mu \mathrm m]-[H]$, the fluxes at 3.6 and 4.5
$\mu$m can be considered to be ``pure'' stellar continuum only for the
very lowest $[8 \mu \mathrm m]-[H]$ ratios; as soon as some PAH
emission appears at 8 $\mu$m, also 3.6 and 4.5 $\mu$m are affected. At
$[8 \mu \mathrm m]-[H]>-0.5$ typical of star-forming galaxies (see
Fig. \ref{fig_stack_SEDcorr}), these bands appear to be heavily
contaminated by the PAH feature at 3.3 $\mu$m, in the 3.6 $\mu$m band,
and by hot dust and extended wings of neighbouring PAH features, in
the 4.5 $\mu$m band \citep[e.g.][]{draine+07,draine_li_07}. This issue
of dust contamination in the first two IRAC bands is also consistent
with work by Meidt et al. (2011, ApJ submitted) and the increase in
contamination with star formation was also found by \cite{mentuch+10}.

The slope of the $[5.7 \mu \mathrm{m}]-[8 \mu \mathrm{m}]$ vs. $[8 \mu
\mathrm m]-[H]$ relation indicates that the 5.7 $\mu$m follows the 8
$\mu$m emission much more closely (as expected). At the lowest $[8 \mu
\mathrm m]-[H]$ ratios, slope does not even reach $-1$, indicating
that nowhere in the five ``regular'' galaxies does the emission at 
5.7 $\mu$m arise purely from stellar continuum. The slope
$\alpha$ quickly increases toward 0, although this value is only
reached at the very highest $[8 \mu \mathrm m]-[H]$ values: only there
does the PAH/dust component fully dominate at 5.7 $\mu$m, while at lower
values the contamination by stellar continuum is still
significant. Interestingly, the scatter in the $[5.7 \mu
\mathrm{m}]-[8 \mu \mathrm{m}]$ at the highest $[8 \mu
\mathrm{m}]-[H]$ values most likely reveals the maximal variation
between the 6.6$\mu$m and 7.7$\mu$m PAH features (and underlying
continuum) that dominate these bands, at least in NGC\,4254 which
dominates these colours. This scatter is small, both
relative to the expected noise (bottom panel of
Fig.~\ref{fig_IRAC_CHXall}\textit{c)}) and to the variation observed
in these PAH features between galaxies, as shown for the SINGS
galaxies in \citet[][especially Fig.~13]{smith+07}.

The exact knowledge of the purely stellar contribution to the
galaxies' mid-IR SED is of great importance in order to gain a more
accurate description of dust emission and its physics. As we showed in
this section, this task is complicated by the fact that stellar and
dust emission in this spectral region are always mixed to some
degree. The commonly adopted approach to decouple these two components
heavily relies on models (both for stellar and for dust emission)
or/and SED fitting techniques \citep[see
e.g.][]{helou+04,dacunha+08}. It has to be stressed, however, that
stellar population synthesis models in the mid-IR are still very
uncertain and poorly constrained by observations. The diagnostic plots
of Fig. \ref{fig_IRAC_CHXall} \textit{a)}, \textit{b)} and \textit{c)}
allow us to derive empirical constraints on the purely stellar SED in
the following way.  In the simple hypothesis that \textit{i)} the SED
at each pixel is just a linear combination of a constant stellar
component and a constant dust component and \textit{ii)} $H$ band is
not contaminated by dust emission, pure stellar emission would
correspond, \textit{in principle}, to the minimum possible value of
$[8 \mu \mathrm m]-[H]$ ($[8 \mu \mathrm m]$ is chosen because it is
the band most sensitive to dust emission and because $[8 \mu \mathrm
m]-[H]$ offers the largest wavelength leverage). However,
\textit{empirically} it is very difficult to determine such a minimum
because it is a limit which is never actually reached (no galaxy is
fully dust-free) and because photometric errors and intrinsic SED
variations of the stellar component (which are predicted by SPS
models) increase the uncertainties. We use the simple analytical
formulae that fit the empirical relations of
Fig. \ref{fig_IRAC_CHXall} \textit{a)}, \textit{b)} and \textit{c)} to
extrapolate the flux ratios to the limit of pure stellar emission,
which we determine as the point ($[8 \mu \mathrm m]-[H]$) at which the
derivative of the relation is $\alpha=-1$, according to the argument
given previously. The determination of the purely stellar $[8 \mu
\mathrm m]-[H]$ depends on the relation one considers: for $[3.6 \mu
\mathrm m]-[8 \mu \mathrm m]$ and $[5.7 \mu \mathrm m]-[8 \mu \mathrm
m]$ the solution is actually found by extrapolation, whereas for $[4.5
\mu \mathrm m]-[8 \mu \mathrm m]$ $\alpha=-1$ occurs within the range
covered by the data\footnote{This is possibly due to the fact that the
  $4.5 \mu \mathrm m$ band is the only band not directly including PAH
  features and therefore might reach a constant ratio to $H$ at higher
  $[8 \mu \mathrm m]-[H]$ than for the two other bands.}. The
variations in this ``stellar point'' among the three different
determinations provides some direct empirical estimate of the
uncertainty of this flux ratio. Flux ratios between the other bands
are obtained by plugging the resulting ``stellar'' $[8 \mu \mathrm
m]-[H]$ colour into the corresponding 4th order polynomials from Table
\ref{tab_fitIRAC}.  The resulting three determinations of the stellar
flux ratios are given in Table \ref{tab_pure_stellar}, along with
their mean and standard deviation. Our determinations are accurate at
$<0.01$~dex (2\%) level for all bands except $[8 \mu \mathrm m]$,
whose ratio with respect to shorter wavelength cannot be estimated
better than 20\%.  We also report the IR flux ratios from SPS models:
the wide-spread \cite{helou+04} values, which were computed using
\texttt{Starburst99} \citep{starburst99}, and those we derive from a
preliminary 2007 version (CB07 hereafter) of the \cite{bc03} SPS
models, which includes an improved treatment of TP-AGB stars. These
models predict little dependence on stellar metallicity and age for
ages larger than 2 Gyr, while variation up to 30\% or more can be
expected for younger populations. For this reason we report the ratios
for simple stellar populations (SSPs) older than 2 Gyr and for an SSP
1 Gyr old, both using solar metallicity and Salpeter IMF. The
\cite{helou+04} values typically over-estimate by $\approx 20$\% the
IR flux ratios to $[3.6 \mu \mathrm m]$ with respect to our empirical
estimates. By contrast, the old CB07 SSPs are in excellent agreement
with observed estimates between $H$ and $4.5 \mu \mathrm m$, while it
appears that fluxes at longer wavelengths are significantly
under-estimated by up to $\approx50$\%. In consideration of the
apparent disagreement between SPS models, the empirical ratios derived
from our analysis are of great importance to improve the SED modeling
of this spectral region.
\begin{table*}
\begin{minipage}{\textwidth}
  \caption{Pure stellar emission flux ratios derived from IR
    colour-colour relations presented in Fig. \ref{fig_IRAC_CHXall}. A
    comparison to the predictions from stellar population synthesis
    models is given in the three bottom rows (see text for
    details).}\label{tab_pure_stellar}
\begin{scriptsize}
\begin{tabular}{lccccccc}

  \hline

  Slope from relation                     & $[3.6]-[H$]&$[4.5]-[H]$ &$[5.7]-[H]$ &$[8]-[H]$   &$[4.5]-[3.6]$ &$[5.7]-[3.6]$ &$[8]-[3.6]$ \\
  (1) & (2) & (3) & (4) & (5) & (6) & (7) & (8)\\
  \hline
  $[3.6 \mu \mathrm m]-[8 \mu \mathrm m]$ & $-0.742$ & $-1.044$ & $-1.194$ & $-1.491$ & $-0.301$   &$-0.453$    &$-0.749$  \\
  $[4.5 \mu \mathrm m]-[8 \mu \mathrm m]$ & $-0.740$ & $-1.046$ & $-1.178$ & $-1.384$ & $-0.306$   &$-0.437$    &$-0.644$  \\
  $[5.7 \mu \mathrm m]-[8 \mu \mathrm m]$ & $-0.741$ & $-1.038$ & $-1.199$ & $-1.572$ & $-0.297$   &$-0.458$    &$-0.832$  \\
  \hline
  mean                                    &$-0.741\pm0.001$&$-1.043\pm0.003$&$-1.190\pm0.009$&$-1.482\pm0.077$&
  $-0.301\pm0.004$&$-0.449\pm0.009$&$-0.742\pm0.077$\\
  \hline
  Helou et al. (2004)                     & $...   $ & $...   $ & $...   $ & $...   $ & $-0.225$   &$-0.399$    &$-0.635$  \\
  CB07 ($>2$Gyr)                          & $-0.745$ & $-1.047$ & $-1.321$ & $-1.687$ & $-0.302$   &$-0.576$    &$-0.942$  \\
  CB07 ($\approx 1$Gyr)                   & $-0.644$ & $-0.893$ & $-1.158$ & $-1.514$ & $-0.249$   &$-0.514$    &$-0.870$  \\
\end{tabular}
\end{scriptsize}
\end{minipage}
\end{table*}

\subsection{Mid-IR colour relations of the ``anomalous'' galaxies
  NGC\,4552 and NGC\,3521}\label{subsec_IR_anolamous}
The two galaxies that stand out due to their anomalous principal
component analyses also have evident anomalies when we look at their
IR colour-colour relations (i.e.~red and blue clouds in
Fig. ~\ref{fig_IRAC_CHXall}).

As expected from the lack of current star-formation, the elliptical
galaxy NGC\,4552 has the lowest ratio of 8 $\mu$m luminosity to $H$
and also the lowest ratios of luminosities in the three shorter IRAC
bands with respect to $[8 \mu$m$]$, as one can see from the red points
in Fig. \ref{fig_IRAC_CHXall}. However, a substantial fraction of
those points appear to lie above the extrapolated colour relations for
all other galaxies. Stellar population synthesis models of these
spectral regions are unfortunately still too inaccurate to provide an
explanation for these deviations of the order of 20--30\%. The
observed phenomenon can be summarized by saying that the luminosity at
shorter IR wavelengths is (increasingly) suppressed with respect to
longer wavelengths. As we verify using the stellar population
synthesis models with dust attenuation of ZCR09, these mid-IR colours
could be explained by invoking a substantial amount of reddening by
dust, which can in principle move the data points above and to the
right of the relations of Fig. \ref{fig_IRAC_CHXall}. However, our
maps show that the outer regions of the galaxy are the most affected
by this effect and that there is a suppression of NIR emission also
with respect to optical wavelengths. In addition, NGC\,4552 is not
detected at 160 $\mu$m with Spitzer \citep{draine+07}, thus implying a
negligible amount of dust. All these facts argue against dust
reddening as a cause of the anomalous IR colours. We also explore the
possibility that a radial variation of the stellar mass function
originates such an effect: using CB07 SPS models we find that for
stellar populations older than a couple of Gyr (as it is the case for
NGC\,4552), the IR colours differ by less than a few per cent going
from a Salpeter to a Chabrier IMF. According to the spectral indices
measured by \cite{trager+00} in different apertures and the Sauron
index maps of \cite{kuntschner+06}, NGC\,4552 is iron-enriched and
$\alpha$-enhanced to supersolar values \citep[by comparison with,
e.g., the][models]{thomas+03}. $[\alpha/\mathrm{Fe}]$ appears to
moderately increase at large radii, while the decrease in the strength
of the iron indices toward larger radii hints at a moderate decrease
in metallicity. According to the CB07 SPS models, the decrease in
metallicity could affect the mid-IR colours in the required
direction. We note however that the anti-correlation between the
strength of the iron indices and $[\alpha/\mathrm{Fe}]$
\citep{thomas+03} makes this metallicity gradient quite uncertain. We
suggest that the increase of $[\alpha/\mathrm{Fe}]$ to values up to
0.35--0.40 dex higher than solar at large radii could be the reason of
the anomalous (near)IR colours. A thorough investigation of the
dependence of these colour on $[\alpha/\mathrm{Fe}]$ and metallicity,
which is beyond the scope of this paper, would be required to provide
a robust explanation of the observed effects.

As already discussed in Sec. \ref{subsec_PCA_anolamous}, anomalies in
the SED of NGC\,3521 result from the complex effects of substantial
dust absorption and scattering. From
Fig. \ref{fig_IRAC_CHXall}\textit{d)} and
Fig. \ref{fig_IRACsynth_N03521} (\textit{online} appendix), there is a
significant fraction of pixels in this galaxy in good agreement with
the ``universal'' relations derived in the previous section. These
pixels are located in the far (north-east) part of the inner disc,
those apparently less affected by extinction and scatter. All other
pixels appear to form a sequence offset from the relations of
Fig. \ref{fig_IRAC_CHXall}, which underestimate the fluxes at 3.6, 4.5
and 5.7$\mu$m with respect to the observed ones. On the near
(south-western) side of the disc the stellar light is heavily dust
obscured, so much so that even the $H$-band (and possibly the shortest
IRAC bands) is attenuated, and this probably explains the offset of
these pixels to the right (and up) of the relations. However, the
reason for the pixels in the halo of the galaxy being similarly offset
is unclear. It is possible that something similar to what happens in
the outer regions of NGC\,4552 occurs here as well, as the continuity
between the sequence of the red and the blue pixels of
Fig. \ref{fig_IRAC_CHXall} suggests.

\section{Summary and concluding remarks}\label{sec_conclusions}
Using multiwavelength data ($u$-band to 8$\mu$m) for seven nearby
galaxies, resolved at scales of $\gtrsim200$\,pc pix$^{-1}$ with a
Signal-to-Noise Ratio of $>10-20$ for each pixel, we have
demonstrated, both qualitatively and quantitatively, the disconnect
that occurs between the optical and infrared colours relative to the
near-IR $H$-band on the small scales within galaxies. A very important
point in the analysis is the choice of the $H$-band point as the
normalization of the pixel SEDs, as this band is the least affected by
dust absorption \textit{and} emission at the same time and divides the
portion of SEDs possibly affected by dust emission (longward of it)
from the one dominated by stellar emission, possibly
\textit{extincted} by dust (shortward of it). Normalizing at other
wavelengths where dust effects are more substantial would likely
destroy the observed correlations by mixing different
emission/absorption mechanisms. In fact, we have checked that using a
slightly shorter wavelength for normalization, such as the $z$ band,
results in similar although more scattered correlations. On the other
hand, if we move the normalization to longer wavelengths, we expect
quite dramatic effects already at 3.6$\mu$m, where we have
demonstrated the contamination by dust emission to be substantial.

Qualitatively, the disconnect between the optical and infrared colours
can be seen in colour--colour plots (figures \ref{fig_SEDcorr_N04254},
\ref{fig_SEDcorr_N04450}, and \ref{fig_SEDcorr_N03521} to
\ref{fig_SEDcorr_N04579}) where correlations can be visibly seen
between optical--optical colours, IR--IR colours, but not optical--IR
colours. This holds true for all but two of the seven galaxies:
NGC\,4552, an elliptical galaxy whose colours at all wavelengths
examined are dominated by stellar light from old stars, and thus all
colours are correlated to a high degree; and NGC\,3521, a more
inclined late-type galaxy with significant scatter in its colours,
with dust causing weird colour correlations due to the effects of
attenuation on the near side of the galaxy, and scattering on the far
side.

The observed disconnect in colours can be investigated using Principal
Component Analysis (PCA) of the distribution in $H$-band normalized
Spectral Energy Distribution (SED) space of the pixels in each
galaxy. This analysis picks out two distinct and uncorrelated
components in each of the five ``typical'' galaxies; a component that
dominates the variance and is strongly correlated to the variations in
the IR, and a component that is mainly correlated to the variations in
the optical (see figures \ref{fig_PCA_N04254}, \ref{fig_PCA_N04450},
and \ref{fig_PCA_N03521} to \ref{fig_PCA_N04579}). These two
components account for most ($>90$\%) of the observed variance in
colours across the galaxies, and show distinct spatial variations in
galaxies, with the dominant primary component associated with the IR
variance (PC1) being stronger in spiral arms and star forming regions,
and the other optically dominated component (PC2) being enhanced in
the interarm regions and outskirts of galaxies.

These two principal components can be associated with physical
properties within the galaxies by comparison with the surface
brightnesses in H$\alpha$ and in $H$-band, themselves proxies for the
star formation rate (SFR) and stellar mass ($M_{*}$). This comparison
(figures \ref{fig_SBPC_N04254}, \ref{fig_SBPC_N04450}, and
\ref{fig_SBPC_N03521} to \ref{fig_SBPC_N04579}) reveals a strong
correlation of PC1 with increasing SB(H$\alpha$)/SB($H$), an indicator
of the specific star formation rate (sSFR), while PC2 variation
correlates with SB($H$), although with significant scatter.

Altogether, these correlations indicate that PC1 is detecting regions
where young stars are dominating the colours, and predominantly in the
IR, while PC2 is picking up variations in the mean stellar age, which
is associated with the stellar mass density, with the most massive
regions generally the oldest on average. It is these two distinct
components that both the colours and PCA technique reveal.

Given the disconnect observed on the small scales, it is surprising
that the global correlations of galaxy colours and types exist
\citep[e.g.][]{kennicutt_98}. However, as figure
\ref{fig_stack_SEDcorr} reveals, while there are distinctions on the
pixel scales ($\gtrsim 200$pc), globally the mean SEDs and colours
trace very well the type of the galaxy, with later-types being bluer
and dustier (IR brighter) than early-types. Within each galaxy, the
scatter (and disconnect of IR and optical) is around this mean SED. In
more physical terms, we can interpret this as the result of sSFR and
stellar mass density (stellar age) being correlated \textit{between}
galaxies, but not (necessarily) \textit{within} galaxies.

Interestingly, the relations between the mid-IR colours are even
tighter when all seven galaxies in our sample are considered together.
When an IR--IR colour (e.g.~$[3.6\mu\rm{m}]-[8\mu\rm{m}]$) is plotted
as a function of a IR--near-IR colour (i.e.~$[8\mu\rm{m}]-[H]$) these
tight relations are apparent (figure \ref{fig_IRAC_CHXall}). Simple
4th-order polynomial fits to these relations (resulting in the
coefficients in Table \ref{tab_fitIRAC}) have a r.m.s.~scatter around
them which is, most of the time, less than the maximum expected
photometric uncertainties. These functional fits are tight enough that
the $[8\mu\rm{m}]-[H]$ colour images can be used as a predictive tool
for the fluxes in the three other IRAC bands with deviation of only a
few percent typically (figure \ref{fig_IRACsynth_N04254} and figures
\ref{fig_IRACsynth_N03521} to \ref{fig_IRACsynth_N04579}).

What these relationships reveal is the varying contribution of dust to
the IRAC bands. The $H$-band is purely stellar, with no dust
contribution, while the 8$\mu$m IRAC band is dominated by the
7.7$\mu$m polycyclic aromatic hydrocarbon (PAH) complex, with only a
weak stellar contribution, and thus the $[8\mu\rm{m}]-[H]$ colour is a
measure of relative dust contribution to the pixel SEDs. The
colour--colour diagrams of figure \ref{fig_IRAC_CHXall} therefore show
the contribution of dust and stars in each band, going from pure
stellar (where the slope is $-1$) to dust dominated (where the slope
is $0$). By extrapolating the function fits to where the slope becomes
$-1$ we are able to empirically determine the pure stellar colours in
the IRAC bands (table \ref{tab_fitIRAC}), at least for the galaxies in
our sample. The plots clearly show the need to correct for dust when
using the IRAC bands as tracers of stellar mass (see e.g. Meidt et
al. 2011 for such a correction). Interestingly, these colours match
that of the 2007 models of \citet{bc03} (not calibrated for these
wavelengths) for old ($>2$Gyr) stars at the shortest wavelengths, but
is much redder (flatter SED) at the long (5.7 and 8$\mu$m) bands and
bluer (steeper SED) than that suggested by \citet{helou+04} in these
bands.  Another interesting note is that the scatter of pixels in the
stellar and dust regimes of these relations gives an indication of the
intrinsic scatter in the stellar SED or the ``dust'' SED respectively
on the scales explored here. This scatter is surprisingly low (figure
\ref{fig_IRAC_CHXall}\textit{a)--b)}, lower panels), given the known
contribution of various stellar evolutionary stages to the $H$-band,
though the scatter is somewhat larger for dust given the observed
variation in PAH features and hot dust contribution in galaxies with
Spitzer-IRS \citep[e.g.][]{smith+07}.

In Summary, while on global scales the IR and optical colours of
galaxies are correlated with their star formation history and
galaxy-type, the correlation of IR and optical colours breaks down
internally in galaxies on local scales. The IR dominated and optical
dominated components can be spatially associated with regions and with
physical processes, with the IR excess coming from regions of high
specific star formation rate, while the optical excess is broadly
associated to low $H$-band surface brightness (stellar mass density)
regions and to unobscured young stellar associations along spiral
arms. The disconnect of optical and IR colours is because of this
spatial distinction of the uniform underlying older stellar
population, and the more concentrated, stochastic star forming
regions.  However, this disconnect of optical and IR colours in
galaxies is around a mean, and the mean colours of galaxies follow the
well-observed trend of bluer optical and stronger IR in later type
galaxies. The IRAC colours within galaxies follow each other closely,
with a tight enough relationship that a IRAC--near-IR colour can be
used to predict the fluxes in the other IRAC bands.

\section*{Acknowledgments}
We thank the anonymous referee for useful comments that have lead to an improved manuscript.\\
The Dark Cosmology Centre is funded by the Danish National Research
Foundation.

\bibliography{all_mybibs}


%
%

%
%
\onecolumn
\appendix
\clearpage
\section{Images in eleven bands and SEDs for the full
  sample}\label{append_images_sed}
\begin{figure*}
\includegraphics[width=\textwidth]{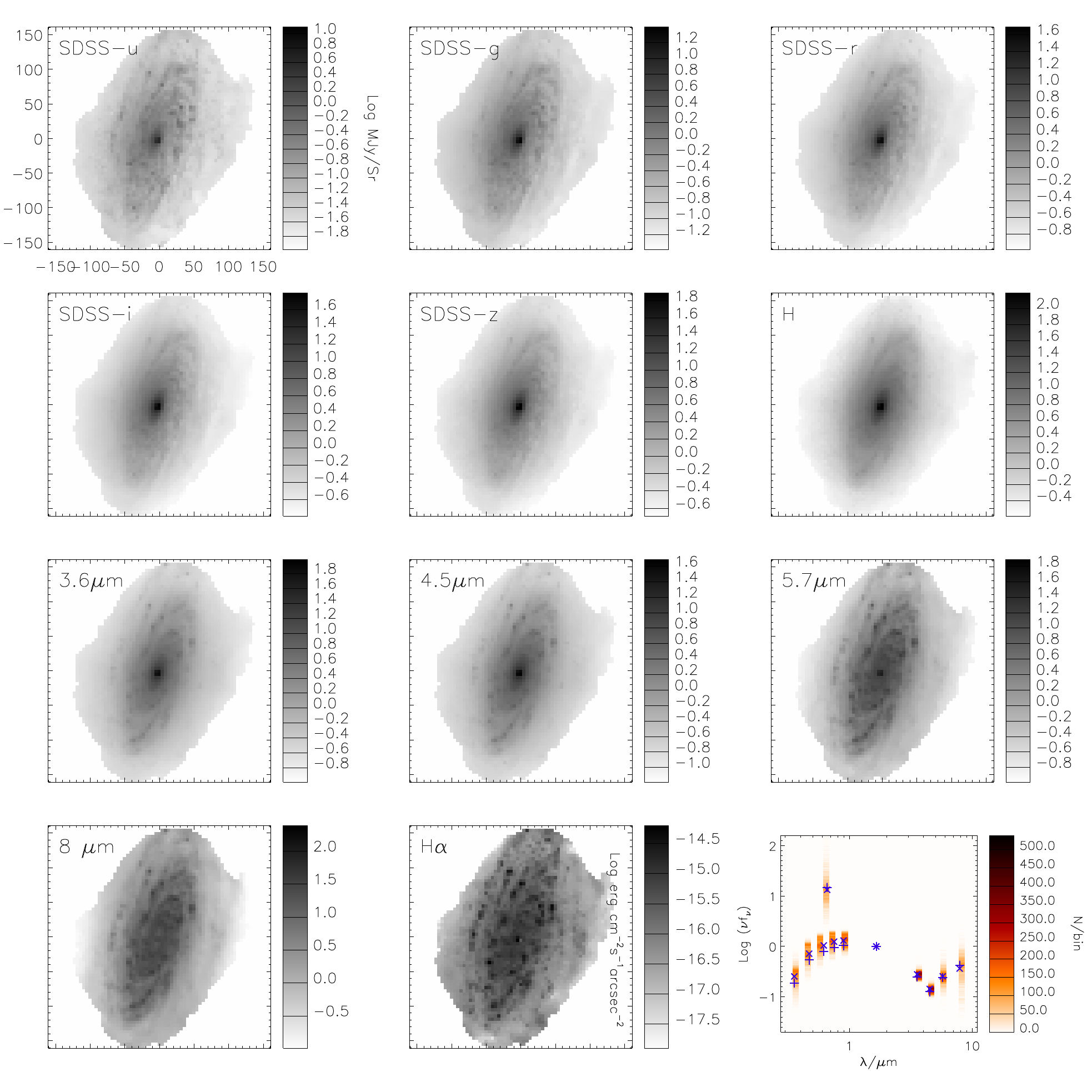}
\caption{Images in the eleven bands used for the analysis in this work
  for NGC\,3521. The greyscale indicates the log of the surface
  brightness in each band in MJy\,Sr$^{-1}$ as defined by the colour
  bar to the right of each panel, except for the H$\alpha$ image which
  is in log erg\,cm$^{-2}$\,s$^{-1}$\,arcsec$^{-2}$. In the lower
  right panel we show the distribution of log$\nu f_{\nu}$ SEDs of the
  pixels for the galaxy, normalized to the $H$-band. The H$\alpha$
  flux is also represented in this plot (at $\lambda=0.6563 \mu$m)
  using a flux density corresponding to the flux divided by a passband
  of 1\AA. The colour scale indicates the number density of pixels (in
  bins of 0.02 dex in normalized log$\nu f_{\nu}$) as labelled in the
  key to the right. The blue symbols indicate the arithmetic mean of
  the pixels ('$\times$' symbol) and the luminosity-weighted mean of
  the pixels (i.e.~ integrated SED of the galaxy, '$+$'
  symbol).}\label{fig_A_N03521}
\end{figure*}
\begin{figure*}
\includegraphics[width=\textwidth]{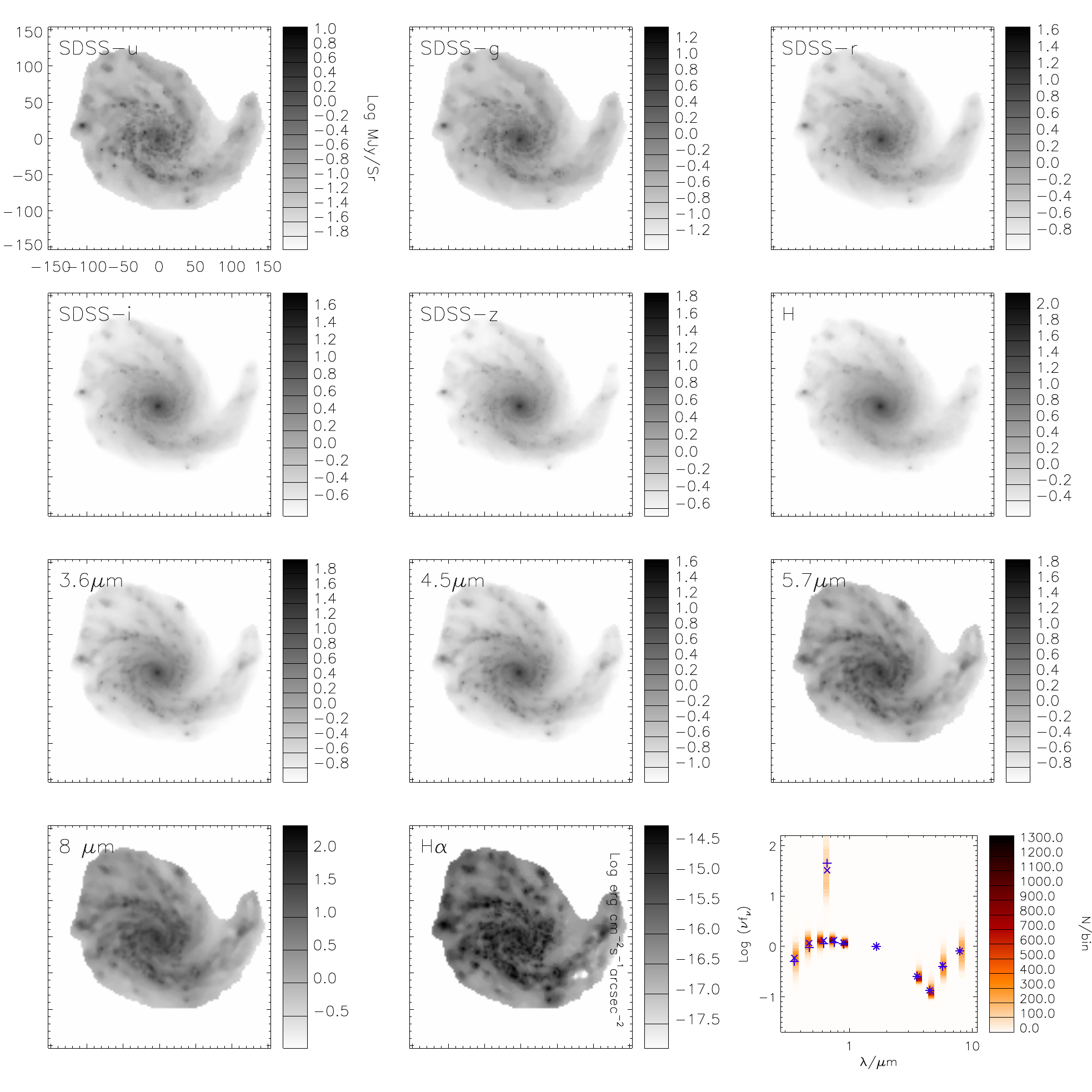}
\caption{As in Fig.\ref{fig_A_N03521}, but for NGC\,4254.}\label{fig_A_N04254}
\end{figure*}
\begin{figure*}
\includegraphics[width=\textwidth]{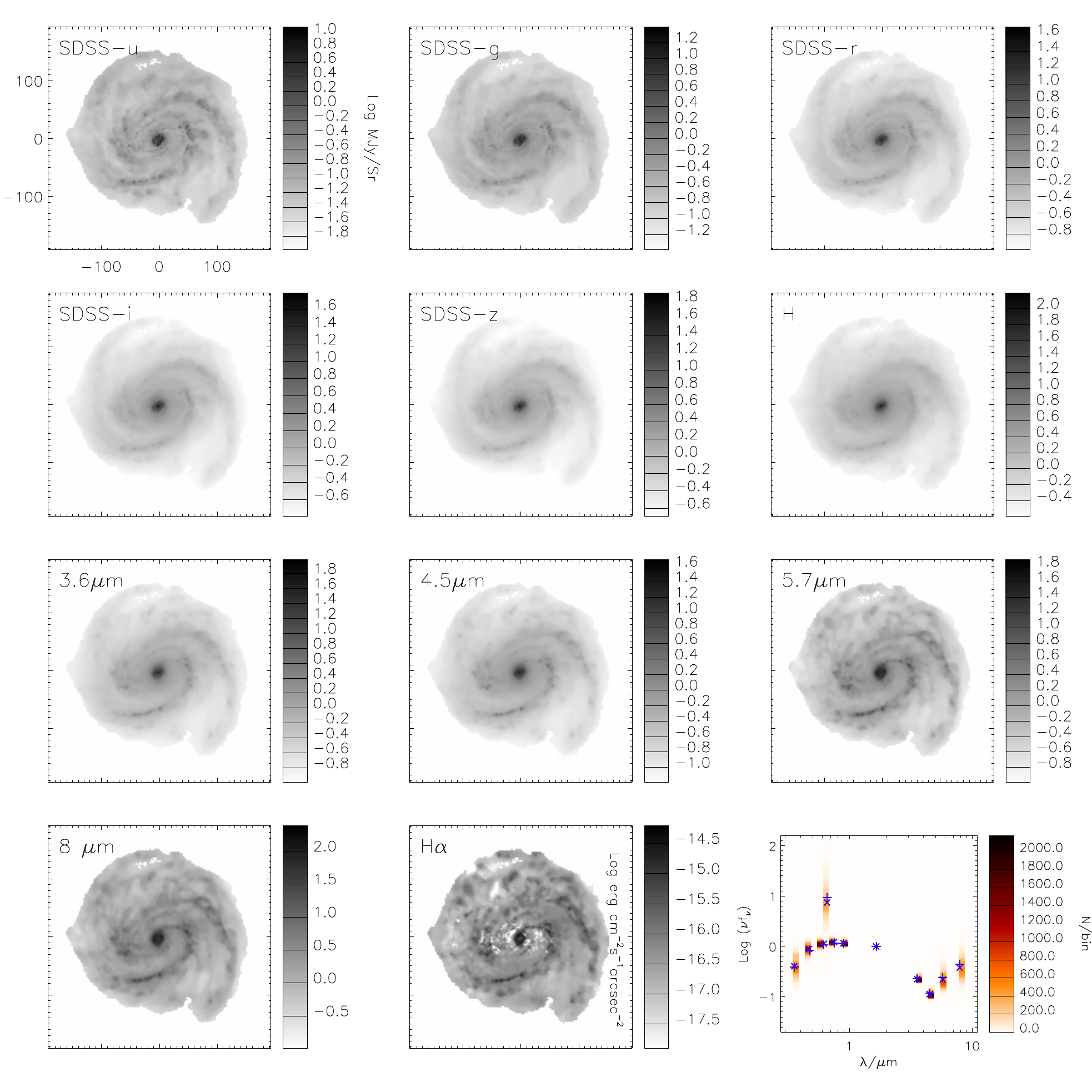}
\caption{As in Fig.\ref{fig_A_N03521}, but for NGC\,4321.}\label{fig_A_N04321}
\end{figure*}
\begin{figure*}
\includegraphics[width=\textwidth]{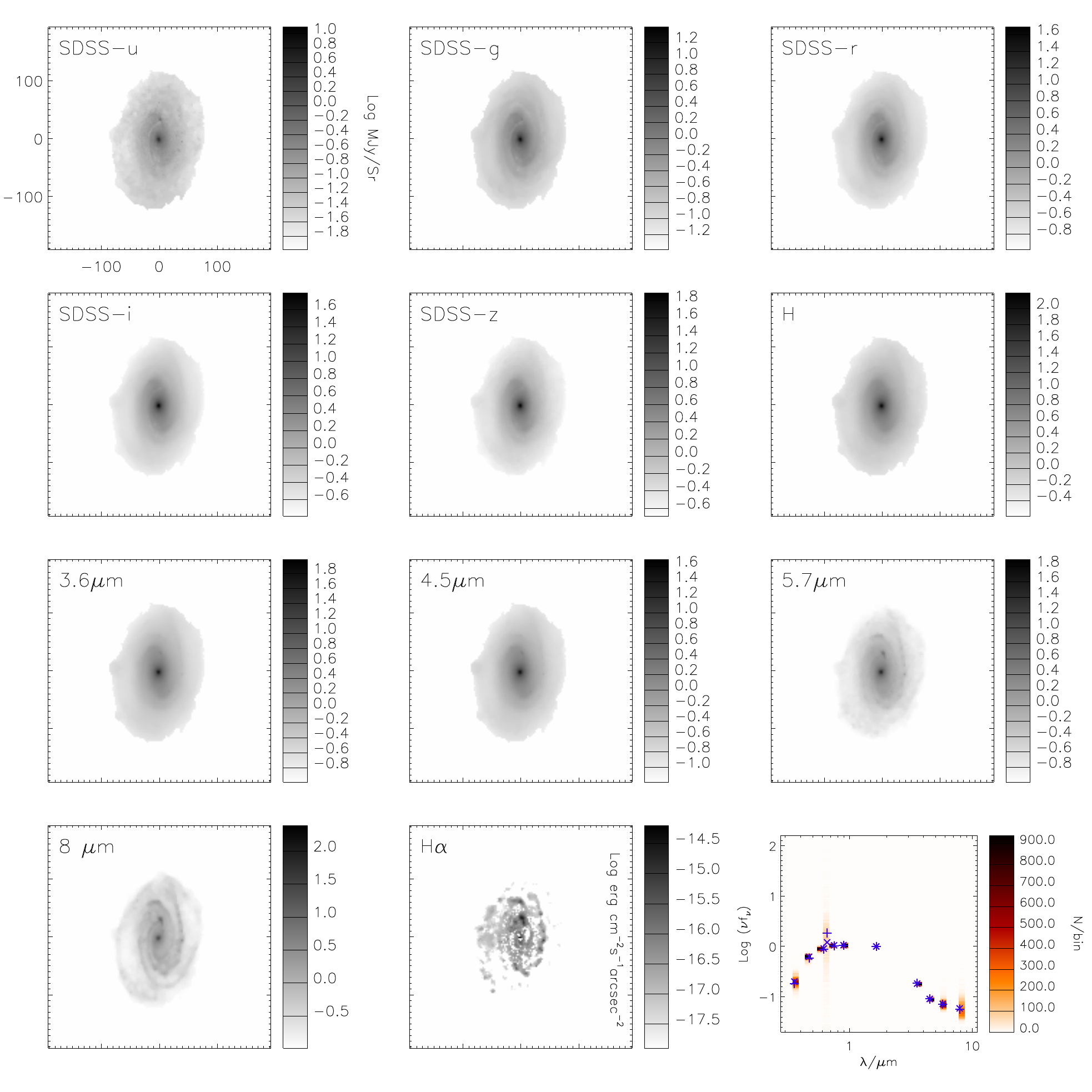}
\caption{As in Fig.\ref{fig_A_N03521}, but for NGC\,4450.}\label{fig_A_N04450}
\end{figure*}
\begin{figure*}
\includegraphics[width=\textwidth]{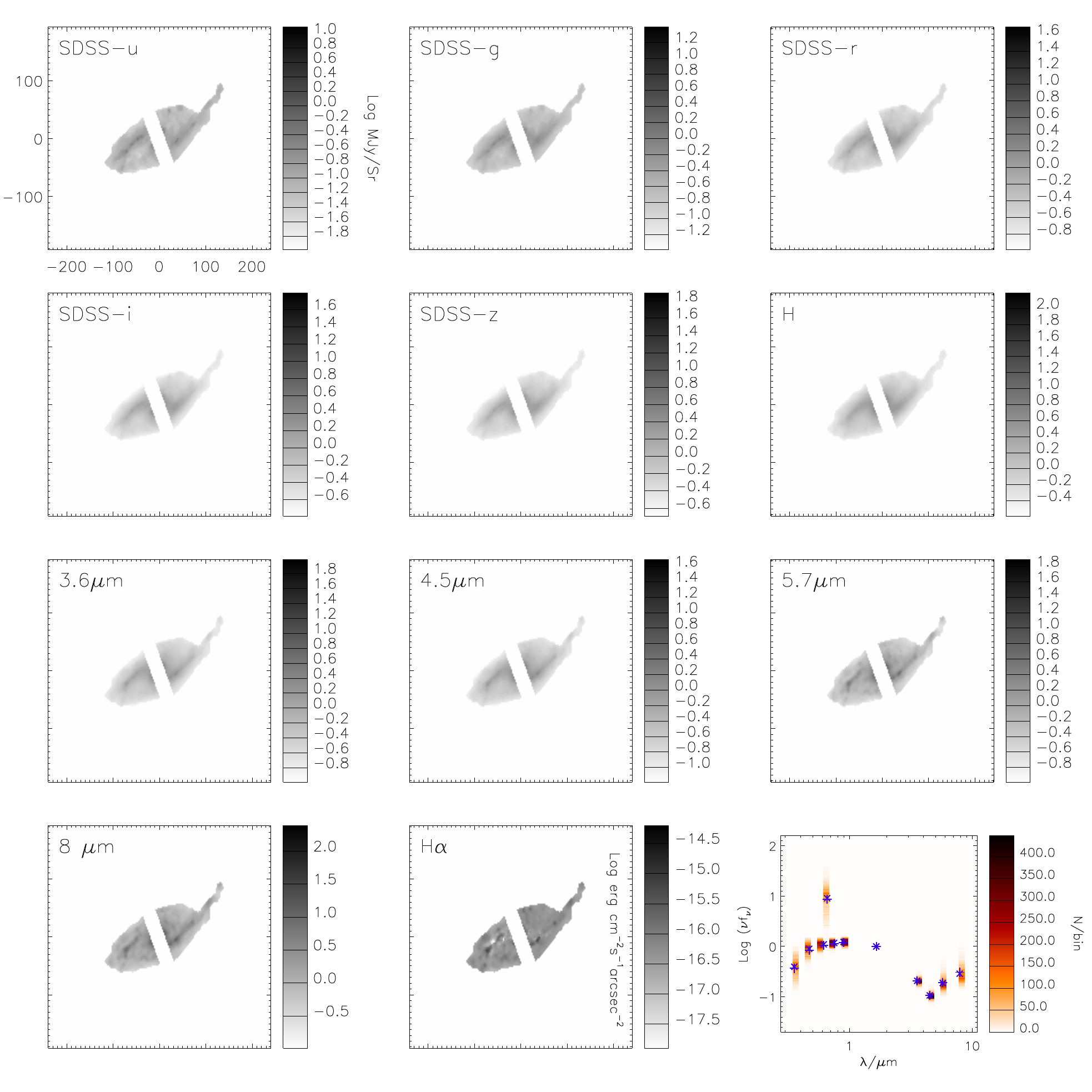}
\caption{As in Fig.\ref{fig_A_N03521}, but for NGC\,4536. The stripe
  of missing data going through the nucleus of the galaxy is due to
  the detector saturation in the IRAC bands.}\label{fig_A_N04536}
\end{figure*}
\begin{figure*}
\includegraphics[width=\textwidth]{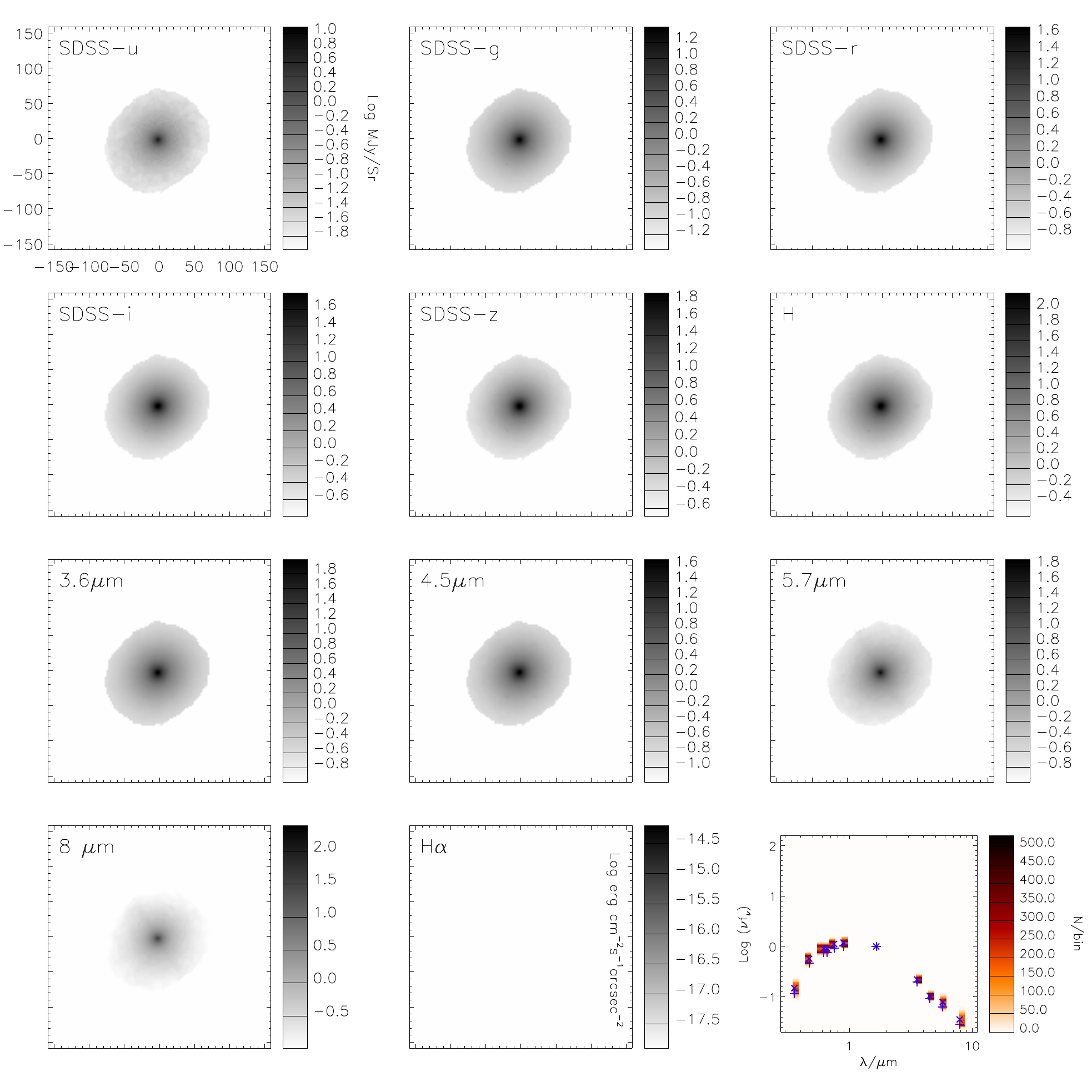}
\caption{As in Fig.\ref{fig_A_N03521}, but for NGC\,4552.}\label{fig_A_N04552}
\end{figure*}
\begin{figure*}
\includegraphics[width=\textwidth]{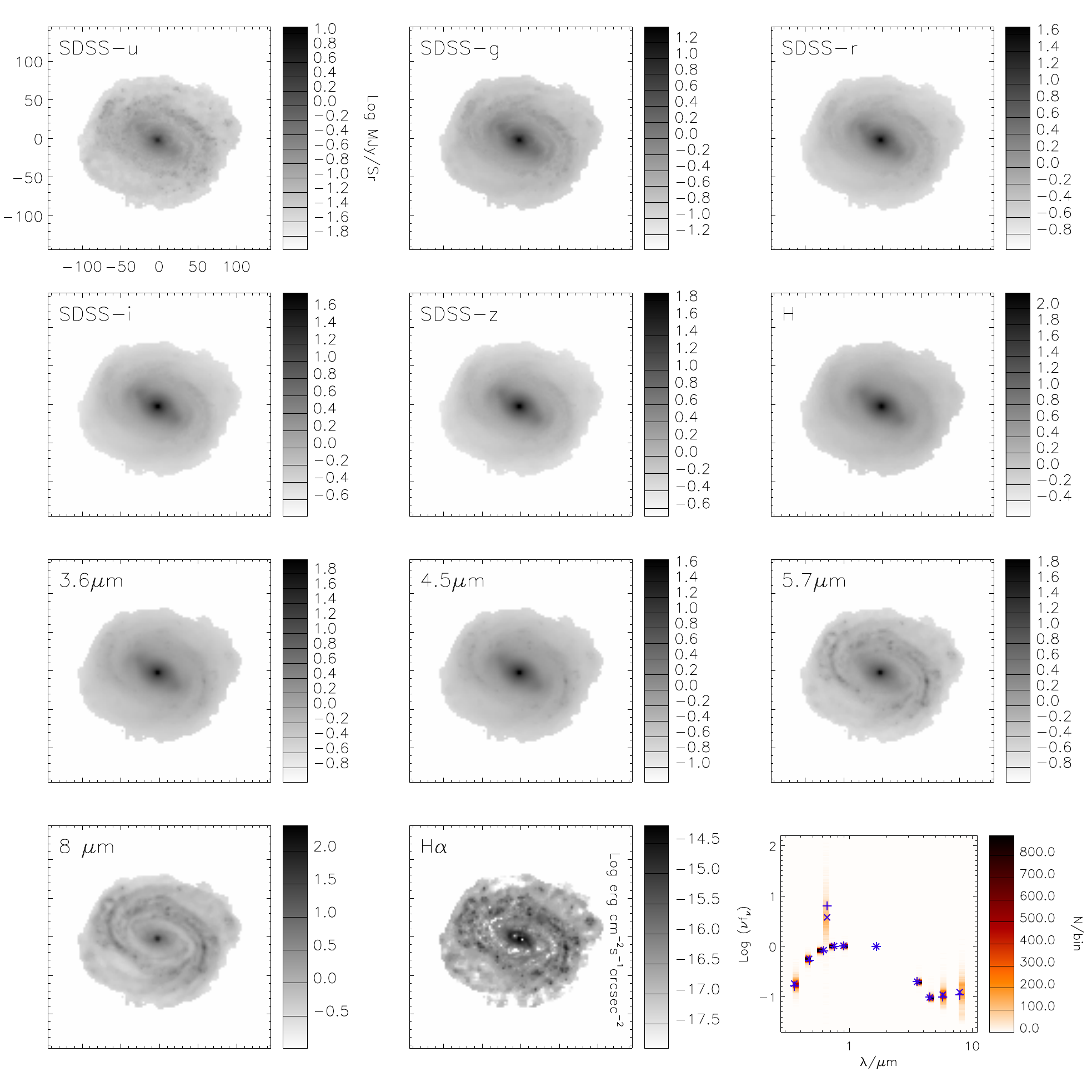}
\caption{As in Fig.\ref{fig_A_N03521}, but for NGC\,4579.}\label{fig_A_N04579}
\end{figure*}
\clearpage
\section{Colour correlations for the five galaxies not in the main
  paper}\label{append_sedcorr}
\begin{figure*}
\includegraphics[width=\textwidth]{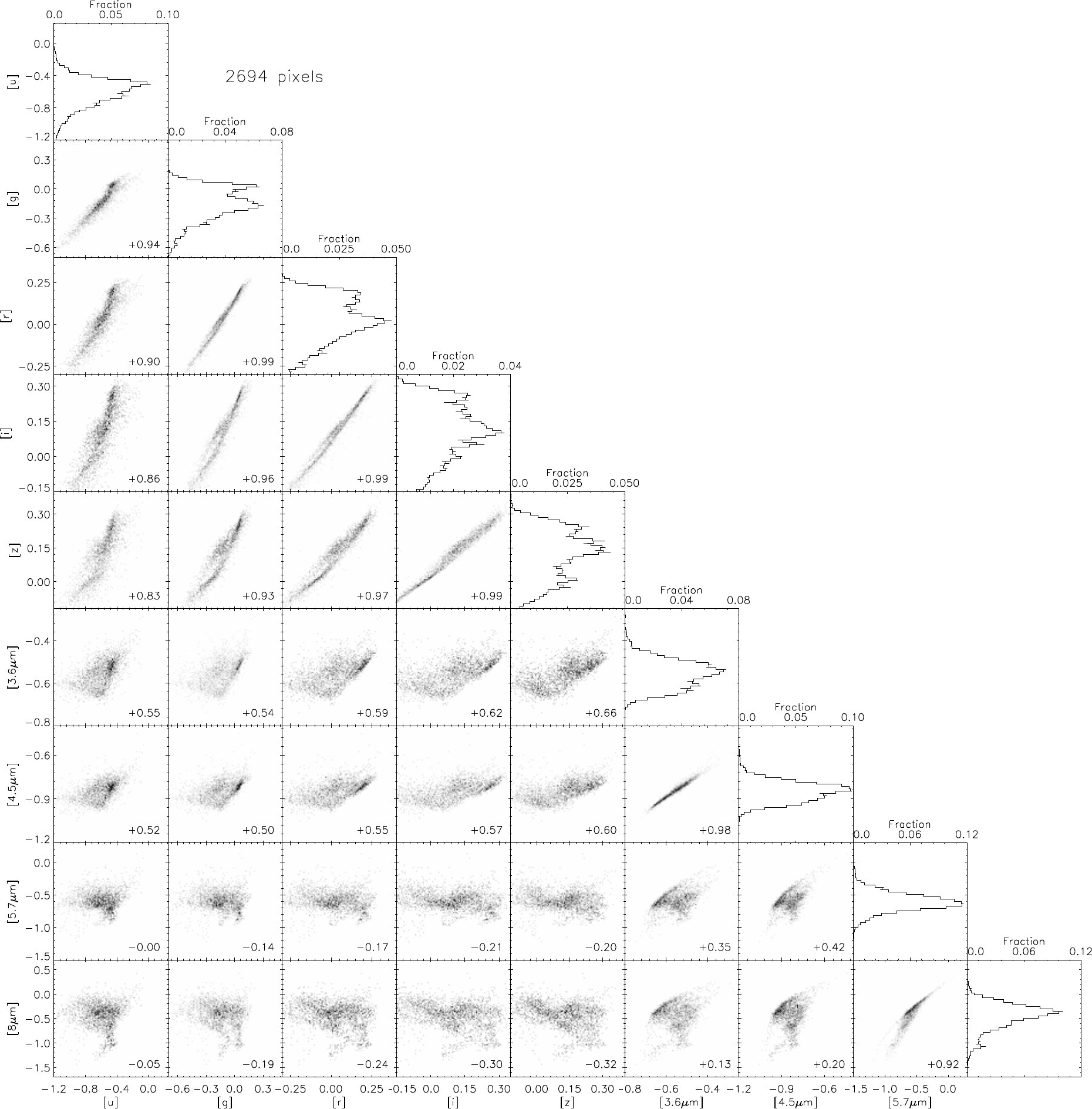}
\caption{As in Fig.\ref{fig_SEDcorr_N04254}, but for NGC\,3521.}\label{fig_SEDcorr_N03521}
\end{figure*}
\begin{figure*}
\includegraphics[width=\textwidth]{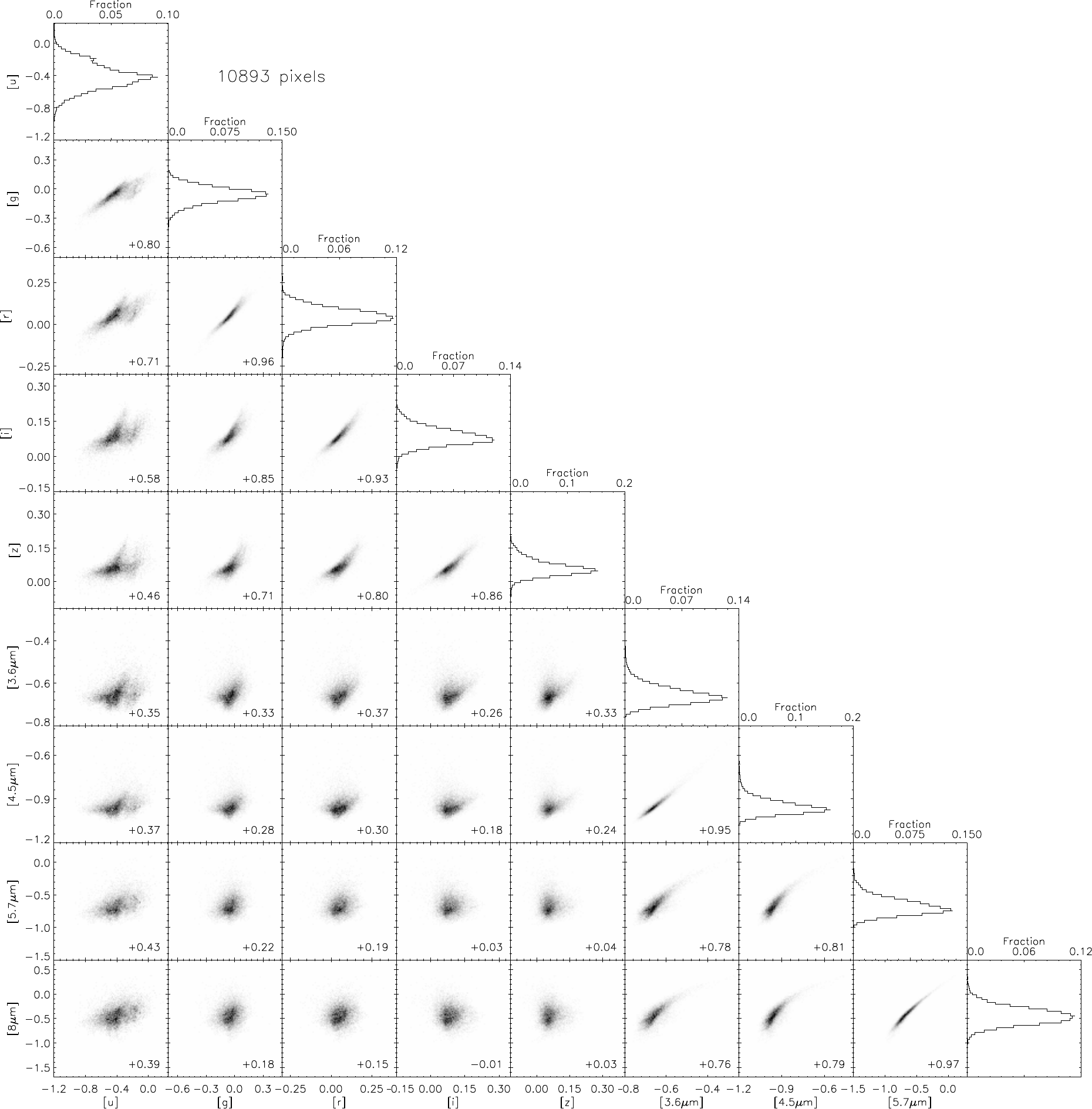}
\caption{As in Fig.\ref{fig_SEDcorr_N04254}, but for NGC\,4321.}\label{fig_SEDcorr_N04321}
\end{figure*}
\begin{figure*}
\includegraphics[width=\textwidth]{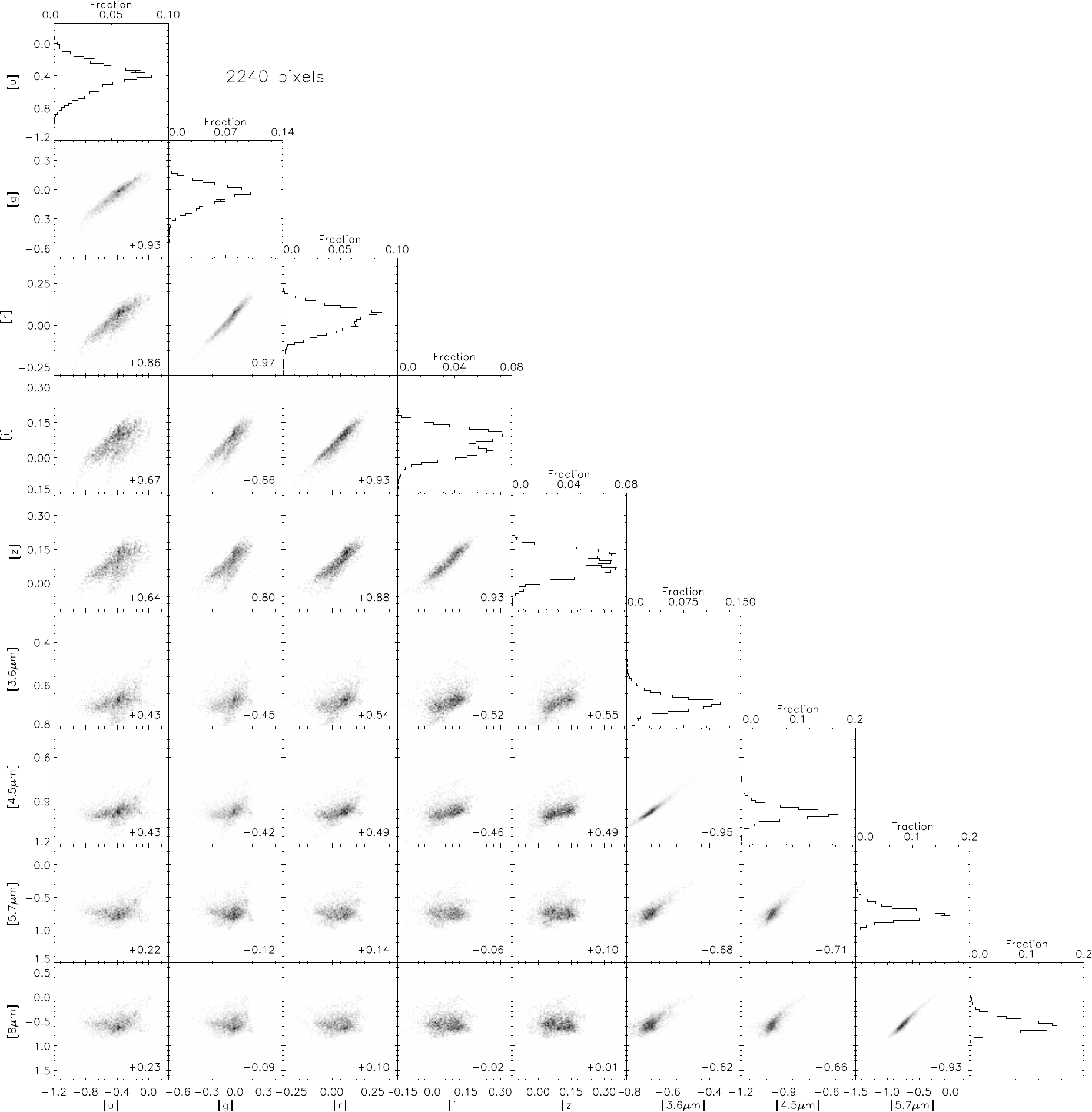}
\caption{As in Fig.\ref{fig_SEDcorr_N04254}, but for NGC\,4536.}\label{fig_SEDcorr_N04536}
\end{figure*}
\begin{figure*}
\includegraphics[width=\textwidth]{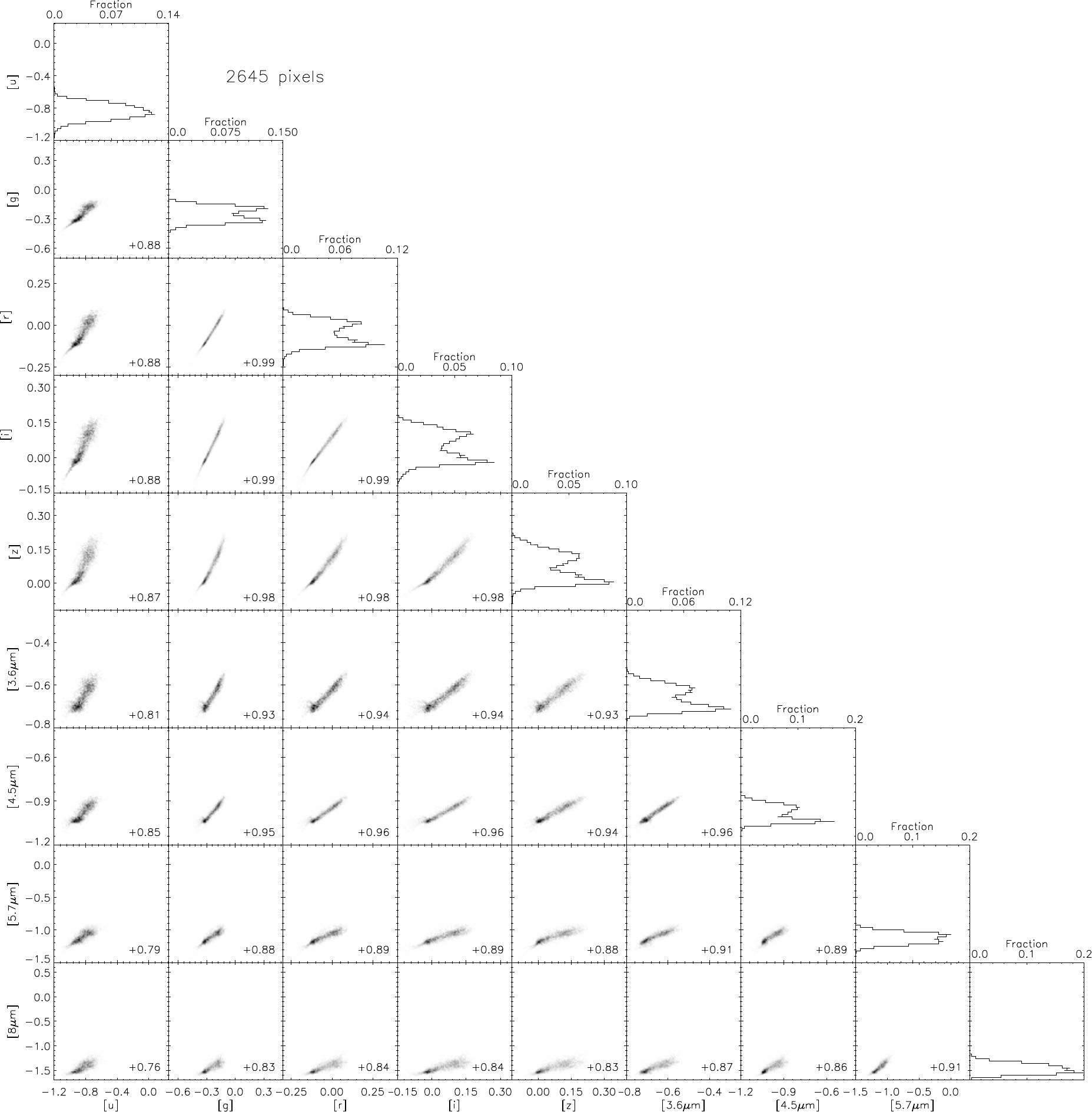}
\caption{As in Fig.\ref{fig_SEDcorr_N04254}, but for NGC\,4552.}\label{fig_SEDcorr_N04552}
\end{figure*}
\begin{figure*}
\includegraphics[width=\textwidth]{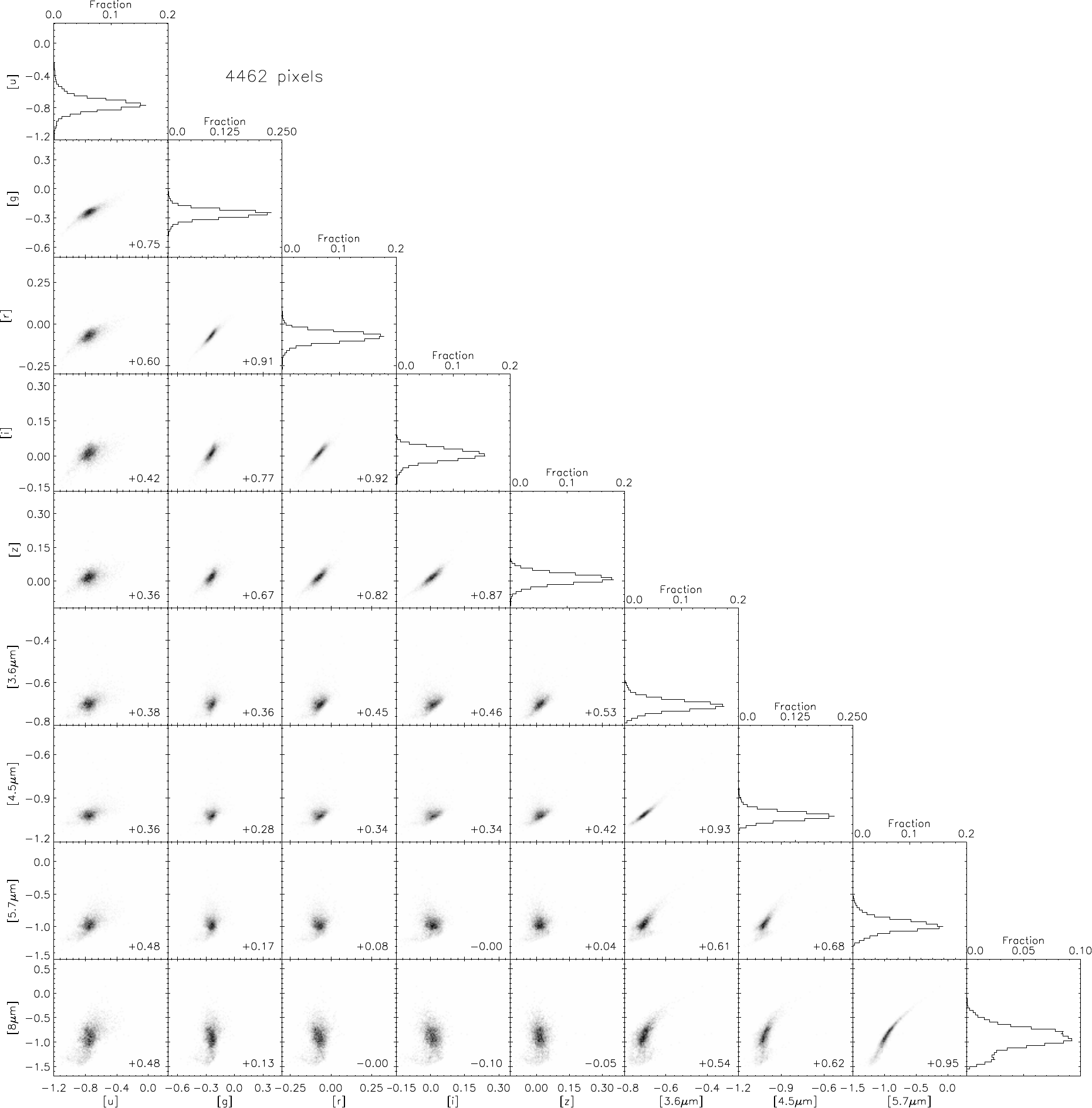}
\caption{As in Fig.\ref{fig_SEDcorr_N04254}, but for NGC\,4579.}\label{fig_SEDcorr_N04579}
\end{figure*}
\clearpage
\section{PCA decomposition for the five galaxies not in the main
  paper}\label{append_PCA}
\begin{figure*}
\includegraphics[width=\textwidth]{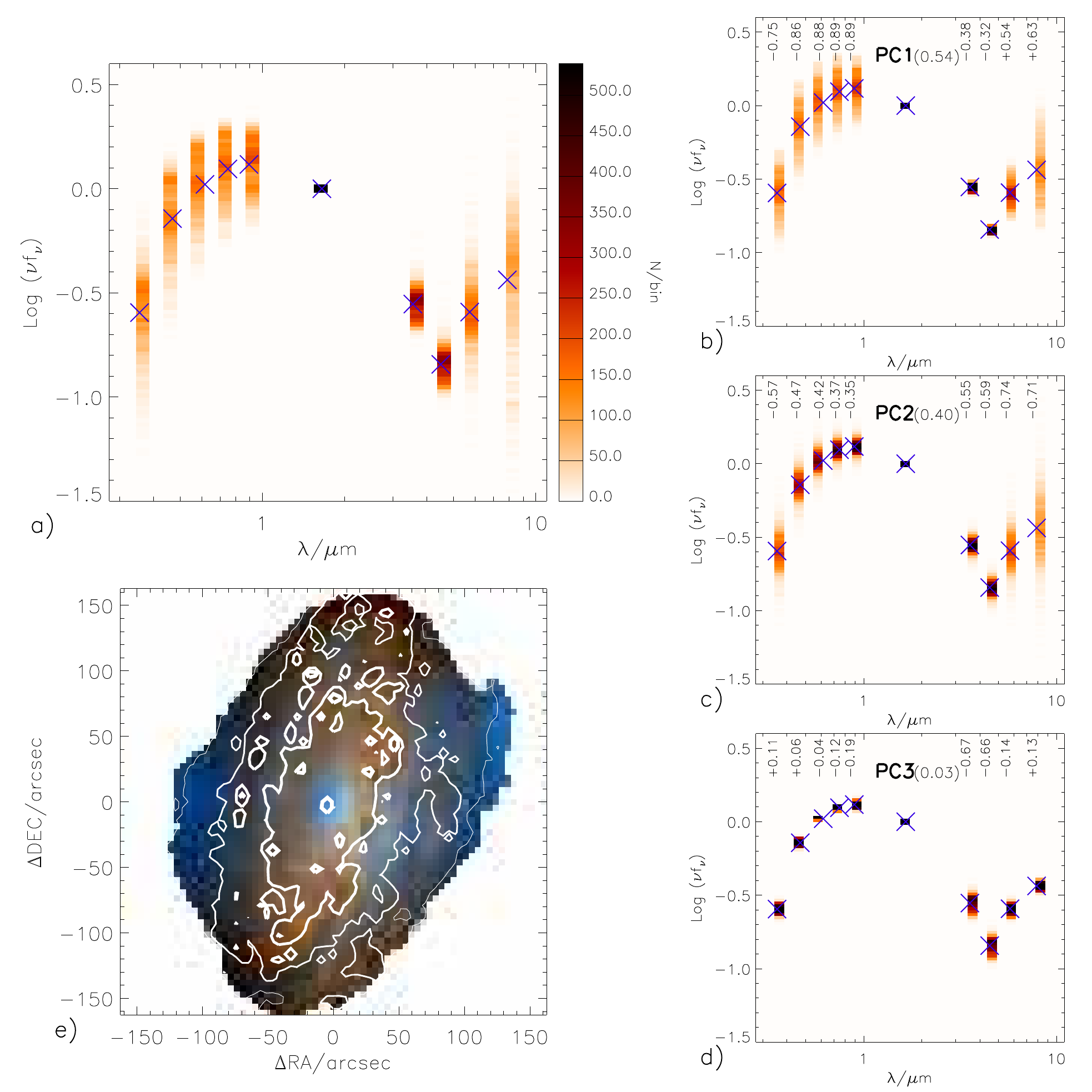}
\caption{As in Fig.\ref{fig_PCA_N04254}, but for the galaxy NGC\,3521.}\label{fig_PCA_N03521}
\end{figure*}
\begin{figure*}
\includegraphics[width=\textwidth]{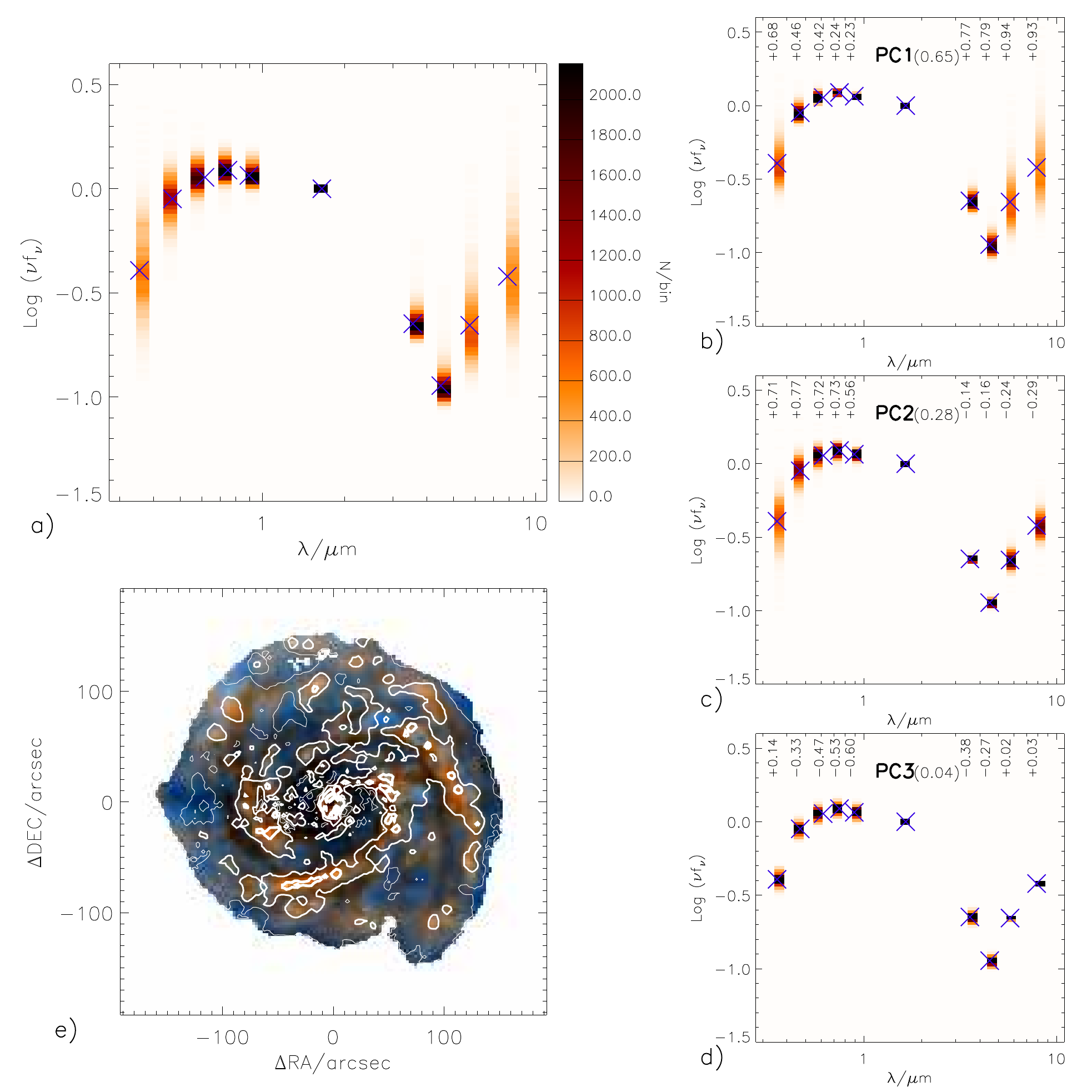}
\caption{As in Fig.\ref{fig_PCA_N04254}, but for the galaxy NGC\,4321.}\label{fig_PCA_N04321}
\end{figure*}
\begin{figure*}
\includegraphics[width=\textwidth]{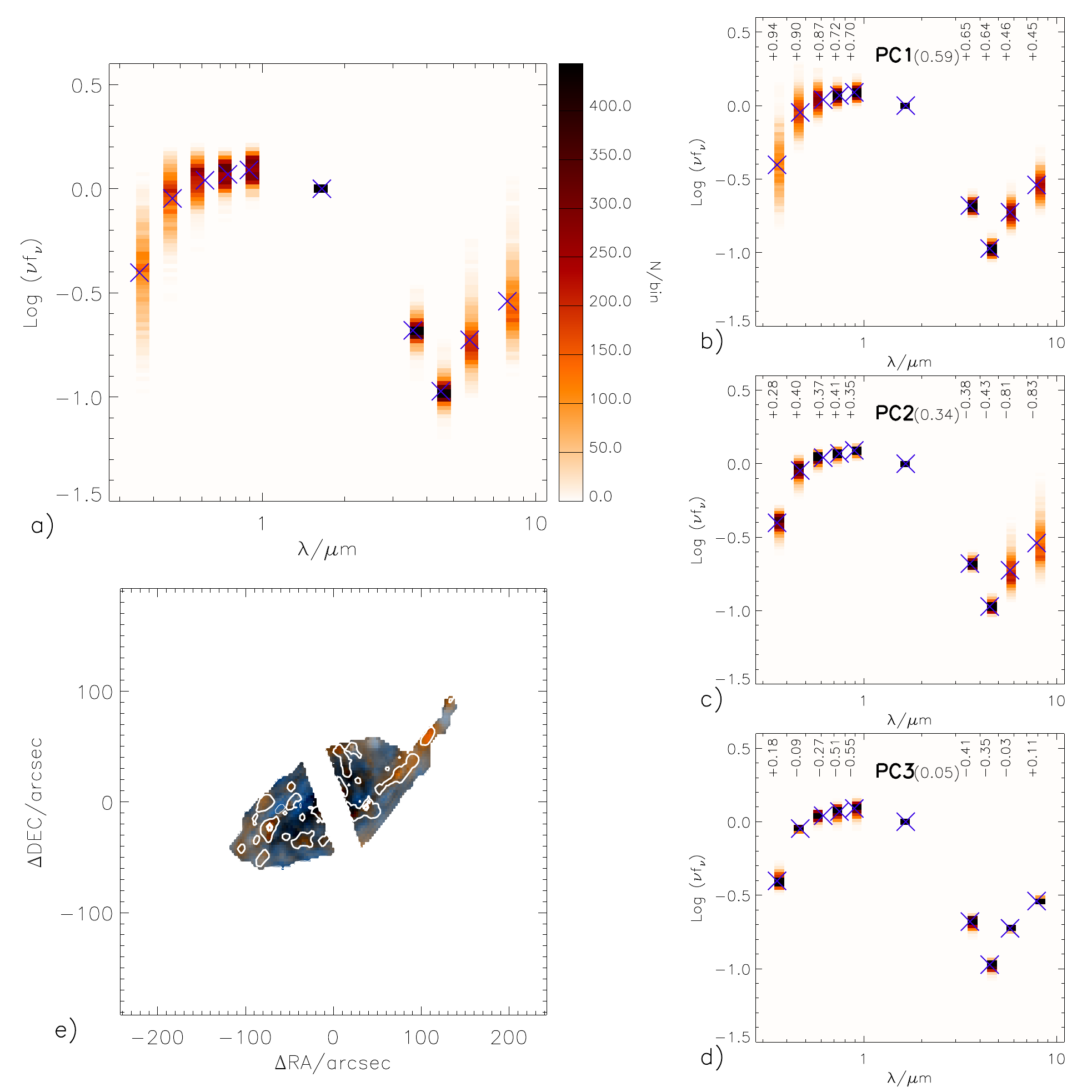}
\caption{As in Fig.\ref{fig_PCA_N04254}, but for the galaxy
  NGC\,4536. The stripe of missing data going through the nucleus of
  the galaxy is due to the detector saturation in the IRAC bands. Note
  that in this case, contrary to the other four ``regular'' galaxies,
  PC1 is mainly correlated to the optical colours and PC2 is
  anti-correlated with the IR colours. Modulo this exchange, main
  conclusions remain unaffected.}\label{fig_PCA_N04536}
\end{figure*}
\begin{figure*}
\includegraphics[width=\textwidth]{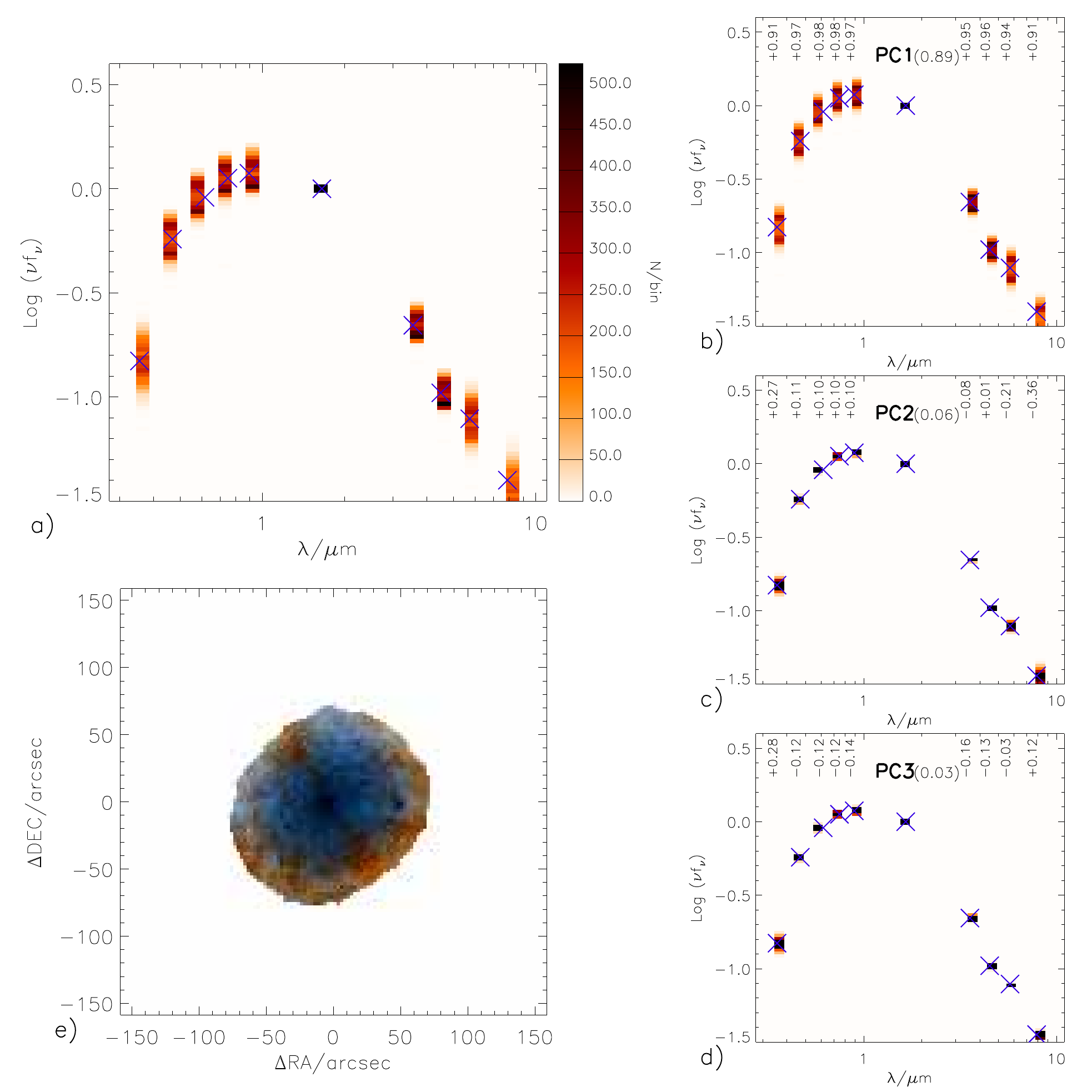}
\caption{As in Fig.\ref{fig_PCA_N04254}, but for the galaxy NGC\,4552.}\label{fig_PCA_N04552}
\end{figure*}
\begin{figure*}
\includegraphics[width=\textwidth]{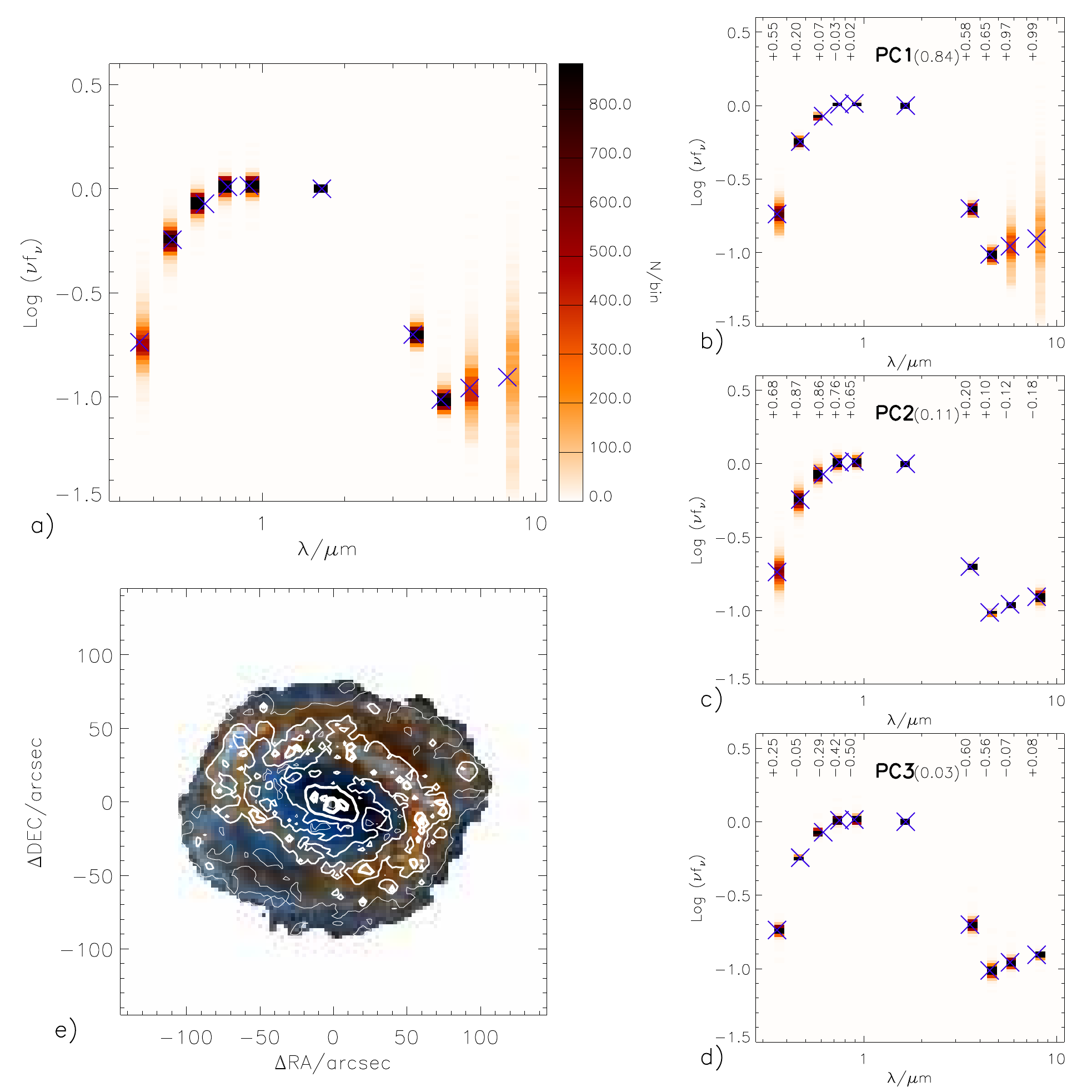}
\caption{As in Fig.\ref{fig_PCA_N04254}, but for the galaxy NGC\,4579.}\label{fig_PCA_N04579}
\end{figure*}
\clearpage
\section{Dependence of principal components on surface brightness and
  star formation for the four galaxies not in the main
  paper}\label{append_SBPCA}
\begin{figure*}
\includegraphics[width=\textwidth]{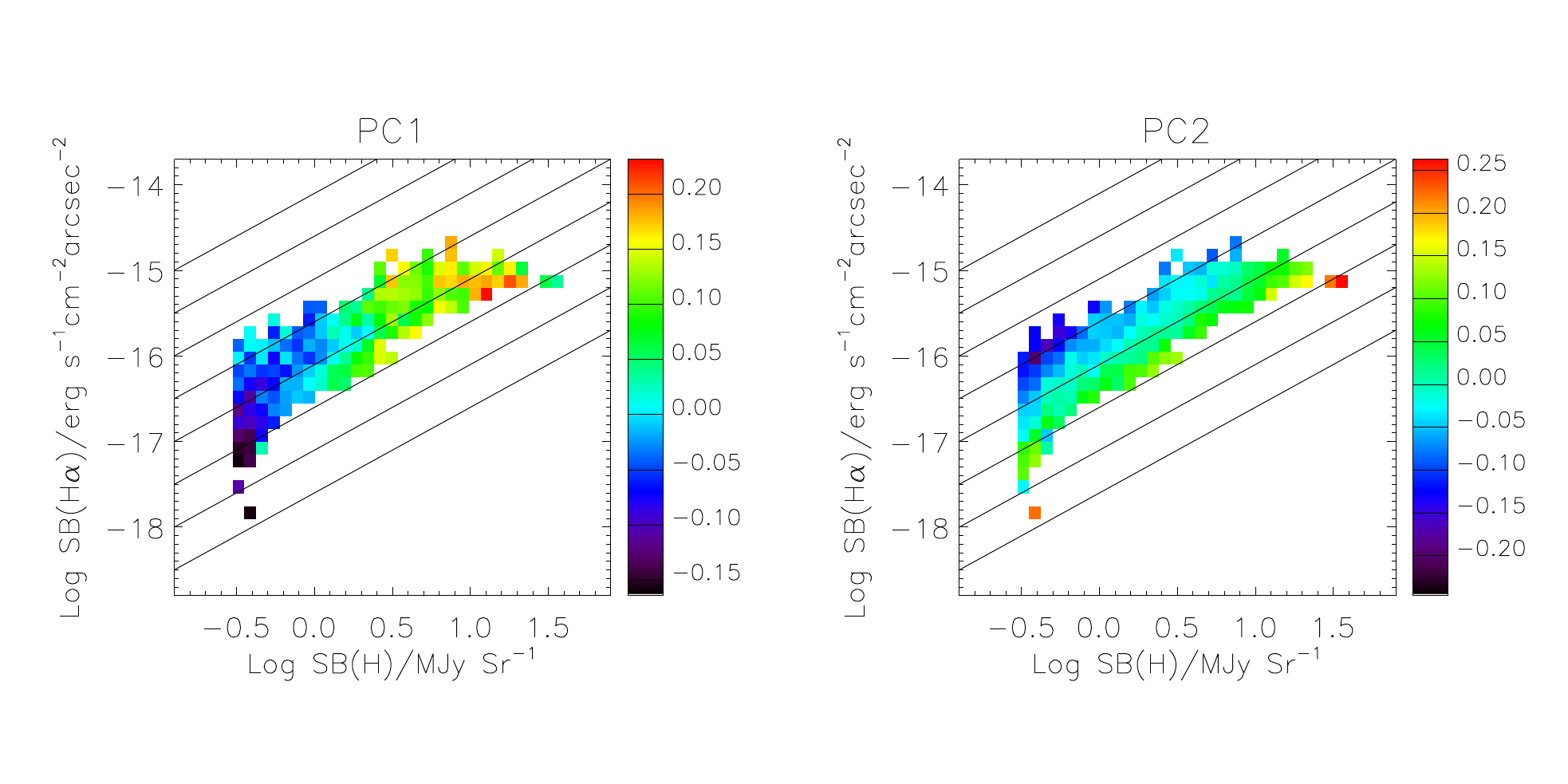}
\caption{As in Fig.\ref{fig_SBPC_N04254}, but for NGC\,3521.}\label{fig_SBPC_N03521}
\end{figure*}
\begin{figure*}
\includegraphics[width=\textwidth]{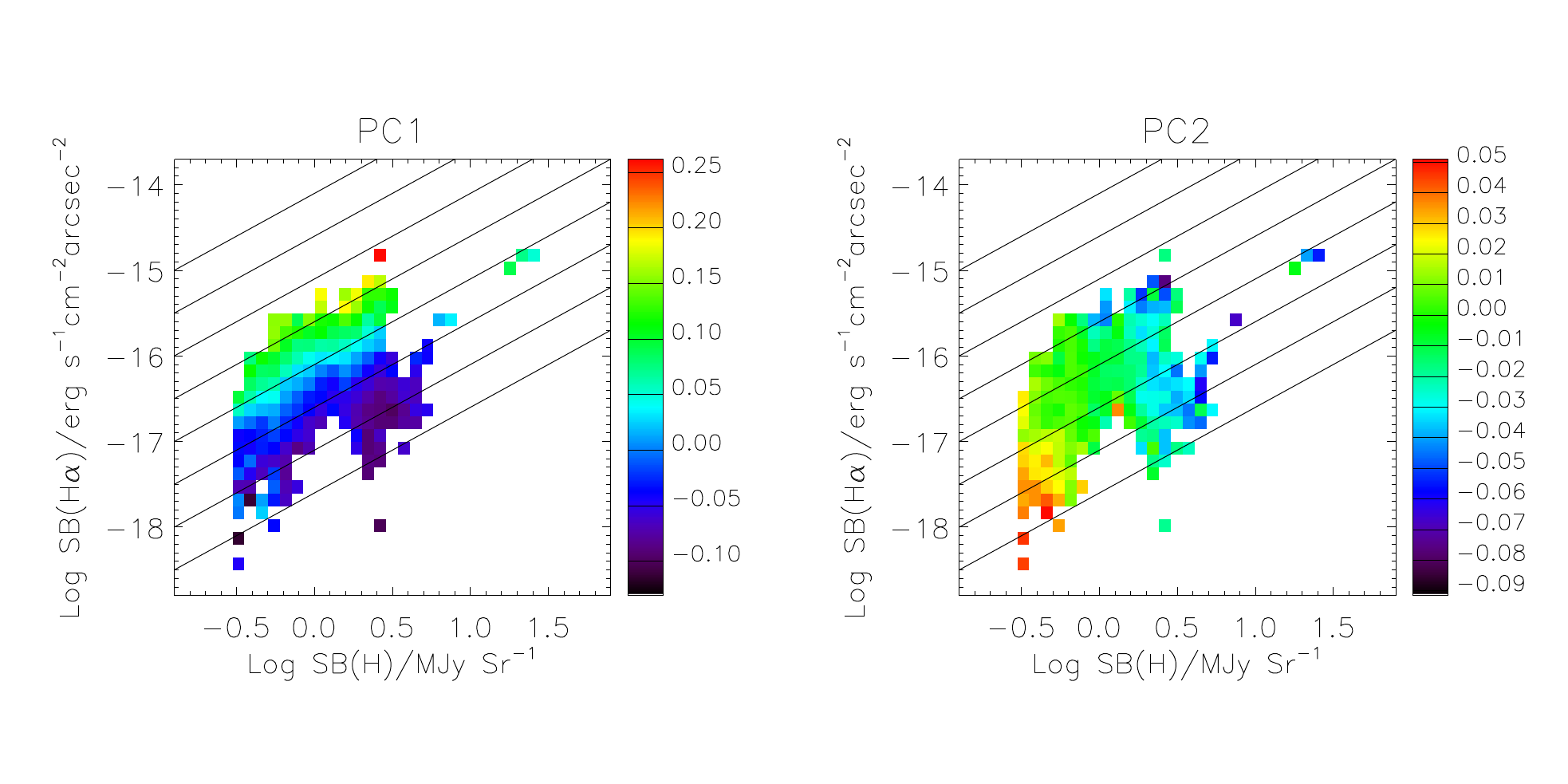}
\caption{As in Fig.\ref{fig_SBPC_N04254}, but for NGC\,4321.}\label{fig_SBPC_N04321}
\end{figure*}
\begin{figure*}
\includegraphics[width=\textwidth]{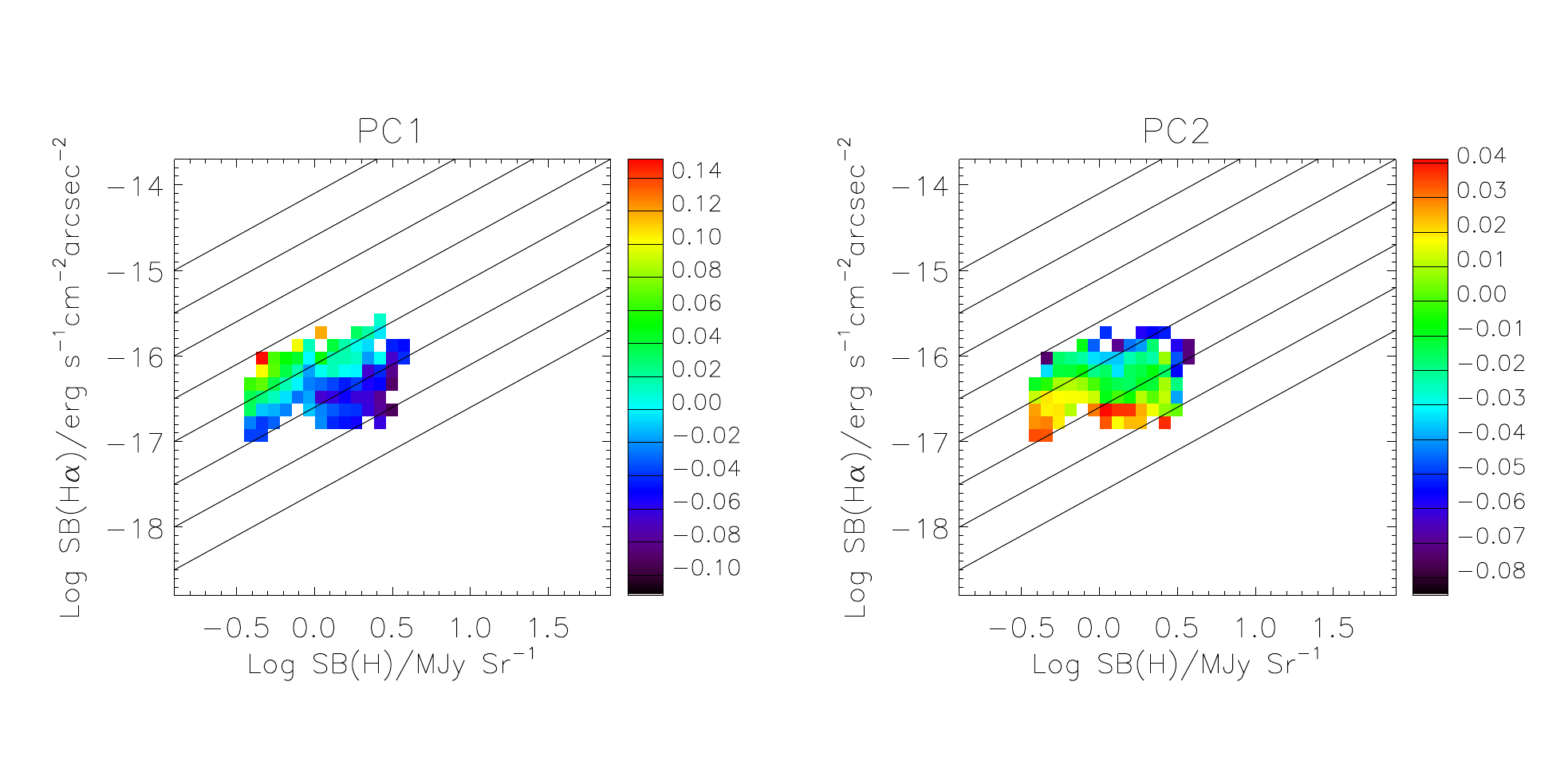}
\caption{As in Fig.\ref{fig_SBPC_N04254}, but for NGC\,4536. Note that
  in this case, contrary to the other four ``regular'' galaxies, PC1
  is mainly correlated to the optical colours and PC2 is
  anti-correlated with the IR colours. Modulo this exchange, main
  conclusions remain unaffected.}\label{fig_SBPC_N04536}
\end{figure*}
\begin{figure*}
\includegraphics[width=\textwidth]{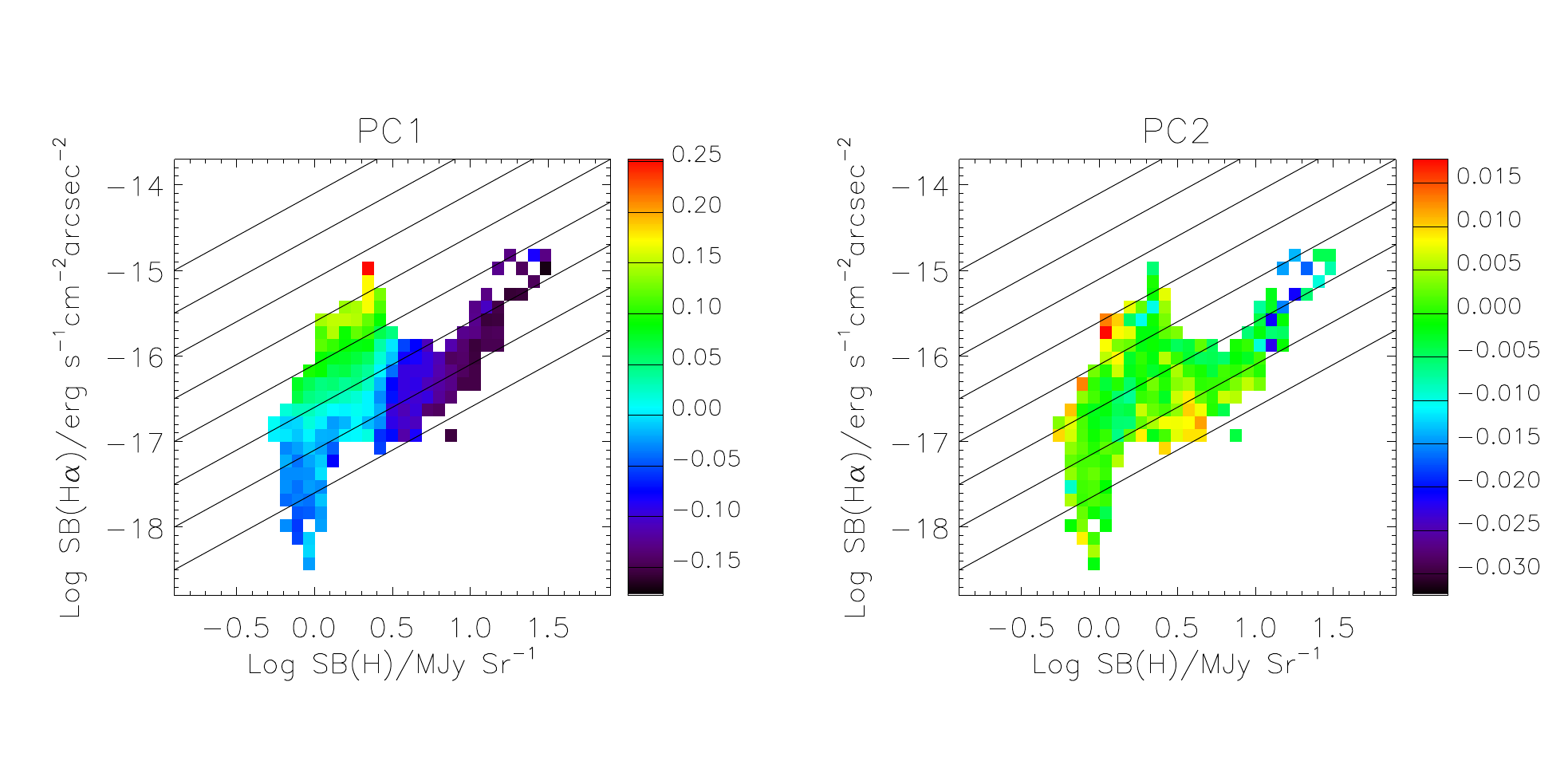}
\caption{As in Fig.\ref{fig_SBPC_N04254}, but for NGC\,4579.}\label{fig_SBPC_N04579}
\end{figure*}
\clearpage
\section{Synthetic IRAC images for the six galaxies not in the main
  paper}\label{append_IRACsynth}
\begin{figure*}
\includegraphics[width=\textwidth]{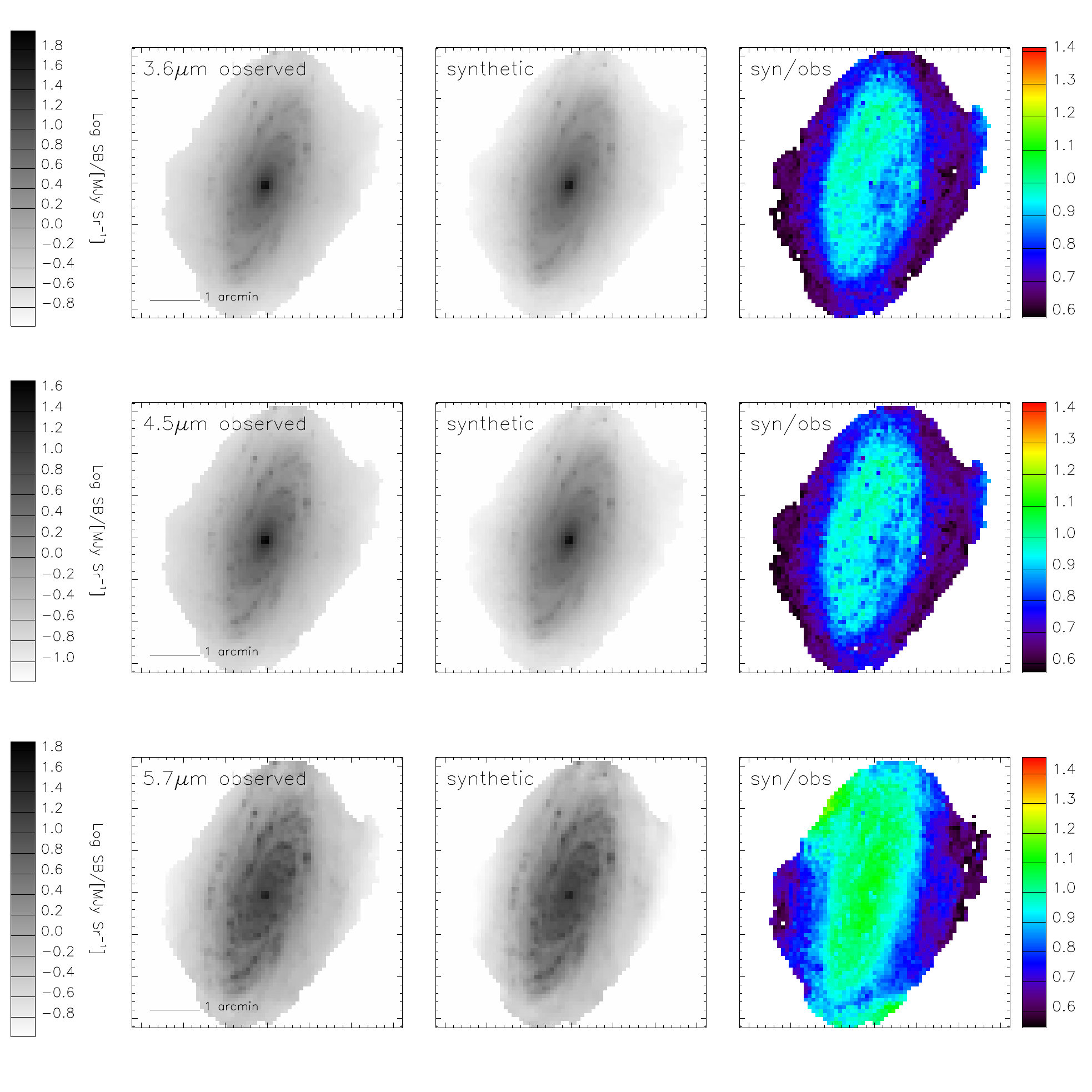}
\caption{As in Fig.\ref{fig_IRACsynth_N04254}, but for NGC\,3521.}\label{fig_IRACsynth_N03521}
\end{figure*}
\begin{figure*}
\includegraphics[width=\textwidth]{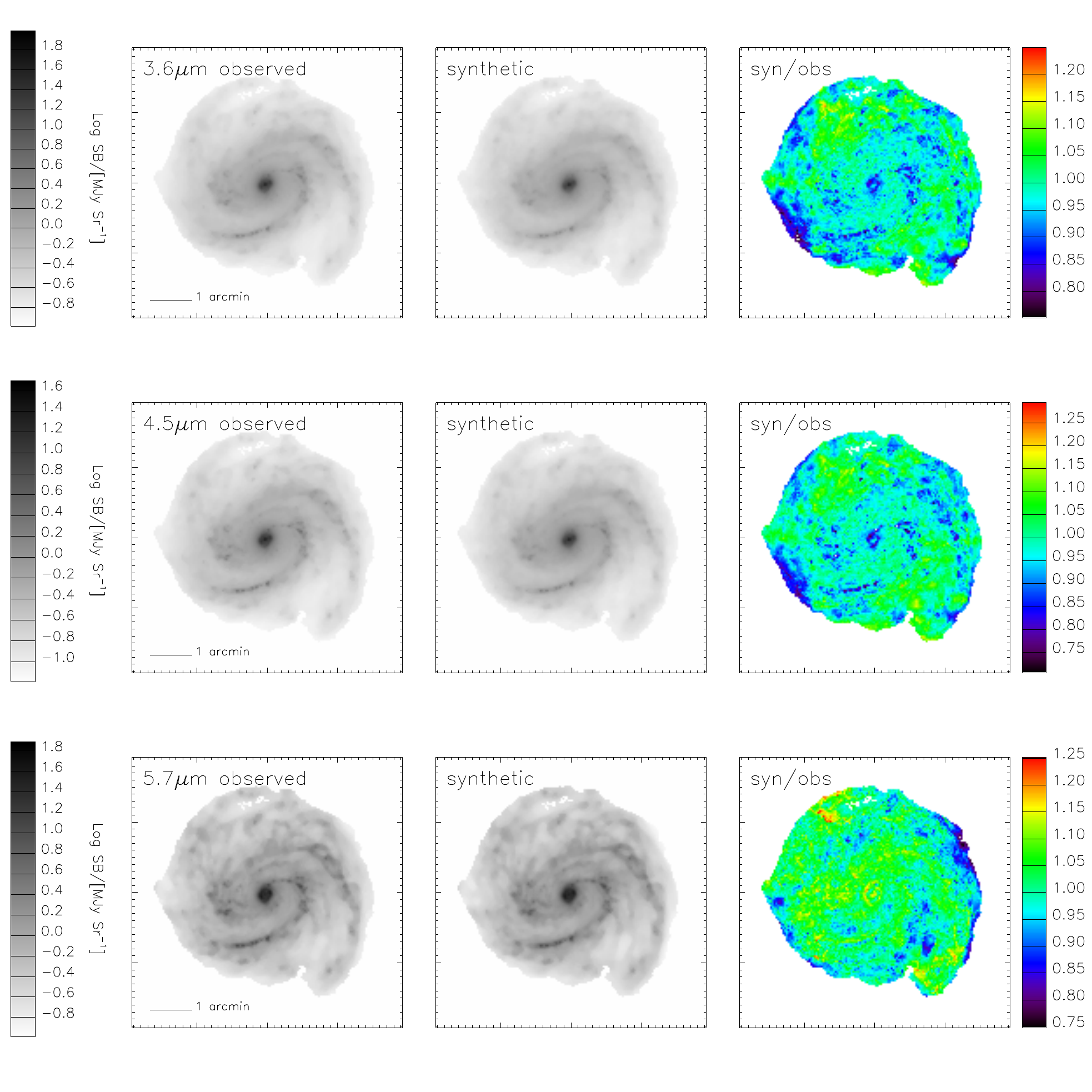}
\caption{As in Fig.\ref{fig_IRACsynth_N04254}, but for NGC\,4321.}\label{fig_IRACsynth_N04321}
\end{figure*}
\begin{figure*}
\includegraphics[width=\textwidth]{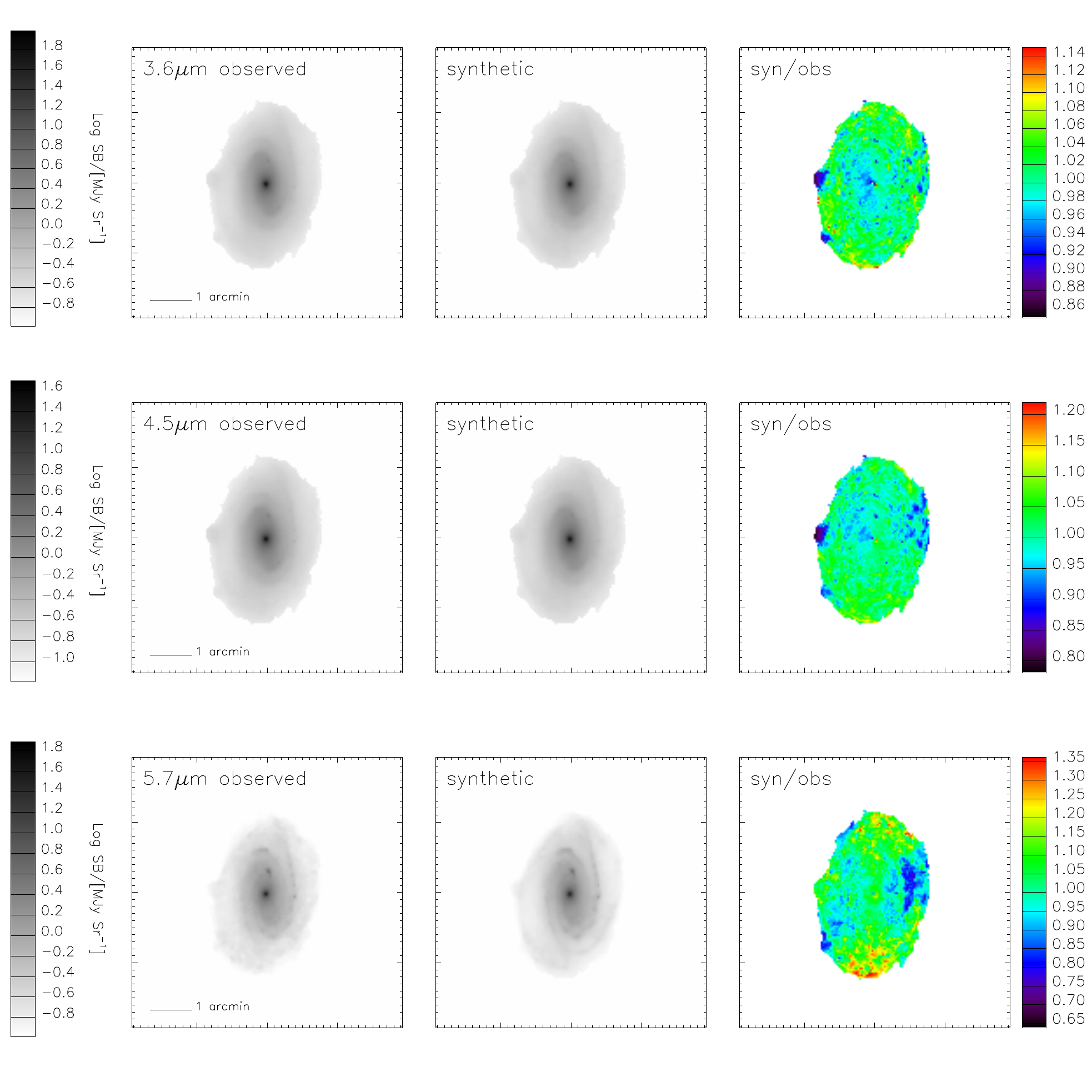}
\caption{As in Fig.\ref{fig_IRACsynth_N04254}, but for NGC\,4450.}\label{fig_IRACsynth_N04450}
\end{figure*}
\begin{figure*}
\includegraphics[width=\textwidth]{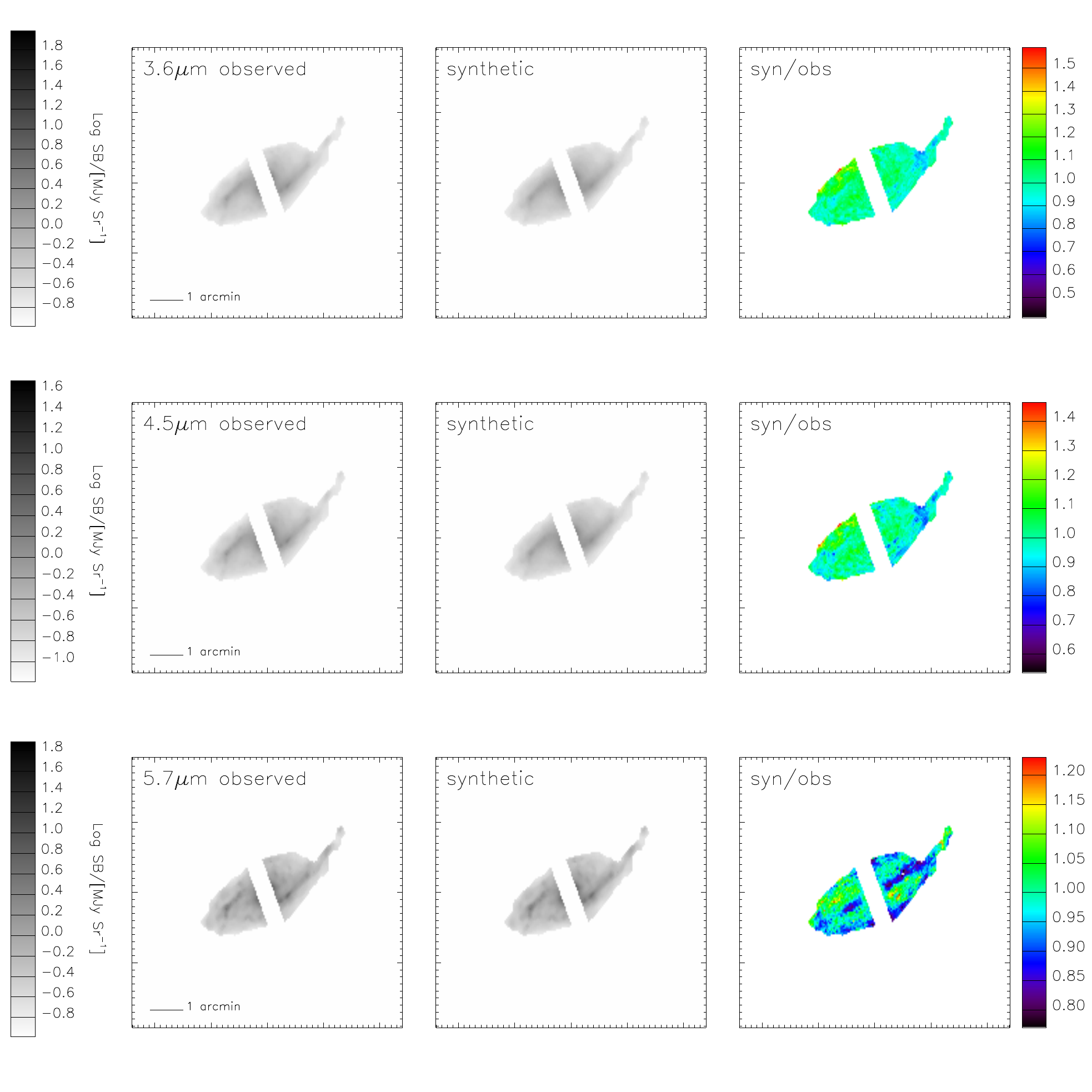}
\caption{As in Fig.\ref{fig_IRACsynth_N04254}, but for NGC\,4536.}\label{fig_IRACsynth_N04536}
\end{figure*}
\begin{figure*}
\includegraphics[width=\textwidth]{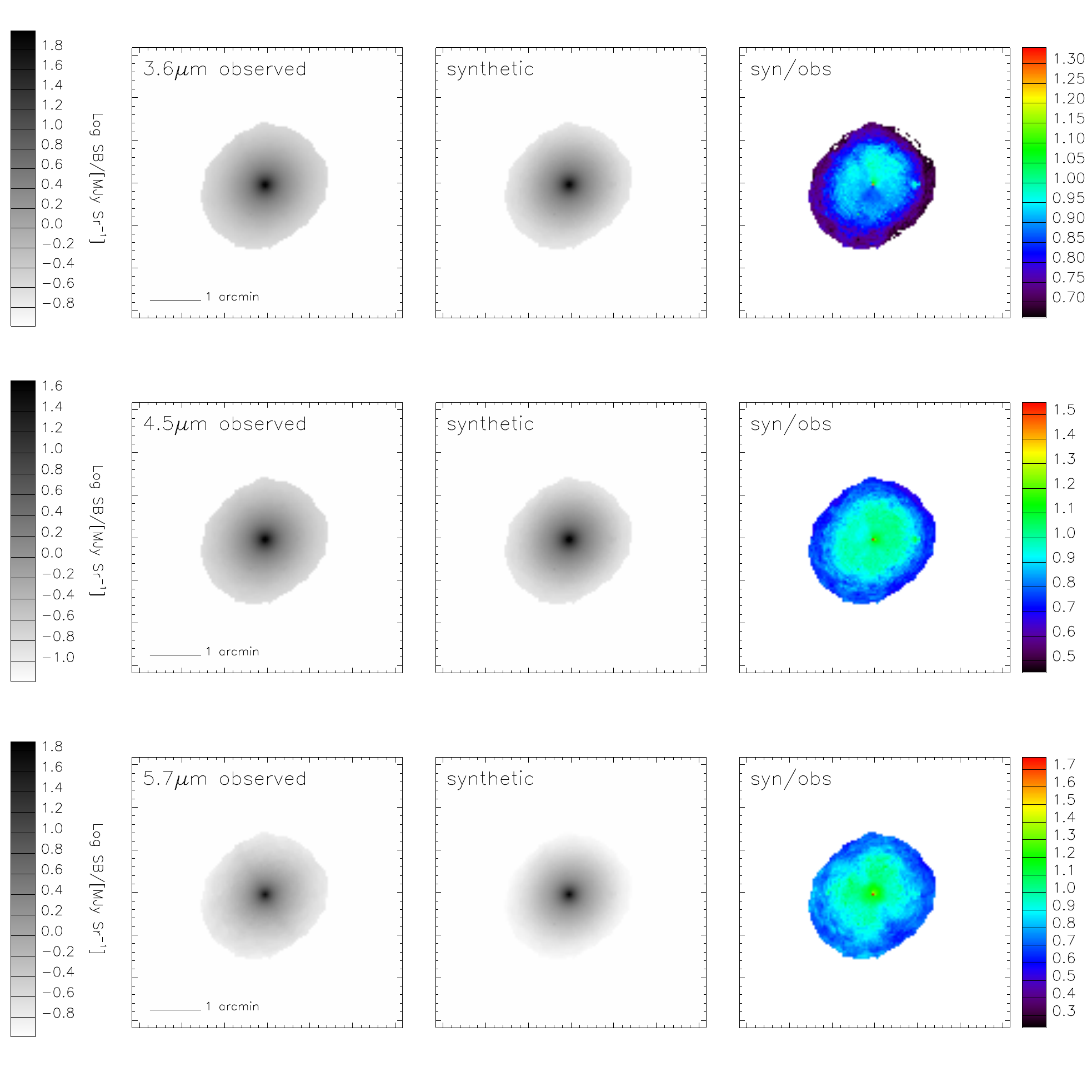}
\caption{As in Fig.\ref{fig_IRACsynth_N04254}, but for NGC\,4552.}\label{fig_IRACsynth_N04552}
\end{figure*}
\begin{figure*}
\includegraphics[width=\textwidth]{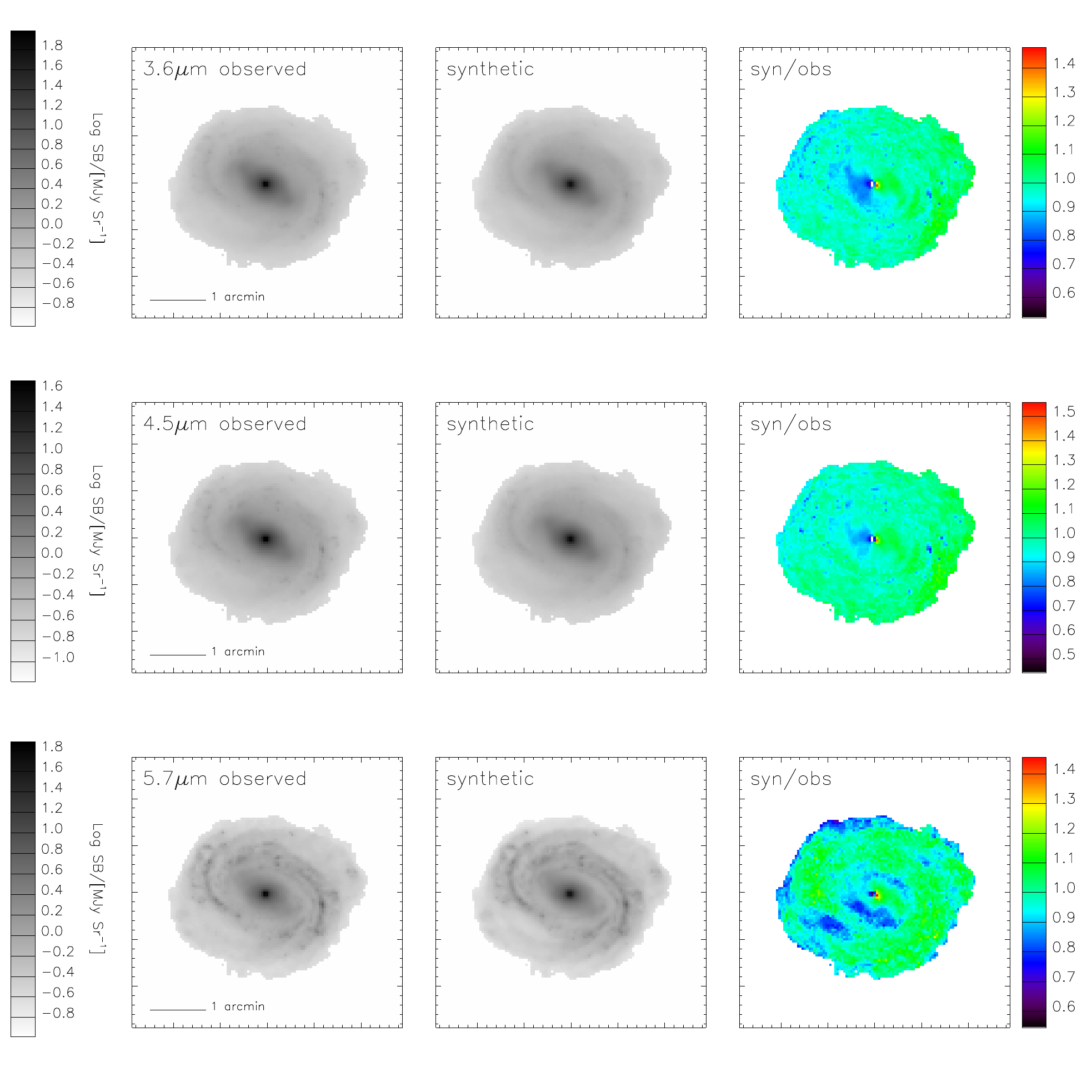}
\caption{As in Fig.\ref{fig_IRACsynth_N04254}, but for NGC\,4579.}\label{fig_IRACsynth_N04579}
\end{figure*}

\label{lastpage}
\end{document}